%% file: main.tex
\documentclass[onecolumn,a4paper,accepted=2026-09-09]{quantumarticle}
\include{preamble}
\include{preamble-thmstyles}

\title{Unitary causal decompositions: a characterisation via lattice theory}
\author{Tein van der Lugt}
\affiliation{Department of Computer Science, University of Oxford, Oxford, United Kingdom}
\affiliation{Perimeter Institute for Theoretical Physics, Waterloo, Ontario, Canada}
\email{teinvdlugt@gmail.com}
\author{Robin Lorenz}
\affiliation{Quantinuum, London, United Kingdom}
\email{robin.lorenz@quantinuum.com}

\begin{document}
    \maketitle

    \begin{abstract}
        \input{sections/abstract}
    \end{abstract}

    \input{sections/introduction}
    \input{sections/prelims}
    \input{sections/result-and-overview}
    \input{sections/canonical-circuit}
    \input{sections/sufficiency}
    \input{sections/discussion}

    \section*{Acknowledgements}
    We thank Augustin Vanrietvelde, Jonathan Barrett, Elie Wolfe, and two anonymous reviewers for helpful discussions and feedback.

    Perimeter Institute is situated on territory of the Anishinaabe, Haudenosaunee, and Neutral peoples.

    Research at Perimeter Institute is supported in part by the Government of Canada through the Department of Innovation, Science and Economic Development and by the Province of Ontario through the Ministry of Colleges, Universities, Research Excellence and Security.

    \section*{Author contributions}
    Both authors contributed significantly to research and writing.
    A first version of the main result was established during an initial phase of equal contribution; subsequently, TvdL was lead researcher and author.
    Large language models were used occasionally for specific research questions and none of the writing was produced by AI.

    \addcontentsline{toc}{section}{References}
    \printbibliography

    \appendix
    \crefalias{section}{appendix}  
    \input{sections/generic-channels}
    \input{sections/app-algebra}
    \input{sections/necessity}
    \input{sections/app-proof-of-soundness}
\end{document}

%% file: preamble.tex
\pdfoutput=1

\usepackage{amsmath}
\usepackage{amsthm}
\usepackage{thmtools}
\usepackage{thm-restate}
\usepackage{amsfonts}
\usepackage{amssymb}
\usepackage{bm}
\usepackage{mathtools}
\usepackage{mathrsfs}
\usepackage{stmaryrd}
\usepackage{physics}
\usepackage[dvipsnames]{xcolor}
\usepackage[pdfusetitle,bookmarksnumbered,hidelinks]{hyperref}
\usepackage[capitalise,nameinlink]{cleveref}
\usepackage[british]{babel}
\usepackage{textcomp}
\usepackage{float}
\usepackage{caption}
\usepackage{subcaption}
\usepackage{ragged2e}

\usepackage[normalem]{ulem}

\usepackage[utf8]{inputenc}
\DeclareFontSeriesDefault[rm]{bf}{b}
\newcommand{\defn}[1]{\textbf{#1}}
\usepackage{packages/tikzit}
\input{figures/tikzit.tikzstyles}
\usetikzlibrary{circuits.ee.IEC}
\usepackage{enumitem}
\setlist[enumerate,1]{label={\normalfont(\roman*)}}
\usepackage{graphicx}
\usepackage{titlesec}

\input{bibsetup}
\newcommand{\labelsA}{\mathtt{A}}
\newcommand{\labelsB}{\mathtt{B}}
\newcommand{\lA}{\mathtt{a}}
\newcommand{\lB}{\mathtt{b}}
\newcommand{\lE}{\mathtt{e}}
\newcommand{\lF}{\mathtt{f}}
\newcommand{\lZ}{\mathtt{Z}}
\newcommand{\lV}{\mathtt{V}}

\newcommand{\aA}{{\bm{A}}}
\newcommand{\aB}{{\bm{B}}}
\newcommand{\aX}{{\bm{X}}}
\newcommand{\aZ}{{\bm{Z}}}
\newcommand{\aW}{{\bm{W}}}
\newcommand{\aC}{{\bm{C}}}
\newcommand{\aD}{{\bm{D}}}
\newcommand{\aE}{{\bm{E}}}
\newcommand{\aF}{{\bm{F}}}
\newcommand{\aM}{{\bm{M}}}
\newcommand{\aN}{{\bm{N}}}

\newcommand{\csP}{\mathtt{C}}
\newcommand{\csQ}{\mathtt{D}}
\newcommand{\circuit}{{\bm{C}}}

\newcommand{\cP}{\mathcal{P}}
\let\a\alpha
\let\b\beta
\let\phi\varphi
\let\la\lambda

\renewcommand{\H}{\mathcal{H}}
\newcommand{\K}{\mathcal{K}}
\newcommand{\J}{\mathcal{J}}
\newcommand{\U}{{\mathcal{U}}}
\newcommand{\V}{\mathcal{V}}
\newcommand{\W}{\mathcal{W}}
\newcommand{\D}{\mathcal{D}}
\newcommand{\E}{\mathcal{E}}
\newcommand{\C}{\mathbb{C}}
\renewcommand{\L}{\mathcal{L}}  
\newcommand{\infl}{\rightarrow}
\newcommand{\ninfl}{\nrightarrow}
\newcommand{\centre}{\mathcal{Z}}
\DeclareMathOperator{\Pa}{Pa}
\DeclareMathOperator{\Ch}{Ch}

\DeclareMathOperator{\SC}{SP}

\DeclareMathOperator{\down}{\downarrow}  
\DeclareMathOperator{\strdown}{\downarrow^*}  
\DeclareMathOperator{\Covp}{Ch}
\DeclareMathOperator{\Covm}{Pa}

\newcommand{\<}{\langle}
\renewcommand{\>}{\rangle}
\newcommand{\cls}[1]{{\mathtt{B}_{#1}}}
\newcommand{\latt}[1]{{\mathtt{L}_{#1}}}

\newcommand{\lattK}{\latt{G}}
\newcommand{\closedA}{{\mathcal{P}^c(\labelsA)}}
\newcommand{\closedB}{{\mathcal{P}^c(\labelsB)}}
\DeclareMathOperator{\supp}{supp}
\let\join\vee

\let\bigjoin\bigvee
\let\bigmeet\bigwedge
\newcommand{\itmref}[1]{\text{{\normalfont\ref{#1}}}}
\newcommand{\cvr}{\mathrel{\lessdot}}
\DeclareMathOperator{\id}{id}
\newcommand{\cccU}{{\U_3}}
\newcommand{\cccG}{{C_3}}
\newcommand{\picG}{{G_1}}
\newcommand{\necessityG}{{C_3'}}
\newcommand{\nA}{{n}}
\newcommand{\nB}{{m}}
\newcommand{\PaG}{\PaGarg{G}}
\newcommand{\PaGarg}[1]{#1^{-1}}
\newcommand{\ChG}{G}
\newcommand{\gP}{\mathfrak{p}}
\newcommand{\gC}{\mathfrak{c}}

\newcommand{\rel}[1]{\mathrel{#1}}
\newcommand{\relG}{\rel{G}}

\newcommand{\uW}{\mathcal{W}}
\newcommand{\uX}{\mathcal{X}}
\newcommand{\uY}{\mathcal{Y}}
\newcommand{\uZ}{\mathcal{Z}}

\newcommand{\quadand}{\quad\text{and}\quad}
\newcommand{\qquadand}{\qquad\text{and}\qquad}

\newcommand{\unichan}{unitary channel}

\newcommand{\cstaralg}{$C^*$-algebra}  

\newenvironment{picc}[1][]{\begin{aligned}
                               \begin{tikzpicture}[font=\tiny,#1]}
                               {\end{tikzpicture}
\end{aligned}}

\newcommand{\discard}[1]{\ensuremath{\tinygroundnew_{#1}}}
\newcommand{\tinygroundnew}{\smash{{\hspace{-3pt}\ensuremath{\begin{picc}[scale=0.5]
                                                                 \node[upgroundsmall, xscale=0.8, yscale=0.7] (1) at (0,0.16) {}; \draw (0,0.03) to (0,-0.25);
\end{picc}}\hspace{-1pt}}}}

\tikzstyle{upgroundsmall}=[circuit ee IEC, thick, ground, rotate=90, scale=1.5, tikzit fill=black, tikzit shape=rectangle]

\newcommand{\vx}{\alpha}
\newcommand{\vxp}{\beta}
\newcommand{\vxpp}{\gamma}

\newcommand{\CEP}{\eqref{eq:C3EP}}
\newcommand{\CEPtext}{$\cccG$ exclusion property}
\newcommand{\PIC}{\CEP}

\let\tns\otimes
\newcommand{\tikzfigpad}[1]{\quad\input{#1.tikz}\quad} 
\newcommand{\smallcdots}{\scalebox{.7}{$\cdots$}}
\newcommand{\one}{\mathbf{1}}  
\newcommand{\incomp}{\mathrel{\#}}

%% file: figures/tikzit.tikzstyles
\tikzstyle{label}=[font={\footnotesize}, text height=1ex, text depth=0.15ex]
\tikzstyle{medium label}=[label, font={\normalsize}, scale=.8]
\tikzstyle{classical label}=[text=gray, font={\footnotesize}, text height=1ex, text depth=0.15ex, tikzit category=classical]
\tikzstyle{map}=[draw, shape=rectangle, inner sep=2pt, minimum height=5mm, fill=white, minimum width=5mm]
\tikzstyle{tiny map}=[draw, shape=rectangle, inner sep=1pt, minimum height=3mm, fill=white, minimum width=3mm, font={\footnotesize}]
\tikzstyle{small map}=[draw, shape=rectangle, inner sep=1pt, minimum height=4mm, fill=white, minimum width=4mm]
\tikzstyle{medium map}=[draw, shape=rectangle, inner sep=2pt, minimum height=5mm, fill=white, minimum width=9mm, tikzit fill=red]
\tikzstyle{large map}=[draw, shape=rectangle, inner sep=2pt, minimum height=5mm, fill=white, minimum width=18mm, tikzit fill=blue]
\tikzstyle{classical map}=[draw=gray, text=gray, shape=rectangle, inner sep=2pt, minimum height=5mm, fill=white, minimum width=5mm, tikzit fill=gray, tikzit category=classical]
\tikzstyle{medium classical map}=[draw=gray, text=gray, shape=rectangle, inner sep=2pt, minimum height=5mm, fill=white, minimum width=9mm, tikzit fill=gray, tikzit category=classical]
\tikzstyle{large classical map}=[draw=gray, text=gray, shape=rectangle, inner sep=2pt, minimum height=5mm, fill=white, minimum width=18mm, tikzit fill=gray, tikzit category=classical]
\tikzstyle{upground}=[circuit ee IEC, thick, ground, rotate=90, scale=1.5, tikzit fill=black, tikzit shape=rectangle]
\tikzstyle{downground}=[circuit ee IEC, thick, ground, rotate=-90, scale=1.5, tikzit fill=black, tikzit shape=rectangle]
\tikzstyle{downgroundnorm}=[circuit ee IEC, thick, ground, rotate=-90, scale=1.5, fill=white, tikzit shape=rectangle, tikzit fill=red]
\tikzstyle{state}=[fill=white, draw, shape=isosceles triangle, shape border rotate=-90, isosceles triangle stretches=true, inner sep=0.2pt, minimum height=4mm, yshift=-0.0mm, tikzit shape=rectangle, minimum width=5mm, font={\footnotesize}]
\tikzstyle{copoint}=[fill=white, draw, shape=isosceles triangle, shape border rotate=90, isosceles triangle stretches=true, inner sep=0.2pt, minimum width=0.5cm, minimum height=0.8mm, yshift=-0.0mm, tikzit shape=rectangle, minimum width=0.5cm]
\tikzstyle{wide point}=[fill=white, draw, shape=isosceles triangle, shape border rotate=-90, isosceles triangle stretches=true, inner sep=0.2pt, minimum height=0.8mm, yshift=-0.0mm, minimum width=12mm, tikzit shape=rectangle, tikzit fill=red]
\tikzstyle{wide copoint}=[fill=white, draw, shape=isosceles triangle, shape border rotate=90, isosceles triangle stretches=true, inner sep=0.2pt, minimum width=0.5cm, minimum height=0.8mm, yshift=-0.0mm, tikzit shape=rectangle, minimum width=12mm, tikzit fill=red]
\tikzstyle{copy}=[circle, draw=gray, inner sep=.4ex, fill=white, tikzit category=classical]
\tikzstyle{decomp}=[fill=white, draw, shape=isosceles triangle, shape border rotate=-90, isosceles triangle stretches=true, inner sep=0pt, minimum width=0.75cm, minimum height=4mm, yshift=-0.0mm, tikzit shape=rectangle]
\tikzstyle{decompwide}=[fill=white, draw, shape=isosceles triangle, shape border rotate=-90, isosceles triangle stretches=true, inner sep=0pt, minimum width=1.5cm, minimum height=4mm, yshift=-0.0mm, tikzit shape=rectangle, tikzit fill=red]
\tikzstyle{decompflip}=[fill=white, draw, shape=isosceles triangle, shape border rotate=90, isosceles triangle stretches=true, inner sep=0pt, minimum width=0.75cm, minimum height=4mm, yshift=-0.0mm, tikzit shape=rectangle]
\tikzstyle{decompwideflip}=[fill=white, draw, shape=isosceles triangle, shape border rotate=90, isosceles triangle stretches=true, inner sep=0pt, minimum width=1.5cm, minimum height=4mm, yshift=-0.0mm, tikzit shape=rectangle, tikzit fill=red]
\tikzstyle{graph placeholder}=[fill=white, draw=gray, shape=rectangle, rounded corners, minimum width=40mm, minimum height=15mm, dashed, thick]
\tikzstyle{small black dot}=[fill=black, draw=black, shape=circle, inner sep=.2ex]
\tikzstyle{small label}=[label, font={\normalsize}, scale=0.6]
\tikzstyle{cnot1}=[fill=black, draw=black, shape=circle, inner sep=.25ex]
\tikzstyle{cnot2}=[fill=none, draw=black, shape=circle, minimum size=7, inner sep=0pt, outer sep=0pt, tikzit fill=none]

\tikzstyle{dir}=[->]
\tikzstyle{graydash}=[-, dashed, draw=gray]
\tikzstyle{triangle}=[->, style={{-{Triangle[open]}}}]
\tikzstyle{norm}=[-, dashed, color=blue, draw={rgb,255: red,51; green,0; blue,255}]
\tikzstyle{maptostyle}=[{|->}]
\tikzstyle{dashdir}=[->, dashed]
\tikzstyle{thickline}=[-, thick]
\tikzstyle{FCMBox}=[-, draw={rgb,255: red,191; green,191; blue,191}, thick]
\tikzstyle{green line}=[-, draw={rgb,255: red,34; green,139; blue,34}]
\tikzstyle{classical}=[-, draw={rgb,255: red,191; green,191; blue,191}]
\tikzstyle{PBT}=[-, thick, color=olive]

%% file: bibsetup.tex
\usepackage[style=numeric-comp, 
    sortcites=true,
    useprefix=true,
    sorting=none,
    maxbibnames=9,
    doi=false,isbn=false,url=false,eprint=true]{biblatex}

\DeclareSortingNamekeyTemplate{
    \keypart{\namepart{family}}
    \keypart{\namepart{prefix}} 
    \keypart{\namepart{given}}
    \keypart{\namepart{suffix}}
}
\renewbibmacro{begentry}{\midsentence}  

\newbibmacro{string+doiurl}[1]{
  \iffieldundef{doi}{%
    \iffieldundef{url}{%
      #1
    }{%
      \href{\thefield{url}}{#1}%
    }%
  }{%
    \href{http://dx.doi.org/\thefield{doi}}{#1}%
  }%
}
\DeclareFieldFormat{title}{\usebibmacro{string+doiurl}{\mkbibemph{#1}}}
\DeclareFieldFormat[thesis]{title}{\usebibmacro{string+doiurl}{\mkbibemph{#1}}} 
\DeclareFieldFormat[article,misc]{title}{\usebibmacro{string+doiurl}{\mkbibquote{#1}}}
\DeclareFieldFormat{titlecase}{#1} 

\AtEveryBibitem{\clearfield{month}} 
\AtEveryBibitem{\clearfield{day}} 
\AtEveryBibitem{\clearfield{urlday}} 
\AtEveryBibitem{\clearfield{urlmonth}} 
\AtEveryBibitem{\clearfield{urlyear}} 

\DeclareFieldFormat{postnote}{#1} 
\renewbibmacro{in:}{\ifentrytype{article}{}{\printtext{\bibstring{in}\intitlepunct}}} 

%% file: preamble-thmstyles.tex
\theoremstyle{definition}
\declaretheorem[name=Definition, refname={Definition,Definitions}, Refname={Definition,Definitions}, qed=$\sslash$]{definition}
\declaretheorem[name=Example, refname={Example,Examples}, Refname={Example,Examples}, qed=$\sslash$, numberlike=definition]{example}
\declaretheorem[name=Remark, refname={Remark,Remarks}, Refname={Remark,Remarks}, numbered=yes, qed=$\sslash$, numberlike=definition]{remark}

\theoremstyle{plain}
\declaretheorem[name=Theorem, refname={Theorem,Theorems}, Refname={Theorem,Theorems}, numberlike=definition]{theorem}
\declaretheorem[name=Proposition, refname={Proposition,Propositions}, Refname={Proposition,Propositions}, numberlike=definition]{proposition}
\declaretheorem[name=Lemma, refname={Lemma,Lemmas}, Refname={Lemma,Lemmas}, numberlike=definition]{lemma}

\counterwithin{definition}{section}
\counterwithin{example}{section}
\counterwithin{remark}{section}
\counterwithin{theorem}{section}
\counterwithin{proposition}{section}
\counterwithin{lemma}{section}
\counterwithin{conjecture}{section}
\counterwithin{corollary}{section}

\newenvironment{innerproof}{\proof}{\endproof} 

%% file: sections/abstract.tex
If a unitary transformation has a circuit representation with no directed path from input $\lA$ to output $\lB$, then $\lA$ does not influence $\lB$ through the overall unitary.
Conversely, if a unitary satisfies a number of no-influence conditions, it is natural to wonder whether a circuit decomposition exists in which all of them are represented by absences of paths.
Such decompositions are known as \emph{causal decompositions}; determining their existence in general is a central open problem in the study of causal structure in quantum theory.
We present progress towards a general solution by considering the special case of \emph{unitary} causal decompositions, i.e.\ decompositions in terms of unitary circuits in the traditional quantum circuit formalism that do not require the generalisation to `extended' or `routed' quantum circuits prompted by earlier research on this topic.
We identify a combinatorial condition that characterises precisely those sets of no-influence constraints $G$ for which any unitary transformation satisfying $G$ admits a unitary causal decomposition representing the constraints.
Our approach, grounded in lattice theory and finite-dimensional operator algebra, is systematic and offers hope for extensions to more general (e.g.\ routed unitary) causal decompositions in the future.


%% file: sections/introduction.tex
\section{Introduction} \label{sec:intro}
The tension between quantum correlations and classical causal explanations, exposed by Bell's theorem, has in recent decades led to a growing literature on causal structure in quantum theory and the potential for quantum causal explanations (see for instance~\cite{WS15,ABH17,BLO19,CS16,HLP14}).
Throughout this literature one encounters two qualitatively distinct kinds of structure imposed on quantum data.
One is top-down in nature, constraining global properties of quantum states or processes: think of no-signalling or no-influence constraints on quantum dynamics, commutation between subalgebras, or statistical independence relationships.
The other is instead bottom-up and compositional, describing how a state or process may be built out of local subprocesses composed into a circuit or network.
A full understanding of causation in quantum theory requires a grasp of the relation between these two types of structural constraints.
This relationship has been investigated in a number of works (see e.g.~\cite{BGNP01,ESW02,SW05,ANW11,SW04,AFNT11,SL13,LB21,VMA25b,vanderlugt2026causalcompositional}); one approach, which we develop further in this work, lies in the study of \emph{causal decompositions} of unitary transformations~\cite{LB21,VKB21,VMA25b,vdL25,vanderlugt2026causalcompositional}.

For a unitary transformation $\U$ whose input and output systems are labelled by the elements of finite sets $\labelsA$ and $\labelsB$, respectively, we will let the \emph{causal structure} of $\U$ refer to a property of the top-down type: it is the binary relation $G_\U\subseteq\labelsA\times\labelsB$ that specifies which outputs $\lB\in\labelsB$ are influenced by (i.e.\ are dynamically dependent on) which inputs $\lA\in\labelsA$ given the dynamics $\U$.
A quantum circuit $\circuit$ with the same inputs and outputs defines, on the other hand, a \emph{connectivity} relation $G_\circuit\subseteq\labelsA\times\labelsB$, indicating the existence of directed paths connecting inputs to outputs.
It encodes one aspect of the circuit's \emph{compositional} structure.
Causal and compositional structure are related in one obvious way: the presence of a causal influence through $\U$ requires a path through its circuit representation that can mediate it.
That is, if $\U$ admits a decomposition into a circuit $\circuit$, then $G_\U\subseteq G_\circuit$.

Conversely, as was shown in Refs.~\cite{ESW02,SW05}, every unitary transformation that satisfies a \mbox{\emph{no}-influence} constraint of the form ${(\lA,\lB)\notin G_\U}$ admits a circuit decomposition $\circuit$ representing that property by the \emph{absence} of a mediating path, i.e.\ a circuit that satisfies ${(\lA,\lB)\notin G_\circuit}$.
A natural question is whether also multiple no-influence constraints can be represented at once in the connectivity of a single circuit representation.
More precisely, if for a given unitary $\U$ and binary relation $G\subseteq\labelsA\times\labelsB$ the causal constraint $G_\U\subseteq G$ is satisfied, then it is appropriate to ask whether a circuit decomposition $\circuit$ of $\U$ exists satisfying $G_\circuit \subseteq G$.
In this case, the circuit $\circuit$ may be said to \emph{represent} the causal constraint $G_\U\subseteq G$, since the latter is implied by the chain of inclusions $G_\U \subseteq G_\circuit \subseteq G$.
In other words, $\circuit$ makes the no-influence conditions expressed by $G_\U \subseteq G$ directly apparent by the absence of paths.
We call circuits achieving such a representational task \emph{causal decompositions}; they form a bridge between causal and compositional structure.

The existence of causal decompositions was first addressed in generality in Ref.~\cite{LB21}, where it was shown for a handful of relations $G\subseteq\labelsA\times\labelsB$ that \emph{any} unitary $\U$ satisfying the causal condition $G_\U\subseteq G$ indeed admits a circuit decomposition with connectivity $G$.%
\footnote{Most results in Ref.~\cite{LB21} were in fact stated in terms of \emph{causally faithful circuit decompositions}, meaning they showed the existence of decompositions of unitaries saturating the causal constraint, i.e.\ satisfying $G_\U = G$. Their proofs however only relied on \emph{no}-influence conditions, not on \emph{influence} conditions, and therefore also go through for unitaries satisfying the constraint $G_\U \subseteq G$.
}
We will call such relations $G$ \emph{(compositionally) representable}.
Moreover, in Ref.~\cite{LB21} the shape of the circuits was uniquely determined by $G$ and independent of the unitary $\U$.
This type of result finds many applications (see~\cite{vanderlugt2026causalcompositional} for more details on some of these).
First of all, compositional structure can act as a powerful handle in proofs; and causal decompositions give access to this tool even when the initial assumptions are in terms of causal structure alone (see e.g.~\cite{BLO21, OVB23, VOKB22}).
Second, results of the type above offer a constructive description of the entire class of unitaries $\U$ with causal structure $G_\U\subseteq G$, parametrising it by the gates and wires in the circuit.
These parameters may be independently varied without leaving the class; in this sense they constitute a generalisation of the \emph{autonomous causal mechanisms} from classical causal modelling~\cite{Pearl09} and thereby find an important application
to the quantum causal modelling approach based on unitary causal influences developed in~\cite{ABH17,BLO19,BLO21}.
Third, causal decompositions are instrumental in comparing the latter with other approaches to quantum causal modelling such as those of Refs.~\cite{HLP14,Fritz16,CS16,Chav+15}, which instead take compositional structure as their starting point.

A fourth application lies in relativistic quantum information, where the existence of causal decompositions can be leveraged to prove equivalence between the (top-down) condition of having \emph{no superluminal influences} and the (bottom-up) \emph{realisability} of multipartite quantum channels in a spacetime context.
It is this connection to relativistic causality that in fact sparked the first results in the spirit of causal decompositions in Refs.~\cite{BGNP01,ESW02,SW05}, where special cases of these two conditions were called \emph{(semi)causality} and \emph{(semi)localisability}, respectively.
The general case and the precise role played by causal decompositions is investigated in detail in Refs.~\cite{vanderlugt2026causalcompositional,losec-pirsa} and upcoming work.
Finally, another application of causal decompositions arises from the perspective of unitary channels not as transformations of physical systems over time but as passive transformations (\mbox{*-isomorphisms}) relating different ways of partitioning the same overall quantum system ({\cstaralg}) into subsystems.
From this viewpoint, causal decompositions clarify how one such partitioning can be obtained from another by a sequence of fine-grainings and coarse-grainings~\cite{VMA25a,VMA25b}.

Despite this wealth of applications, the general existence of causal decompositions remains unknown.
For many of the motivations mentioned above it is natural, or even essential, to require that the gates in the causal decompositions themselves be unitary, too; we refer to circuits satisfying this requirement as \emph{unitary causal decompositions}.
Ref.~\cite{LB21} showed that not all causal conditions $G$ are representable in terms of unitary causal decompositions.
The authors addressed this issue by introducing a generalised kind of unitary quantum circuit that incorporates direct-sum structure and non-factor algebras.
This has been developed into the framework of \emph{routed quantum circuits}~\cite{VKB21}, which has since found significant relevance outside the study of causal decompositions as well.
As far as causal decompositions are concerned, strictly more unitary transformations admit them in terms of \emph{routed unitary circuits} than in terms of traditional unitary circuits~\cite{LB21}.

The focus of this work will, however, be on decompositions into traditional, non-routed unitary circuits---i.e.\ those whose wires represent tensor-product factors of Hilbert spaces and whose gates are unitary transformations between them.
Our main result is the identification of a combinatorial condition on a relation $G\subseteq\labelsA\times\labelsB$---the \emph{\CEPtext}---which we prove is satisfied if and only if $G$ is representable in (non-routed) unitary circuits; that is, if and only if \emph{every} unitary $\U$ satisfying $G_\U \subseteq G$ admits a non-routed unitary circuit decomposition $\circuit$ that has connectivity $G_\circuit \subseteq G$.
The {\CEPtext} amounts to the statement that $G$ restricts nowhere to $\cccG$, a particular relation between three inputs and three outputs.
We make use of the `concept lattice' construction studied in formal concept analysis~\cite{Birk67, GW24} that was shown in Ref.~\cite{vdL25} to provide a canonical shape $\latt{G}$ for any circuit with connectivity relation $G_\circuit \subseteq G$.
We also show that the {\CEPtext} can be characterised in terms of this canonical shape through various insightful syntactic and diagrammatic properties---in particular, it corresponds to the requirement that there be no more than one path between any given input and output.

The remainder of this paper is structured as follows.
\Cref{sec:prelims} introduces causal structure $G_\U$ and the connectivity relation $G_\circuit$ using the order-theoretic formalism for circuit syntax from Ref.~\cite{vdL25} and recalls existing results about causal decompositions.
In \cref{sec:main-result} we state our main result, \cref{thm:main-thm}, and outline its proof.
\Cref{sec:canonical-circuit} deals with the purely syntactical, combinatorial aspects of the proof, recalling the construction of the concept lattice $\latt{G}$, the canonical shape for causal decompositions, and characterising the {\CEPtext} in terms of it (\cref{thm:pic-equivalent-conditions}).
\Cref{sec:sufficiency} combines these combinatorial results with relevant operator-algebraic ones, proven in \cref{app:algebra}, in order to show \emph{sufficiency} of the {\CEPtext} for representability in unitary circuits (the most important direction of our main theorem): that is, if $G$ satisfies the {\CEPtext}, then any unitary transformation $\U$ satisfying the causal constraint $G_\U \subseteq G$ admits a unitary causal decomposition of the canonical shape that represents the constraint.
Causal constraints that fail the {\CEPtext} may still be representable in terms of \emph{routed} unitary circuits as in Ref.~\cite{LB21}, and we believe that the lattice-theoretic approach employed in this work may in the future offer insights into the existence of those more general decompositions as well.
Finally, \cref{sec:discussion} summarises and discusses our results; \cref{app:generic-channels} explains the reason that this work focusses on unitary transformations $\U$, rather than generic quantum channels; and \cref{app:faithful} makes the connection to the related notion of causally \emph{faithful} circuit decompositions.

%% file: sections/prelims.tex
\section{Preliminaries and background}\label{sec:prelims}

\subsection{Quantum systems and transformations}\label{subsec:prelims-algebra}
We start with some terminology and notation.
By a \defn{(quantum) system} we mean the full algebra $\aA = \L(\H_\aA)$ of operators on some positive- and finite-dimensional Hilbert space $\H_\aA$.
The identity operator is denoted by $\one_\aA$, and the commutant of a subset $\aX\subseteq\aA$ is the subalgebra
\begin{equation}
    \aX' \coloneqq \{a\in\aA \mid \forall x\in\aX: ax = xa\}.
\end{equation}
Given quantum systems $\aA_1$ and $\aA_2$, we will often denote the composite system $\aA_1\tns\aA_2 = \L(\H_{\aA_1}) \tns \L(\H_{\aA_2}) \cong \L(\H_{\aA_1}\tns\H_{\aA_2})$ by $\aA_1\aA_2$, and, when clear from context, we will abuse notation by letting $\aA_1$ also denote the subalgebra $\aA_1\tns\{\one_{\aA_2}\} = \{a_1 \tns \one_{\aA_2} \mid a_1 \in \aA_1\} \subseteq \aA_1\aA_2$.
Adopting this notation, we have $\aA_1' = \aA_2$.

We regard transformations in the Schrödinger picture: thus, \defn{(quantum) channels} are linear maps $\E: \aA \to \aB$ between operator algebras $\aA = \L(\H_\aA)$ and $\aB = \L(\H_\aB)$ that are completely positive and trace-preserving ($\Tr_\aB \circ\, \E = \Tr_\aA$).
The latter condition is equivalent to unitality of the Hilbert-Schmidt adjoint $\E^\dagger : \aB \to \aA$ (that is, that $\E^\dagger(\one_\aB) = \one_\aA$), which would describe the channel in the Heisenberg picture instead.
We call $\E:\aA\to\aB$ \defn{unitary} if it is in fact a \mbox{*-isomorphism} of the operator algebras; this is the case precisely if it is of the form $U(\cdot)U^*$ for some unitary map $U:\H_\aA\to\H_\aB$, where $U^*$ is its adjoint with respect to the inner products on $\H_\aA$ and $\H_\aB$ (see \cref{prop:isomorphisms-are-unitaries}).

Throughout the paper we will strive to keep a healthy separation between \emph{syntactic} notions (such as labels of quantum systems, relations between labels, shapes of circuits) and \emph{semantic} ones (such as quantum systems, unitary channels, commutation of subalgebras).
The former are combinatorial objects, the latter operator-algebraic.
We will often use a typewriter font for symbols relating to syntactic notions and a bold font for semantic ones.
On the syntactic side, our general setup will consist of two finite sets $\labelsA$ and $\labelsB$, whose elements should be viewed as labels of input and output quantum systems, respectively.
Semantically, meanwhile, we consider quantum systems $\aA_\lA$ labelled by $\lA\in\labelsA$ and $\aB_{\lB}$ labelled by $\lB\in\labelsB$.
Given such a choice of quantum systems and a subset $\a\subseteq\labelsA$, we write
\begin{equation}
    \aA_\a \coloneqq \bigotimes_{\lA\in\a} \aA_\lA, \qquad\text{and thus}\qquad \aA_\labelsA \coloneqq \bigotimes_{\lA\in\labelsA} \aA_{\lA};
\end{equation}
$\aB_\b$ and $\aB_\labelsB$ are defined similarly.
In particular, $\aA_\emptyset = \aB_\emptyset = \mathbb C$, the trivial quantum system.
As above, we will often use the same notation $\aA_\a$ and $\aB_\b$ to refer instead to the subalgebras $\aA_{\a}\tns\{\one_{\aA_{\labelsA\setminus\a}}\} \subseteq \aA_\labelsA$ and $\aB_{\b}\tns\{\one_{\aB_{\labelsB\setminus\b}}\} \subseteq \aB_\labelsB$, respectively, whose commutants are then given by
\begin{equation}
    \aA_\a' = \aA_{\labelsA\setminus\a} \qquadand \aB_\b' = \aB_{\labelsB\setminus\b}.
\end{equation}

Our main objects of study are the multi-input, multi-output {\unichan}s $\U : \aA_\labelsA \to \aB_\labelsB$.

\subsection{Causal structure} \label{subsec:causal-structure}
The term \emph{causal structure} is used for many distinct concepts across the quantum causality literature.
In this work, we follow Refs.~\cite{ABH17,BLO19,BLO21,LB21,OVB23} and let the term refer to the collection of influence relations between a unitary's global inputs and outputs.
Other works use it to refer to a circuit or network notion of structure~\cite{HLP14,Fritz16,CS16} that we instead call \emph{compositional structure} and which is the subject of \cref{subsec:prelims-compositional-structure}.

\begin{definition}
    \label{def:causal-influence}
    Let $\U: \aA_\labelsA \to \aB_\labelsB$ be a {\unichan} and $\a\subseteq\labelsA$ and $\b\subseteq\labelsB$.
    Say that the composite input system $\aA_\a$ \defn{does not (causally) influence} the composite output system $\aB_\b$ through $\U$, and write $\aA_\a \ninfl_\U \aB_\b$, precisely if
    \begin{equation}\label{eq:no-influence}
        \U^\dagger (\aB_\b) \subseteq \aA_\a' = \aA_{\labelsA\setminus\a}.
    \end{equation}
    Write $\aA_\a \infl_\U \aB_\b$ if this is not the case.
    In the case of singletons $\a = \{\lA\}$ and $\b = \{\lB\}$, we simply write $\aA_\lA \ninfl_\U \aB_\lB$ or $\aA_\lA \infl_\U \aB_\lB$.
    The \defn{causal structure} of $\U$ is the relation $G_\U \subseteq \labelsA \times \labelsB$ defined by
    \begin{equation}
        \forall \lA\in\labelsA,\lB\in\labelsB: \quad \lA \rel{G_\U} \lB \iff \aA_\lA\infl_\U \aB_\lB.
    \end{equation}
    (As customary for a binary relation $G\subseteq\labelsA\times\labelsB$, we write $\lA \relG \lB$ for $(\lA,\lB)\in G$.)
\end{definition}

A few remarks are in order.
First, various alternative ways of formulating the condition of no-influence $\aA_\a \ninfl_\U \aB_\b$ exist and appear in the literature; see e.g.~\cite{SW05,OVB23} for overviews and proofs of equivalence.
A commonly used one is the existence of some quantum channel $\D: \aA_{\labelsA\setminus\a} \to \aB_\b$ so that $\Tr_{\aB_{\labelsB\setminus\b}} \circ\, \U = \Tr_{\aA_\a} \tns\, \D$---or, diagrammatically,
\begin{equation}
    \label{eq:no-influence-operational}
    \input{circuits/no-influence-operational.tikz}\ .
\end{equation}
(Circuit diagrams are always to be read from bottom to top and the symbol $\discard{}$ denotes the partial trace.)
In other words, the marginal output on system $\aB_\b$ is independent of the input state prepared on $\aA_\a$.

Second, \cref{def:causal-influence} may also be applied to generic quantum channels $\E : \aA_\labelsA \to \aB_\labelsB$, yielding a notion usually referred to as \emph{(no) signalling}, the terminology \emph{(no) influence} being reserved for the unitary case.
We will discuss the relation between signalling and causal influence in more detail in \cref{app:generic-channels} and focus only on the unitary case in the main text.

Third, in the unitary case, multisystem influence relations are always completely determined by the single-system ones, a fact that has been called \emph{causal atomicity}.
Though simple to prove, this fact is noteworthy seeing as many properties of quantum systems do not reduce to properties of their subsystems.
\begin{proposition}[Causal atomicity~\cite{BLO19,WV21,OVB23}]
    \label{prop:causal-atomicity}
    Let $\U: \aA_\labelsA \to \aB_\labelsB$ be a {\unichan}.
    For any $\a\subseteq\labelsA$ and $\b\subseteq\labelsB$, $\aA_\a \infl_\U \aB_\b$ if and only if there are $\lA\in\a$ and $\lB\in\b$ such that $\aA_\lA\infl_\U \aB_\lB$.
\end{proposition}
The assumption of unitarity is essential here; the analogous statement for signalling through generic quantum channels does not go through (see \cref{app:generic-channels}).
In the unitary case, causal atomicity allows us to focus on just the single-system influence relations, captured by the object $G_\U\subseteq\labelsA\times\labelsB$ defined above.

\subsection{Compositional structure}\label{subsec:prelims-compositional-structure}
We begin by outlining the purely syntactical aspects of quantum circuits and their connectivity, adopting the order-theoretic formalism from Ref.~\cite{vdL25}.
This approach will prove useful in \cref{sec:canonical-circuit} due to the connection with lattice theory discussed there.

\begin{definition}[\cite{vdL25}]
    \label{def:circuit-shape}
    Let $\labelsA$ and $\labelsB$ be finite sets.
    A \defn{circuit shape} with inputs $\labelsA$ and outputs $\labelsB$ is a quadruple $(\csP,\leq,\la,\mu)$ consisting of a finite set $\csP$, a partial order $\leq$ on $\csP$, and maps $\la:\labelsA \to \csP$ and $\mu:\labelsB\to \csP$.
    We will often use $\csP$ to denote this quadruple in its entirety.
\end{definition}

Each element of the partial order $\csP$ is to be thought of as a box, and each edge in its Hasse diagram as a wire in a circuit diagram.

\begin{definition}[\cite{vdL25}]
    \label{def:covering-relation-and-circuit-diagram}
    For $p,q\in \csP$, write $p\cvr q$ if $p$ \defn{covers} $q$ from below: that is, if $p < q$ and no $r\in \csP$ satisfies $p < r < q$.
    The \defn{Hasse diagram} of the poset $\csP$ is the directed acyclic graph on $\csP$ that has a directed edge from $p$ to $q$ iff $p\cvr q$.
    The \defn{circuit diagram} of a circuit shape $(\csP,\leq,\la,\mu)$ is obtained by adding to the Hasse diagram a vertex for each $\lA\in\labelsA$ and $\lB\in\labelsB$, as well as directed edges from $\lA$ to $\la(\lA)$ and from $\mu(\lB)$ to $\lB$.
    All vertices are arranged on the page such that edges point upwards; the latter are then drawn as undirected edges, so as to look like wires connecting up boxes.
    See below for an example.
    We refer to elements $p\in\csP$ of a circuit shape using the syntactical term \emph{box}, reserving the term `gate' for their semantic counterparts (i.e.\ quantum channels composed into a quantum circuit).
\end{definition}

\begin{definition}[\cite{vdL25}]
    \label{def:faithfulness}
    Let $\csP$ be a circuit shape with inputs $\labelsA$ and outputs $\labelsB$.
    A \defn{path from $\lA\in\labelsA$ to $\lB\in\labelsB$ through $\csP$} is a sequence $p_1,p_2,\dots,p_n \in \csP$ such that $\la(\lA) = p_1 \cvr p_2 \cvr\cdots\cvr p_n = \mu(\lB)$.
    Since $\csP$ is finite, there exists a path from $\lA$ to $\lB$ iff $\la(\lA) \leq \mu(\lB)$.
    The \defn{(input-output) connectivity} of $\csP$ is the relation $G_\csP \subseteq \labelsA\times\labelsB$ satisfying
    \begin{equation}
        \label{eq:connectivity}
        \forall \lA\in\labelsA,\lB\in\labelsB: \quad \lA \rel{G_\csP} \lB \quad\iff\quad \la(\lA) \leq \mu(\lB). \qedhere
    \end{equation}
\end{definition}

\begin{example}
    \label{ex:connectivity}
    Let $\labelsA = \{\lA_1,\lA_2,\lA_3\}$ and $\labelsB = \{\lB_1,\lB_2,\lB_3\}$ and let $\latt{\cccG}$ and $\cls{\cccG}$ be the circuit shapes defined by the circuit diagrams below.
    Both of these circuit shapes have connectivity $\cccG$.
    \begin{equation}
        \label{eq:ccc-circuits}
        \latt{\cccG} \coloneqq \input{circuits/ccc-diamond-labelled.tikz}; \qquad \cls{\cccG} \coloneqq \input{circuits/ccc-classical.tikz} ; \qquad \cccG  \coloneqq \input{structures/ccc-structure-numbers.tikz}.
    \end{equation}
    Explicitly, $\latt{\cccG}$ is the circuit $(\latt{\cccG},\leq,\la,\mu)$ whose underlying set is $\{p,q,r,s\}$, whose partial order satisfies $p < q < s$, $p<r<s$, $q \nleq r$ and $r \nleq q$, and whose input and output maps are given by $\la: \lA_1\mapsto q, \lA_2\mapsto p, \lA_3\mapsto r$ and $\mu: \lB_1\mapsto q, \lB_2\mapsto s, \lB_3\mapsto r$, respectively.%
    \footnote{
        In the literature on (quantum) circuits, the syntax of circuits is often modelled by directed acyclic graphs, rather than partial orders.
        This allows one to distinguish, for example, the circuit shape $\latt{\cccG}$ as depicted in \cref{eq:ccc-circuits} from a circuit shape that features an additional wire directly from $p$ to $s$.
        For our purposes, however, this distinction is not relevant: not to input-output connectivity nor to the class of channels that can be implemented by circuits of these shapes.
        After all, any such wire from $p$ to $s$ can always be incorporated into either of the gates $q$ and $r$.
        For simplicity we therefore focus on circuits with transitively reduced graphs as in \cref{eq:ccc-circuits}.}

    As another example, the circuit shape
    \begin{equation}
        \label{eq:pic-example}
        \latt{\picG} \coloneqq \input{circuits/pic-example_non_hat.tikz} \quad\text{has connectivity}\quad \picG \coloneqq \input{structures/pic-example.tikz}.
    \end{equation}
    We will see in \cref{sec:canonical-circuit} that $\latt{\cccG}$ and $\latt{\picG}$ are special cases of a general construction of a circuit shape $\latt{G}$ with a given connectivity relation $G$ (see also \cref{prop:canonicity-prelims} below).
    Note that the top box of $\latt{\picG}$ serves not much purpose in the context of quantum circuits defined below (it is always necessarily a discarding operation) and can be ignored; but it arises from this general construction.
\end{example}

A quantum circuit is a circuit shape equipped with quantum semantics.
In the below we write
\begin{equation}
    \label{eq:covm-and-covp}
    \Covm(p) \coloneqq\{q \in \csP \mid q \cvr p \} \qquad\text{and}\qquad \Covp(p) \coloneqq\{r\in \csP \mid p \cvr r \}.
\end{equation}

\begin{definition}
    \label{def:circuit-and-circuit-decomposition}
    Let $\aA_\lA$ for $\lA\in\labelsA$ and $\aB_\lB$ for $\lB\in\labelsB$ be quantum systems.
    A \defn{(quantum) circuit} $\circuit = (\csP, \{\aZ_p^q\}_{p\cvr q}, \{\E_q\}_q)$ with inputs $\{\aA_\lA\}_{\lA\in\labelsA}$ and outputs $\{\aB_\lB\}_{\lB\in\labelsB}$ is a circuit shape $\csP \equiv (\csP,\leq,\la,\mu)$ on $\labelsA$, $\labelsB$ together with a choice of
    \begin{align}
        \text{quantum systems } \quad &\aZ_p^q \quad \text{ for each } p,q\in \csP \text{ so that } p\cvr q \text{ and }\\
        \text{CPTP maps } \quad &\E_p : \aA_{\la^{-1}(p)} \aZ_{\Covm(p)}^p \to \aZ_p^{\Covp(p)} \aB_{\mu^{-1}(p)} \quad \text{ for each } p\in \csP.
    \end{align}
    Here $\la^{-1}(p)$ and $\mu^{-1}(p)$ are the preimages of $p$ under $\la$ and $\mu$, respectively, and $\aZ_{\Covm(p)}^p \coloneqq \bigotimes_{q\in\Covm(p)} \aZ_q^p$ and $\aZ_p^{\Covp(p)} \coloneqq \bigotimes_{r\in\Covp(p)} \aZ_p^r$.
    The systems $\aZ_p^q$ may of course be trivial (one-dimensional).

    Each quantum circuit $\circuit$ defines a quantum channel $\aA_\labelsA \to \aB_\labelsB$ obtained by composing all gates $\E_p$ in the obvious way (i.e.\ by composing them along $\aZ_q^r$ systems, which each appear precisely once as an input and once as an output of the gates $\E_p$).
    A \defn{circuit decomposition} of a {\unichan} $\U$ is a quantum circuit that so composes to $\U$ and a \defn{unitary circuit decomposition} is one in which each gate $\E_p$ is itself a unitary channel.
    Finally, the \defn{connectivity} $G_\circuit \subseteq\labelsA\times\labelsB$ of a circuit $\circuit$ is the connectivity of its underlying circuit shape $G_\csP$.
\end{definition}

It will be useful to already mention the following result, which is a consequence of Ref.~\cite{vdL25} and will be considered in more detail in \cref{sec:canonical-circuit} (\cref{thm:canonicity}).

\begin{proposition}[\cite{vdL25}]
    \label{prop:canonicity-prelims}
    Let $\labelsA,\labelsB$ be finite sets and let $G\subseteq\labelsA\times\labelsB$.
    There exists a circuit shape $\latt{G}$ with connectivity $G_{\latt{G}} = G$ such that every quantum channel that admits a circuit decomposition $\circuit$ with connectivity $G_\circuit \subseteq G$ also admits a circuit decomposition of shape $\latt{G}$.
\end{proposition}

\subsection{Causal decompositions}\label{subsec:prelims-causal-decos}
Causal decompositions relate the causal and compositional structure of unitary transformations.
We start by noting the straightforward direction: the absence of a path through a circuit decomposition implies absence of causal influence.

\begin{restatable}[Soundness of no-connectivity for no-influence]{proposition}{soundness}
    \label{prop:soundness}
    If $\U : \aA_\labelsA \to \aB_\labelsB$ is a unitary quantum channel with circuit decomposition $\circuit$, then $G_\U \subseteq G_\circuit$.
\end{restatable}

This is a consequence, amongst other things, of the assumption that all gates in a circuit decomposition are trace-preserving.
The formal proof is simple except for some involved bookkeeping and is relegated to \cref{app:proof-of-soundness}.

A circuit decomposition can thus be regarded as a (possible) more fine-grained description of the process giving rise to $\U$ in which the lack of causal influence between certain input and output systems is made immediately clear by the lack of mediating paths.
We now turn to the question: when can causal conditions satisfied by a unitary channel be made apparent in that way?

First of all, it is a special case of the result of Ref.~\cite{ESW02} that any unitary channel $\U$ satisfying the constraint $\aA_\lA \ninfl_\U \aB_\lB$ for some $\lA\in\labelsA$ and $\lB\in\labelsB$ (i.e.\ $(\lA,\lB)\notin G_\U$) indeed admits a circuit decomposition $\circuit$ with no path from $\lA$ to $\lB$ (i.e.\ $(\lA,\lB)\notin G_\circuit$).
More specifically, it can be decomposed as
\begin{equation}
    \label{eq:no-influence-semilocalisable}
    \input{circuits/no-influence-semilocalisable2.tikz}\ .
\end{equation}
for some quantum system $\aZ$ and quantum channels $\V : \aA_{\labelsA\setminus\{\lA\}} \to \aZ \aB_{\lB}$ and $\W : \aA_\lA \aZ \to \aB_{\labelsB\setminus\{\lB\}}$.
$\V$ and $\W$ can in fact be chosen unitary~\cite{LB21}.

Every individual no-influence condition $(\lA,\lB)\notin G_\U$ satisfied by $\U$ can thus be represented in the compositional structure of an appropriately chosen circuit decomposition.
Causal decompositions address the more general problem of representing a given collection of no-influence conditions---expressed by the inclusion $G_\U \subseteq G$ for some relation $G\subseteq\labelsA\times\labelsB$---at once in the compositional structure of a single circuit.

\begin{definition}
    \label{def:causal-decomposition}
    Let $G\subseteq\labelsA\times\labelsB$ be a relation and $\U: \aA_\labelsA\to\aB_\labelsB$ a unitary channel that satisfies the causal constraint $G_\U \subseteq G$.
    We say that a circuit decomposition $\circuit$ of $\U$ \defn{represents} this constraint if its connectivity satisfies $G_\circuit \subseteq G$.
    After all, in this case it directly implies the causal constraint $G_\U\subseteq G$ by the chain of inclusions $G_\U \subseteq G_\circuit \subseteq G$, appealing to soundness (\cref{prop:soundness}).
    A \defn{causal decomposition} is a circuit decomposition representing some given causal constraint; a \defn{unitary causal decomposition} is one all of whose gates are themselves unitary.

    Moreover, we say that a relation $G\subseteq\labelsA\times\labelsB$ is \defn{(compositionally) representable} if for \emph{each} choice of quantum systems $\aA_\lA$ for $\lA\in\labelsA$ and $\aB_\lB$ for $\lB\in\labelsB$, \emph{every} unitary channel $\U:\aA_\labelsA\to\aB_\labelsB$ satisfying the causal constraint $G_\U \subseteq G$ admits of a circuit decomposition representing the constraint (or equivalently, by \cref{prop:canonicity-prelims}, a circuit decomposition of shape $\latt{G}$).
    If the gates in this circuit decomposition can always be chosen unitary, then $G$ is \defn{(compositionally) representable in unitary circuits}.
\end{definition}

In most use cases for causal decompositions such as described in the introduction, one is given no further information about the unitary $\U$ than the fact that it satisfies a causal constraint $G_\U \subseteq G$.
This makes the above concept of an abstract relation $G$ being \emph{representable} relevant, and we will focus on it in the rest of this paper.

Specifically, our main result is a characterisation of those $G$ that are representable \emph{in unitary circuits}.
Our motivation for considering unitary circuits is twofold.
First of all, unitary transformations $\U$ play a fundamental role in physics by describing the evolution of isolated quantum systems.
If the purpose of a circuit decomposition of $\U$ is to describe the physics of such an evolution on a more fine-grained level, then it is natural to require that its constituent gates also correspond to evolutions of isolated systems.
Secondly, if $G$ is indeed representable in unitary circuits, then the systems $\aZ_p^q$ and unitary gates $\E_p$ constituting the unitary circuit decompositions with shape $\latt{G}$ \emph{parametrise} the class of unitary channels $\U$ satisfying $G_\U \subseteq G$, thereby providing generalised autonomous causal mechanisms for the quantum causal models of~\cite{ABH17,BLO19}, as described in the introduction.
For this application of causal decompositions it is of course crucial that any combination of parameters in fact specifies a channel that belongs to the class---i.e.\ one that is, in particular, unitary.
Since inserting generic channels into the boxes of $\latt{G}$ generally yields non-unitary channels, it makes sense to focus on circuit decompositions with unitary gates only.

\subsubsection{Causally faithful decompositions}\label{subsec:faithful-decos}
In the above, we have regarded the abstract relation $G\subseteq\labelsA\times\labelsB$ as a `negative' constraint on causal structure: it specifies a set of input-output pairs between which influence is known \emph{not} to occur, $G$ itself being the complement of that set.
In the literature, the term `causal structure' itself is often interpreted in this purely negative sense: in a spacetime setting, for instance, relativistic causal structure imposes no-influence conditions between spacelike-separated systems while timelike separation does not necessitate influence.
Similarly, compositional structure is best understood as constraining causal influence negatively: an absence of paths through a circuit implies an absence of influence (\cref{prop:soundness}) but the presence of a path does, without any further assumptions, in no way guarantee presence of causal influence through the overall implemented unitary.

The notion of causal influence from~\cite{ABH17,BLO19,BLO21,LB21,RobinThesis} and \cref{subsec:causal-structure} however enables a more positive notion of causal structure: `the' causal structure $G_\U$ of a unitary transformation, defined in \cref{def:causal-influence}, is a specification of precisely those input-output pairs between which causal influences \emph{are actually} present.
Rather than as something imposed by an external causal framework such as a relativistic spacetime or the topology of a circuit or network, this enables a more primitive view of causal structure as a property of unitary dynamics itself, which is helpful in foundational contexts (see e.g.\ \cite{LB21,Orm24,ormrod2026decoherencestate}).

The following special type of causal decomposition, which was the focus of Ref.~\cite{LB21}, then becomes relevant.

\begin{definition}
    \label{def:causally-faithful-deco}
    A circuit decomposition $\circuit$ of $\U$ is \defn{causally faithful} if $G_\U = G_\circuit$; that is, if it represents precisely all no-influence relations satisfied by $\U$.
    We say that the relation $G\subseteq\labelsA\times\labelsB$ \defn{implies (unitary) causally faithful decompositions} if for each choice of quantum systems $\aA_\lA$ for $\lA\in\labelsA$ and $\aB_\lB$ for $\lB\in\labelsB$, every unitary channel $\U:\aA_\labelsA\to\aB_\labelsB$ with causal structure $G_\U = G$ admits a (unitary) causally faithful circuit decomposition, or, equivalently, a (unitary) circuit decomposition with shape $\latt{G}$.
\end{definition}

This condition on a relation $G$ is subtly distinct from, but implied by the \emph{representability} condition of \cref{def:causal-decomposition}.
The focus of this paper is on the stronger notion of representability, but in \cref{thm:main-thm} and \cref{app:faithful} we also comment on causally faithful decompositions.

\subsubsection{Existing results and routed quantum circuits}\label{subsubsec:existing-results}
The question now becomes: which relations $G\subseteq\labelsA\times\labelsB$ are representable in unitary circuits?
\textcite{LB21} proved a number of positive results, but also found counterexamples.
A simple counterexample, which will turn out crucial for the treatment of the general case in \cref{sec:main-result} and onwards, is the following.

\begin{example}
    \label{ex:ccc-has-no-unitary-deco}
    Let $\labelsA = \labelsB = \{1,2,3\}$ and consider the 3-qubit {\unichan}
    \begin{equation}
        \label{eq:two-cnots-unitary-definition}
        \input{circuits/u-bar.tikz} \coloneqq \quad \input{circuits/two-cnots.tikz},
    \end{equation}
    consisting of two CNOT gates, where $\aA_i\cong\aB_i\cong\L(\mathbb C^2)$ for $i\in\{1,2,3\}$.
    The causal structure of this {\unichan} is $G_\cccU = \cccG$ from \cref{eq:ccc-circuits}, which we reproduce here for convenience (and with appropriate labels):
    \begin{equation}
        \cccG\ = \input{structures/ccc-structure-numbers-only.tikz}.
    \end{equation}
    Indeed, it is clear that the no-influence relation $\aA_3 \ninfl_\cccU \aB_1$ holds since there is no path between those systems in the circuit representation of $\cccU$ given above
    (cf.\ \cref{prop:soundness}).
    Since the two CNOTs commute, we also have $\aA_1\ninfl_\cccU \aB_3$.
    There is influence between all other systems.

    However, $\cccU$ does not admit a unitary circuit decomposition with connectivity $\cccG$.
    If it did, then it would have to have one of the particular shape $\latt{\cccG}$ from \cref{prop:canonicity-prelims}.
    We will see in \cref{sec:canonical-circuit} that this is the circuit shape already defined in \cref{eq:ccc-circuits}.
    A unitary decomposition into this shape is however easily seen to be impossible, i.e.
    \begin{equation}
        \label{eq:ccc-faithful-circuit}
        \input{circuits/two-cnots.tikz} \quad \neq \quad \quad \input{circuits/ccc-supposed-decomposition.tikz},
    \end{equation}
    no matter the choice of systems $\aC,\aD,\aE,\aF$ and unitary channels $\uW,\uX,\uY,\uZ$.
    Indeed: since $\aA_2$ is a qubit and $\uW$ is unitary, either of the two systems $\aC,\aD$ would have to be trivial, contradicting the fact that $\cccU$ has influence $\aA_2 \infl_\cccU \aB_1$ as well as $\aA_2 \infl_\cccU \aB_3$.

    $\cccG$ is therefore not representable in unitary circuits.
\end{example}

It was however shown in~\cite{LB21} that the situation changes when considering an extended quantum circuit paradigm that, in addition to tensor products and sequential composition, incorporates direct sums.
This extended paradigm has been formalised in the framework of \emph{routed quantum circuits}~\cite{VKB21,Vanr22}, which turn out to naturally model phenomena outside the context of causal decompositions as well~\cite{VKB21,VC21,VOKB22,OVB23,Orm24,Vanr22,WV21}.
Routed circuits come with a graphical circuit language in which the direct-sum structure is represented by indices on the wires.
This circuit language satisfies a notion of soundness of no-connectivity for no-influence analogous to \cref{prop:soundness} (see~\cite[App.~A.10]{LB21} and \cite{VKB21}).
It therefore makes sense to consider causal and causally faithful decompositions in terms of routed circuits as well, and in particular---for the same reasons as above---in terms of \emph{unitary} routed circuits.
\textcite{LB21} showed that some unitaries $\U$ admit unitary routed circuit decompositions of a given connectivity $G$ even if they do not admit unitary circuit decompositions with that connectivity in the traditional, non-routed circuit formalism: as an example, $\cccU$ from \cref{eq:two-cnots-unitary-definition} admits a unitary routed circuit decomposition of the shape in \cref{eq:ccc-faithful-circuit}.
In fact, they showed that every tripartite unitary channel $\U$ with $G_\U\subseteq\cccG$ does~\cite[Theorem~3]{LB21}: in other words, $\cccG$ is representable in unitary \emph{routed} circuits even though (as we showed above) the same does not hold for the traditional (non-routed) unitary circuit formalism.

It is worth noting that any unitary routed circuit may be converted into a traditional quantum circuit of the same shape that expresses the same overall channel; however, in the process, one generally has to give up unitarity of the gates.
Indeed, as far as causal decompositions are concerned, the merit of generalising to routed circuits predominantly lies in preserving unitarity of the gates in the decomposition.
The generalisation makes much sense from a mathematical perspective, too:
traditional unitary quantum circuits can be regarded as decompositions of \mbox{*-isomorphisms} with the special property that the intermediate {\cstaralg}s (inner wires in the circuit) are all factors, while \emph{routed} unitary decompositions allow for non-factor algebras, too.

Summarising, for each relation $G\subseteq\labelsA\times\labelsB$ and {\unichan} $\U:\aA_\labelsA\to\aB_\labelsB$ satisfying $G_\U \subseteq G$, one may consider the statements
\begin{enumerate}
    \item\label{itm:deco-unitary} $\U$ admits a \emph{unitary circuit decomposition} with connectivity $G_\circuit\subseteq G$ (i.e.\ into unitary gates in the traditional, non-routed circuit setting);
    \item\label{itm:deco-routed-unitary} $\U$ admits a \emph{unitary routed circuit decomposition} with connectivity $G_\circuit\subseteq G$;
    \item\label{itm:deco-generic} $\U$ admits a \emph{circuit decomposition} with connectivity $G_\circuit\subseteq G$ (i.e.\ into generic CPTP gates in the traditional circuit setting).
\end{enumerate}
We have $\itmref{itm:deco-unitary}\Rightarrow\itmref{itm:deco-routed-unitary}\Rightarrow\itmref{itm:deco-generic}$ and, as witnessed by \cref{ex:ccc-has-no-unitary-deco} and~\cite[Theorem~3]{LB21}, $\itmref{itm:deco-unitary}\nLeftarrow\itmref{itm:deco-routed-unitary}$.
Future work will show that also $\itmref{itm:deco-routed-unitary}\nLeftarrow\itmref{itm:deco-generic}$.
Whether all $\U$ with $G_\U\subseteq G$ satisfy $\itmref{itm:deco-generic}$ is an open problem.

In the rest of this work we shall not discuss routed circuits further and only focus on decompositions of type \itmref{itm:deco-unitary}.

\subsubsection{Notes on further generalisations} \label{subsubsec:beyondunitary}

\paragraph{Generic channels}
Even prior to the question of whether to allow the gates in causal decompositions to be generic quantum channels or only unitary ones, one might wonder why we require that the overall channel $\U:\aA_\labelsA\to\aB_\labelsB$---that which is to be decomposed---be unitary to begin with.
As we discuss in detail in \cref{app:generic-channels}, the notions of causal influence and causal decompositions as defined above are not directly applicable in a meaningful sense to generic channels.
Moreover, we show that appropriate analogous problems about circuit decompositions of generic channels can always be brought back to the unitary case by considering appropriate unitary dilations of the channels.
For these reasons, the rest of this work focusses on the purely unitary case.

\paragraph{DAGs and higher-order quantum maps}
In the literature on classical and quantum causality, `causal structure' is usually expressed in terms of directed acyclic graphs (DAGs); this is also the case for the quantum causal modelling framework based on causal influences through \emph{higher-order} unitary transformations developed in~\cite{ABH17,BLO19,BLO21}.
In this work, we essentially restrict our attention to causal models on DAGs whose vertices happen to partition into a set $\labelsA$ of `causes' (vertices with no incoming edges) and a set $\labelsB$ of `effects' (vertices with no outgoing edges).
Indeed, any binary relation $G\subseteq\labelsA\times\labelsB$ uniquely corresponds to such a bipartitioned DAG and vice versa.

One might wonder whether the study of causal decompositions of unitary channels $\U$ can be generalised to an appropriate problem for quantum causal models on arbitrary DAGs not admitting of such a partition.
The answer is in the positive, and that crucially, the general problem completely reduces to the one for unitary channels studied here.
This reduction is detailed in~\cite[\S7.3]{LB21}; for our present purposes, suffice it to say that it is achieved by a procedure employed across research on classical and quantum causality---known in different contexts as `splitting nodes'~\cite{BLO19,BLO21}, constructing the `single-world intervention graph'~\cite{RR13}, `maximal interruption'~\cite{Wolfe+21}, or `plugging SWAPs into the higher-order quantum map'~\cite{OVB23}---which converts any DAG-based causal model in terms of higher-order quantum maps to one of the particular type we consider here.

%% file: figures/circuits/no-influence-operational.tikz
\begin{tikzpicture}
	\begin{pgfonlayer}{nodelayer}
		\node [style=large map] (0) at (-2.5, 0) {$\U$};
		\node [style=upground] (3) at (-3.95, 1.25) {};
		\node [style=none] (6) at (-3.95, -1.25) {};
		\node [style=none] (9) at (0.5, 0) {$=$};
		\node [style=label] (15) at (-3.5, 1.875) {$\aB_{\labelsB\setminus\b}$};
		\node [style=label] (17) at (-3.5, -1.75) {$\aA_\a$};
		\node [style=none] (23) at (-3.05, -1.25) {};
		\node [style=upground] (24) at (-3.05, 1.25) {};
		\node [style=none] (25) at (-3.45, -1.1) {\smallcdots};
		\node [style=none] (26) at (-3.45, 0.9) {\smallcdots};
		\node [style=none] (27) at (-1.95, 1.25) {};
		\node [style=none] (28) at (-1.95, -1.25) {};
		\node [style=label] (29) at (-1.5, 1.875) {$\aB_{\b}$};
		\node [style=label] (30) at (-1.5, -1.75) {$\aA_{\labelsA\setminus\a}$};
		\node [style=none] (31) at (-1.05, -1.25) {};
		\node [style=none] (32) at (-1.05, 1.25) {};
		\node [style=none] (33) at (-1.45, -1.1) {\smallcdots};
		\node [style=none] (34) at (-1.45, 1.05) {\smallcdots};
		\node [style=medium map] (35) at (4.625, 0) {$\D$};
		\node [style=upground] (36) at (1.925, 0) {};
		\node [style=none] (37) at (1.925, -1.25) {};
		\node [style=label] (39) at (2.375, -1.75) {$\aA_\a$};
		\node [style=none] (40) at (2.825, -1.25) {};
		\node [style=upground] (41) at (2.825, 0) {};
		\node [style=none] (42) at (2.425, -1.1) {\smallcdots};
		\node [style=none] (44) at (4.1, 1.25) {};
		\node [style=none] (45) at (4.1, -1.25) {};
		\node [style=label] (46) at (4.625, 1.875) {$\aB_{\b}$};
		\node [style=label] (47) at (4.625, -1.75) {$\aA_{\labelsA\setminus\a}$};
		\node [style=none] (48) at (5.15, -1.25) {};
		\node [style=none] (49) at (5.15, 1.25) {};
		\node [style=none] (50) at (4.675, -1.1) {\smallcdots};
		\node [style=none] (51) at (4.675, 1.05) {\smallcdots};
	\end{pgfonlayer}
	\begin{pgfonlayer}{edgelayer}
		\draw (6.center) to (3);
		\draw (24) to (23.center);
		\draw (28.center) to (27.center);
		\draw (32.center) to (31.center);
		\draw (37.center) to (36);
		\draw (41) to (40.center);
		\draw (45.center) to (44.center);
		\draw (49.center) to (48.center);
	\end{pgfonlayer}
\end{tikzpicture}

%% file: figures/circuits/ccc-diamond-labelled.tikz
\begin{tikzpicture}
	\begin{pgfonlayer}{nodelayer}
		\node [style=tiny map] (0) at (1.75, 1.25) {$s$};
		\node [style=none] (1) at (1.575, 1) {};
		\node [style=none] (2) at (1.925, 1) {};
		\node [style=none] (3) at (0.575, 0.25) {};
		\node [style=none] (4) at (0.925, 0.25) {};
		\node [style=none] (5) at (2.575, 0.25) {};
		\node [style=none] (6) at (2.925, 0.25) {};
		\node [style=label] (8) at (1.75, 2.5) {$\lB_2$};
		\node [style=label] (11) at (2.925, 2.5) {$\lB_3$};
		\node [style=label] (12) at (0.575, 2.5) {$\lB_1$};
		\node [style=tiny map] (13) at (1.75, -1.25) {$p$};
		\node [style=tiny map] (14) at (0.75, 0) {$q$};
		\node [style=tiny map] (15) at (2.75, 0) {$r$};
		\node [style=none] (16) at (1.575, -1) {};
		\node [style=none] (17) at (1.925, -1) {};
		\node [style=none] (18) at (0.575, -0.25) {};
		\node [style=none] (19) at (0.925, -0.25) {};
		\node [style=none] (20) at (2.575, -0.25) {};
		\node [style=none] (21) at (2.925, -0.25) {};
		\node [style=label] (23) at (1.75, -2.5) {$\lA_2$};
		\node [style=label] (26) at (2.925, -2.5) {$\lA_3$};
		\node [style=label] (27) at (0.575, -2.5) {$\lA_1$};
	\end{pgfonlayer}
	\begin{pgfonlayer}{edgelayer}
		\draw [in=-90, out=90] (4.center) to (1.center);
		\draw [in=90, out=-90] (2.center) to (5.center);
		\draw [in=90, out=-90] (19.center) to (16.center);
		\draw [in=-90, out=90] (17.center) to (20.center);
		\draw (13) to (23);
		\draw (26) to (21.center);
		\draw (27) to (18.center);
		\draw (0) to (8);
		\draw (11) to (6.center);
		\draw (12) to (3.center);
	\end{pgfonlayer}
\end{tikzpicture}

%% file: figures/circuits/ccc-classical.tikz
\begin{tikzpicture}
	\begin{pgfonlayer}{nodelayer}
		\node [style=tiny map] (0) at (-1.75, 1.125) {};
		\node [style=tiny map] (1) at (0, 1.125) {};
		\node [style=tiny map] (2) at (1.75, 1.125) {};
		\node [style=tiny map] (3) at (-1.75, -1.125) {};
		\node [style=tiny map] (4) at (0, -1.125) {};
		\node [style=tiny map] (5) at (1.75, -1.125) {};
		\node [style=none] (6) at (-1.925, 0.875) {};
		\node [style=none] (7) at (-1.575, 0.875) {};
		\node [style=none] (8) at (-0.175, 0.875) {};
		\node [style=none] (9) at (0, 0.875) {};
		\node [style=none] (10) at (0.175, 0.875) {};
		\node [style=none] (11) at (1.575, 0.875) {};
		\node [style=none] (12) at (1.925, 0.875) {};
		\node [style=none] (13) at (-1.925, -0.875) {};
		\node [style=none] (14) at (-1.575, -0.875) {};
		\node [style=none] (15) at (-0.175, -0.875) {};
		\node [style=none] (16) at (0, -0.875) {};
		\node [style=none] (17) at (0.175, -0.875) {};
		\node [style=none] (18) at (1.575, -0.875) {};
		\node [style=none] (19) at (1.925, -0.875) {};
		\node [style=label] (26) at (-1.75, -2.625) {$\lA_1$};
		\node [style=label] (27) at (0, -2.625) {$\lA_2$};
		\node [style=label] (28) at (1.75, -2.625) {$\lA_3$};
		\node [style=label] (29) at (-1.75, 2.625) {$\lB_1$};
		\node [style=label] (30) at (0, 2.625) {$\lB_2$};
		\node [style=label] (31) at (1.75, 2.625) {$\lB_3$};
	\end{pgfonlayer}
	\begin{pgfonlayer}{edgelayer}
		\draw (6.center) to (13.center);
		\draw [in=-90, out=90] (14.center) to (8.center);
		\draw [in=90, out=-90] (7.center) to (15.center);
		\draw (16.center) to (9.center);
		\draw [in=90, out=-90] (10.center) to (18.center);
		\draw [in=-90, out=90] (17.center) to (11.center);
		\draw (12.center) to (19.center);
		\draw (26) to (3);
		\draw (27) to (4);
		\draw (28) to (5);
		\draw (2) to (31);
		\draw (1) to (30);
		\draw (0) to (29);
	\end{pgfonlayer}
\end{tikzpicture}

%% file: figures/structures/ccc-structure-numbers.tikz
\begin{tikzpicture}
	\begin{pgfonlayer}{nodelayer}
		\node [style=label] (0) at (0, -1) {$\lA_1$};
		\node [style=label] (1) at (2, -1) {$\lA_2$};
		\node [style=label] (2) at (4, -1) {$\lA_3$};
		\node [style=label] (3) at (0, 1) {$\lB_1$};
		\node [style=label] (4) at (2, 1) {$\lB_2$};
		\node [style=label] (5) at (4, 1) {$\lB_3$};
	\end{pgfonlayer}
	\begin{pgfonlayer}{edgelayer}
		\draw [style=dir] (0) to (3);
		\draw [style=dir] (0) to (4);
		\draw [style=dir] (1) to (3);
		\draw [style=dir] (1) to (4);
		\draw [style=dir] (1) to (5);
		\draw [style=dir] (2) to (5);
		\draw [style=dir] (2) to (4);
	\end{pgfonlayer}
\end{tikzpicture}

%% file: figures/circuits/pic-example_non_hat.tikz
\begin{tikzpicture}
	\begin{pgfonlayer}{nodelayer}
		\node [style=tiny map] (7) at (-1.25, 0) {};
		\node [style=tiny map] (8) at (0, -1.5) {};
		\node [style=none] (10) at (-0.175, -1.25) {};
		\node [style=none] (11) at (0.175, -1.25) {};
		\node [style=none] (12) at (1.125, -0.25) {};
		\node [style=none] (13) at (1.325, -0.25) {};
		\node [style=none] (14) at (1.525, -0.25) {};
		\node [style=none] (15) at (-1.425, -0.25) {};
		\node [style=none] (16) at (-1.075, -0.25) {};
		\node [style=none] (17) at (-1.45, 0.25) {};
		\node [style=none] (18) at (-1.25, 0.25) {};
		\node [style=label] (19) at (0, -2.75) {$\lA_2$};
		\node [style=label] (20) at (-2, -1.75) {$\lA_1$};
		\node [style=label] (22) at (1.325, -1.75) {$\lA_3$};
		\node [style=label] (23) at (2.25, -1.75) {$\lA_4$};
		\node [style=label] (24) at (-2.25, 1.75) {$\lB_1$};
		\node [style=label] (25) at (-1.225, 1.75) {$\lB_2$};
		\node [style=tiny map] (26) at (0, 1.5) {};
		\node [style=none] (27) at (1.5, 0.25) {};
		\node [style=label] (28) at (2.075, 1.75) {$\lB_3$};
		\node [style=none] (29) at (1.15, 0.25) {};
		\node [style=none] (30) at (0.185, 1.25) {};
		\node [style=none] (33) at (-0.2, 1.25) {};
		\node [style=none] (34) at (-1.05, 0.25) {};
		\node [style=tiny map] (35) at (1.325, 0) {};
	\end{pgfonlayer}
	\begin{pgfonlayer}{edgelayer}
		\draw [in=90, out=-90] (16.center) to (10.center);
		\draw [in=-90, out=90] (11.center) to (12.center);
		\draw (13.center) to (22);
		\draw [in=90, out=-90] (14.center) to (23);
		\draw (19) to (8);
		\draw [in=-90, out=90] (20) to (15.center);
		\draw [in=270, out=90] (17.center) to (24);
		\draw (18.center) to (25);
		\draw [in=270, out=90] (27.center) to (28);
		\draw [in=-90, out=90] (29.center) to (30.center);
		\draw [in=-90, out=90] (34.center) to (33.center);
	\end{pgfonlayer}
\end{tikzpicture}

%% file: figures/structures/pic-example.tikz
\begin{tikzpicture}
	\begin{pgfonlayer}{nodelayer}
		\node [style=label] (0) at (0, -1) {$\lA_1$};
		\node [style=label] (1) at (2, -1) {$\lA_2$};
		\node [style=label] (2) at (4, -1) {$\lA_3$};
		\node [style=label] (3) at (1, 1) {$\lB_1$};
		\node [style=label] (4) at (3, 1) {$\lB_2$};
		\node [style=label] (5) at (5, 1) {$\lB_3$};
		\node [style=label] (6) at (6, -1) {$\lA_4$};
	\end{pgfonlayer}
	\begin{pgfonlayer}{edgelayer}
		\draw [style=dir] (0) to (3);
		\draw [style=dir] (0) to (4);
		\draw [style=dir] (1) to (3);
		\draw [style=dir] (1) to (4);
		\draw [style=dir] (1) to (5);
		\draw [style=dir] (2) to (5);
		\draw [style=dir] (6) to (5);
	\end{pgfonlayer}
\end{tikzpicture}

%% file: figures/circuits/no-influence-semilocalisable2.tikz
\begin{tikzpicture}
	\begin{pgfonlayer}{nodelayer}
		\node [style=large map] (0) at (-2.5, 0) {$\U$};
		\node [style=none] (3) at (-2.7, -0.25) {};
		\node [style=none] (6) at (-2.7, -1.25) {};
		\node [style=none] (9) at (0.5, 0) {$=$};
		\node [style=label] (15) at (-3.125, 1.875) {$\aB_{\labelsB\setminus\{\lB\}}$};
		\node [style=label] (17) at (-3.975, -1.75) {$\aA_\lA$};
		\node [style=none] (23) at (-1.05, -1.25) {};
		\node [style=none] (24) at (-1.05, -0.25) {};
		\node [style=none] (25) at (-1.825, -1.1) {\smallcdots};
		\node [style=none] (27) at (-3.95, 1.25) {};
		\node [style=none] (28) at (-3.95, 0.25) {};
		\node [style=label] (29) at (-1.05, 1.875) {$\aB_\lB$};
		\node [style=label] (30) at (-1.875, -1.75) {$\aA_{\labelsA\setminus\{\lA\}}$};
		\node [style=none] (31) at (-2.3, 0.25) {};
		\node [style=none] (32) at (-2.3, 1.25) {};
		\node [style=none] (34) at (-3.075, 1.05) {\smallcdots};
		\node [style=medium map] (35) at (4.625, -1.1) {$\V$};
		\node [style=label] (39) at (1.95, -2.6) {$\aA_\lA$};
		\node [style=none] (40) at (1.875, -2.1) {};
		\node [style=medium map] (52) at (2.4, 1.15) {$\W$};
		\node [style=none] (54) at (1.875, 0.65) {};
		\node [style=none] (55) at (2.95, 0.65) {};
		\node [style=none] (56) at (4.15, -0.6) {};
		\node [style=none] (57) at (1.85, 2.15) {};
		\node [style=label] (58) at (2.375, 2.775) {$\aB_{\labelsB\setminus\{\lB\}}$};
		\node [style=none] (59) at (2.9, 2.15) {};
		\node [style=none] (60) at (2.425, 1.95) {\smallcdots};
		\node [style=none] (61) at (1.85, 1.65) {};
		\node [style=none] (62) at (2.9, 1.65) {};
		\node [style=none] (63) at (4.1, -2.1) {};
		\node [style=label] (64) at (4.625, -2.6) {$\aA_{\labelsA\setminus\{\lA\}}$};
		\node [style=none] (65) at (5.15, -2.1) {};
		\node [style=none] (66) at (4.675, -1.95) {\smallcdots};
		\node [style=none] (67) at (4.1, -1.6) {};
		\node [style=none] (68) at (5.15, -1.6) {};
		\node [style=label] (70) at (5.15, 2.775) {$\aB_{\lB}$};
		\node [style=none] (71) at (5.15, -0.6) {};
		\node [style=none] (74) at (5.15, 2.15) {};
		\node [style=label] (75) at (3.2, -0.275) {$\aZ$};
		\node [style=none] (76) at (-3.975, -0.25) {};
		\node [style=none] (77) at (-3.975, -1.25) {};
		\node [style=none] (78) at (-1.05, 0.25) {};
		\node [style=none] (79) at (-1.05, 1.25) {};
	\end{pgfonlayer}
	\begin{pgfonlayer}{edgelayer}
		\draw (6.center) to (3.center);
		\draw (24.center) to (23.center);
		\draw (28.center) to (27.center);
		\draw (32.center) to (31.center);
		\draw (54.center) to (40.center);
		\draw [in=90, out=-90] (55.center) to (56.center);
		\draw (57.center) to (61.center);
		\draw (59.center) to (62.center);
		\draw (67.center) to (63.center);
		\draw (68.center) to (65.center);
		\draw (74.center) to (71.center);
		\draw (76.center) to (77.center);
		\draw (78.center) to (79.center);
	\end{pgfonlayer}
\end{tikzpicture}

%% file: figures/circuits/u-bar.tikz
\begin{tikzpicture}
	\begin{pgfonlayer}{nodelayer}
		\node [style=large map] (0) at (1.5, 0) {$\cccU$};
		\node [style=none] (6) at (3, 1.25) {};
		\node [style=label] (18) at (0.5, 1) {$\aB_1$};
		\node [style=label] (19) at (2, 1) {$\aB_2$};
		\node [style=label] (20) at (3.5, 1) {$\aB_3$};
		\node [style=label] (21) at (0.5, -1) {$\aA_1$};
		\node [style=label] (22) at (2, -1) {$\aA_2$};
		\node [style=label] (23) at (3.5, -1) {$\aA_3$};
		\node [style=none] (34) at (1.5, 1.25) {};
		\node [style=none] (36) at (0, 1.25) {};
		\node [style=none] (40) at (3, -1.25) {};
		\node [style=none] (48) at (1.5, -1.25) {};
		\node [style=none] (50) at (0, -1.25) {};
	\end{pgfonlayer}
	\begin{pgfonlayer}{edgelayer}
		\draw (40.center) to (6.center);
		\draw (36.center) to (50.center);
		\draw (34.center) to (48.center);
	\end{pgfonlayer}
\end{tikzpicture}

%% file: figures/circuits/two-cnots.tikz
\begin{tikzpicture}
	\begin{pgfonlayer}{nodelayer}
		\node [style=none] (6) at (3, 1.5) {};
		\node [style=label] (18) at (0.5, 1.25) {$\aB_1$};
		\node [style=label] (19) at (2, 1.25) {$\aB_2$};
		\node [style=label] (20) at (3.5, 1.25) {$\aB_3$};
		\node [style=label] (21) at (0.5, -1) {$\aA_1$};
		\node [style=label] (22) at (2, -1) {$\aA_2$};
		\node [style=label] (23) at (3.5, -1) {$\aA_3$};
		\node [style=none] (34) at (1.5, 1.5) {};
		\node [style=none] (36) at (0, 1.5) {};
		\node [style=none] (40) at (3, -1.25) {};
		\node [style=none] (48) at (1.5, -1.25) {};
		\node [style=none] (50) at (0, -1.25) {};
		\node [style=none] (51) at (-0.2, -0.25) {};
		\node [style=none] (52) at (1.5, -0.25) {};
		\node [style=none] (53) at (1.5, 0.5) {};
		\node [style=none] (54) at (3.2, 0.5) {};
		\node [style=small black dot] (56) at (1.5, -0.25) {};
		\node [style=small black dot] (57) at (1.5, 0.5) {};
		\filldraw[fill=none](0, -0.25) circle (0.2);
		\filldraw[fill=none](3, 0.5) circle (0.2);
	\end{pgfonlayer}
	\begin{pgfonlayer}{edgelayer}
		\draw (40.center) to (6.center);
		\draw (36.center) to (50.center);
		\draw (34.center) to (48.center);
		\draw (51.center) to (52.center);
		\draw (53.center) to (54.center);
	\end{pgfonlayer}
\end{tikzpicture}

%% file: figures/structures/ccc-structure-numbers-only.tikz
\begin{tikzpicture}
	\begin{pgfonlayer}{nodelayer}
		\node [style=label] (0) at (0, -1) {$1$};
		\node [style=label] (1) at (2, -1) {$2$};
		\node [style=label] (2) at (4, -1) {$3$};
		\node [style=label] (3) at (0, 1) {$1$};
		\node [style=label] (4) at (2, 1) {$2$};
		\node [style=label] (5) at (4, 1) {$3$};
	\end{pgfonlayer}
	\begin{pgfonlayer}{edgelayer}
		\draw [style=dir] (0) to (3);
		\draw [style=dir] (0) to (4);
		\draw [style=dir] (1) to (3);
		\draw [style=dir] (1) to (4);
		\draw [style=dir] (1) to (5);
		\draw [style=dir] (2) to (5);
		\draw [style=dir] (2) to (4);
	\end{pgfonlayer}
\end{tikzpicture}

%% file: figures/circuits/ccc-supposed-decomposition.tikz
\begin{tikzpicture}
	\begin{pgfonlayer}{nodelayer}
		\node [style=medium map] (104) at (1.7, 0) {$\uX$};
		\node [style=medium map] (105) at (5.7, 0) {$\uY$};
		\node [style=medium map] (106) at (3.7, 2.25) {$\uZ$};
		\node [style=medium map] (107) at (3.7, -2.25) {$\uW$};
		\node [style=none] (108) at (3.075, 1.75) {};
		\node [style=none] (109) at (4.325, 1.75) {};
		\node [style=none] (110) at (2.2, 0.5) {};
		\node [style=none] (111) at (5.2, 0.5) {};
		\node [style=none] (112) at (3.075, -1.75) {};
		\node [style=none] (113) at (4.325, -1.75) {};
		\node [style=none] (114) at (5.2, -0.5) {};
		\node [style=none] (115) at (2.2, -0.5) {};
		\node [style=none] (116) at (3.7, -3.5) {};
		\node [style=none] (117) at (3.7, -2.5) {};
		\node [style=none] (122) at (1.325, 3.5) {};
		\node [style=none] (123) at (1.325, -3.5) {};
		\node [style=none] (124) at (6.075, -3.5) {};
		\node [style=none] (125) at (6.075, 3.5) {};
		\node [style=none] (128) at (3.7, 2.5) {};
		\node [style=none] (129) at (3.7, 3.5) {};
		\node [style=medium label] (140) at (2.95, -1) {$\aC$};
		\node [style=medium label] (141) at (4.45, -1) {$\aD$};
		\node [style=medium label] (146) at (3.075, 0.875) {$\aE$};
		\node [style=medium label] (147) at (4.325, 0.875) {$\aF$};
		\node [style=label] (149) at (4.275, 3.25) {$\aB_2$};
		\node [style=label] (150) at (6.65, 3.25) {$\aB_3$};
		\node [style=label] (152) at (4.275, -3.25) {$\aA_2$};
		\node [style=label] (153) at (6.65, -3.25) {$\aA_3$};
		\node [style=label] (154) at (1.825, 3.25) {$\aB_1$};
		\node [style=label] (155) at (1.825, -3.25) {$\aA_1$};
	\end{pgfonlayer}
	\begin{pgfonlayer}{edgelayer}
		\draw [in=-90, out=90] (110.center) to (108.center);
		\draw [in=90, out=-90] (109.center) to (111.center);
		\draw [in=-90, out=90] (112.center) to (115.center);
		\draw (117.center) to (116.center);
		\draw [in=90, out=-90] (114.center) to (113.center);
		\draw (122.center) to (123.center);
		\draw (124.center) to (125.center);
		\draw (129.center) to (128.center);
	\end{pgfonlayer}
\end{tikzpicture}

%% file: sections/result-and-overview.tex
\section{Main result}\label{sec:main-result}
In this paper, we characterise those relations $G\subseteq\labelsA\times\labelsB$ that are representable in unitary circuits.
Recall from \cref{ex:ccc-has-no-unitary-deco} that $\cccG$ does not have this property.
The general condition in fact turns out to boil down to the absence of a copy of $\cccG$.

\begin{definition}
    \label{def:C3EP}
    A relation $G \subseteq \labelsA\times\labelsB$ satisfies the \defn{\CEPtext}, abbreviated \CEP{}, if it restricts nowhere to the relation $\cccG$ from \cref{eq:ccc-circuits}; that is, if for all $\lA_1,\lA_2,\lA_3\in\labelsA$ and $\lB_1,\lB_2,\lB_3\in\labelsB$, we have
    \begin{equation}
        \label{eq:C3EP}
        \tag{\textbf{$\cccG$-EP}}
        G \cap \left(\{\lA_1,\lA_2,\lA_3\} \times \{\lB_1,\lB_2,\lB_3\}\right) \quad\neq\quad \input{structures/ccc-structure-numbers.tikz}.
    \end{equation}
\end{definition}

\begin{theorem}
    \label{thm:main-thm}
    Let $\labelsA$ and $\labelsB$ be finite sets and $G\subseteq\labelsA\times\labelsB$.
    The following are equivalent.
    \begin{enumerate}[series=picmainthm]
        \item\label{itm:main-thm-pic} $G$ satisfies the {\CEPtext}.
        \item\label{itm:main-thm-unitary-decos} $G$ is representable in unitary circuits. That is, for all choices of quantum systems $\aA_\lA$ for $\lA\in\labelsA$ and $\aB_\lB$ for $\lB\in\labelsB$, every unitary channel $\U: \aA_\labelsA\to\aB_\labelsB$ with causal structure $G_\U \subseteq G$ admits a unitary circuit decomposition $\circuit$ with connectivity $G_\circuit \subseteq G$.\footnote{The statement obtained by substituting $G_\circuit = G$ for $G_\circuit \subseteq G$ is equivalent: one can always artificially increase the connectivity of a circuit decomposition by adding trivial wires and (if necessary) trivial gates.}
        \item \label{itm:main-thm-causally-faithful} $G$ implies unitary causally faithful decompositions. That is, for all choices of quantum systems $\aA_\lA$ for $\lA\in\labelsA$ and $\aB_\lB$ for $\lB\in\labelsB$, every unitary channel $\U: \aA_\labelsA\to\aB_\labelsB$ with causal structure $G_\U = G$ admits a unitary circuit decomposition $\circuit$ with connectivity $G_\circuit = G$.
    \end{enumerate}
\end{theorem}
\begin{proof}
    $\itmref{itm:main-thm-pic}\Rightarrow\itmref{itm:main-thm-unitary-decos}$ is the focus of the rest of the main text and will be proven in \cref{thm:inductive-semantics-thm}.
    An overview of the proof is given in \cref{subsec:proof-overview} below.
    To prove that conversely $\neg\itmref{itm:main-thm-pic}\Rightarrow\neg\itmref{itm:main-thm-unitary-decos}$, we can adapt the counterexample from \cref{ex:ccc-has-no-unitary-deco} by padding it with trivial systems.
    \label{page:proof-of-ii-implies-i}
    More precisely, assuming that $G$ violates the {\CEPtext}, let $\lA_1,\lA_2,\lA_3\in\labelsA$ and $\lB_1,\lB_2,\lB_3\in\labelsB$ be the system labels where $G$ restricts to $\cccG$ (i.e.\ such that there is equality in \cref{eq:C3EP} above).
    To construct the required counterexample to \itmref{itm:main-thm-unitary-decos}, let $\aA_{\lA_1},\aA_{\lA_2},\aA_{\lA_3},\aB_{\lB_1},\aB_{\lB_2},\aB_{\lB_3}$ all be qubits and let all other input and output systems be trivial, one-dimensional systems.
    Take for $\U:\aA_\labelsA\to\aB_\labelsB$ simply the tripartite unitary $\cccU$ from \cref{eq:two-cnots-unitary-definition} uniquely extended to the remaining (trivial) systems.
    This unitary satisfies the causal constraint $G_\U \subseteq G$, but it cannot be represented in a unitary circuit: if it could, then $\cccU$ itself would have a causal decomposition representing the constraint $G_\cccU \subseteq \cccG$, which we showed in \cref{ex:ccc-has-no-unitary-deco} is not possible.

    That $\itmref{itm:main-thm-unitary-decos}\Rightarrow\itmref{itm:main-thm-causally-faithful}$ is immediate from their statements.
    It thus remains to prove that $\neg\itmref{itm:main-thm-pic}\Rightarrow\neg\itmref{itm:main-thm-causally-faithful}$.
    This can also be done by constructing a counterexample based on $\cccU$; but since the counterexample unitary is now required to satisfy the saturated causal condition $G_\U = G$, simply padding with trivial systems won't do.
    The proof is more involved and is given in \cref{app:faithful}.
\end{proof}

\begin{remark}
    It is important to stress what \cref{thm:main-thm} does not show.
    For one, it is not the case that if $G$ fails \PIC{}, then \emph{no} {\unichan} $\U$ with $G_\U\subseteq G$ admits a unitary causal decomposition representing that constraint.
    Indeed, there exist {\unichan}s with causal structure $G_\U = \cccG$ that, in contrast to $\cccU$ in \cref{ex:ccc-has-no-unitary-deco}, do have a unitary causally faithful decomposition~\cite{LB21}: consider for example the circuit
    \begin{equation}
        \label{eq:some-loose-wires}
        \input{circuits/some-loose-wires.tikz}
    \end{equation}
    consisting only of qubits and identity gates.
    It is a unitary causally faithful decomposition for the {\unichan} that it defines.

    Moreover, \cref{thm:main-thm} characterises which relations $G$ are representable in non-routed unitary circuits, specifically.
    As discussed in \cref{subsubsec:existing-results}, every unitary with $G_\U \subseteq \cccG$ still admits a \emph{routed} unitary causal decomposition representing that constraint (i.e.\ of type~\itmref{itm:deco-routed-unitary} in \cref{subsubsec:existing-results}), and therefore also a causally faithful decomposition into generic CPTP maps (i.e.\ of type~\itmref{itm:deco-generic}).
\end{remark}

\subsection{Overview of the proof of $\itmref{itm:main-thm-pic}\Rightarrow\itmref{itm:main-thm-unitary-decos}$}\label{subsec:proof-overview}
Giving a causal decomposition of a {\unichan} $\U$ involves providing its syntax, i.e.\ a circuit shape with the appropriate connectivity $G$, as well as a semantics: an interpretation of its wires and boxes as quantum systems $\aZ_p^q$ and {\unichan}s $\V_q$ so that the target unitary $\U$ is recovered.
However, there may be multiple circuit shapes with the same connectivity $G$, raising the question which of these one should hope admits appropriate quantum semantics.

We address this problem---whose solution was already previewed in \cref{prop:canonicity-prelims}---in \cref{sec:canonical-circuit}.
We recall a construction from lattice theory, studied in particular in formal concept analysis, that takes any given binary relation $G\subseteq\labelsA\times\labelsB$ to a complete lattice~\cite{Birk67,GW24}.
Together with two natural maps $\la$ and $\mu$ into it, this lattice defines a circuit shape $\latt{G}$ whose connectivity relation is $G$.
Moreover, we recall from~\cite{vdL25} that any circuit with connectivity $G_\circuit\subseteq G$ can be rewritten into this lattice shape $\latt{G}$ using purely syntactical operations.
As a consequence, if a {\unichan} has a circuit decomposition with this connectivity, then it also has a circuit decomposition of shape $\latt{G}$ (\cref{prop:canonicity-prelims} and \cref{thm:canonicity}).
We will therefore call $\latt{G}$ the \defn{canonical circuit shape with connectivity $G$}.
We have already seen the special cases $\latt{\cccG}$ and $\latt{\picG}$ in \cref{ex:connectivity}.

This result goes through whether or not $G$ satisfies \PIC{}.
In \cref{thm:pic-equivalent-conditions} we show that \PIC{} however admits a natural reformulation in terms of properties of $\latt{G}$: in particular, $G$ satisfies \PIC{} if and only if $\latt{G}$ has no more than one path between each input $\lA\in\labelsA$ and output $\lB\in\labelsB$.
Note that $\latt{\picG}$ satisfies this property while $\latt{\cccG}$ does not, thus restating the facts that $\picG$ satisfies \PIC{} while $\cccG$ does not.

In \cref{sec:sufficiency} we use these results to prove the direction $\itmref{itm:main-thm-pic}\Rightarrow\itmref{itm:main-thm-unitary-decos}$ in \cref{thm:main-thm}: we prove that if a unitary $\U$ satisfies $G_\U\subseteq G$ and $G$ satisfies \PIC{}, then $\U$ admits a decomposition with connectivity $G$.
We do this by induction on the circuit shape $\latt{G}$, providing it with appropriate quantum semantics from the bottom upwards.
Each induction step relies on an operator-algebraic result (\cref{lma:the-algebraic-lemma} proven in \cref{app:algebra}) about the representation of commuting subalgebras on tensor-product factors; after all, a unitary transformation $\V_q$ is nothing but a refactorisation of its input space into a tensor product of output systems.
The canonical circuit shape $\latt{G}$ turns out essential here, telling us at each step which commuting subalgebras need to be considered.
Moreover, it is the combinatorial properties of $\latt{G}$ implied by \PIC{} from \cref{sec:canonical-circuit} that guarantee the premises of the operator-algebraic \cref{lma:the-algebraic-lemma}, which in turn yields the desired decompositions into tensor products of factor algebras---as opposed to direct sums over tensor products, which may in some cases instead be used for \emph{routed} circuit decompositions.
\Cref{sec:sufficiency} is the mathematical core of the paper, where the lattice-theoretic and operator-algebraic results meet, together providing the syntax and semantics of the quantum circuit, respectively.

%% file: figures/circuits/some-loose-wires.tikz
\begin{tikzpicture}
	\begin{pgfonlayer}{nodelayer}
		\node [style=none] (6) at (-2, 0.875) {};
		\node [style=none] (7) at (-1.75, 0.875) {};
		\node [style=none] (8) at (-0.25, 0.875) {};
		\node [style=none] (9) at (0, 0.875) {};
		\node [style=none] (10) at (0.25, 0.875) {};
		\node [style=none] (11) at (1.75, 0.875) {};
		\node [style=none] (12) at (2, 0.875) {};
		\node [style=none] (13) at (-2, -0.875) {};
		\node [style=none] (14) at (-1.75, -0.875) {};
		\node [style=none] (15) at (-0.25, -0.875) {};
		\node [style=none] (16) at (0, -0.875) {};
		\node [style=none] (17) at (0.25, -0.875) {};
		\node [style=none] (18) at (1.75, -0.875) {};
		\node [style=none] (19) at (2, -0.875) {};
		\node [style=label] (26) at (-1.875, -2.375) {$\aA_{\lA_1}$};
		\node [style=label] (27) at (0, -2.375) {$\aA_{\lA_2}$};
		\node [style=label] (28) at (1.875, -2.375) {$\aA_{\lA_3}$};
		\node [style=label] (29) at (-1.875, 2.375) {$\aB_{\lB_1}$};
		\node [style=label] (30) at (0, 2.375) {$\aB_{\lB_2}$};
		\node [style=label] (31) at (1.875, 2.375) {$\aB_{\lB_3}$};
		\node [style=none] (32) at (-2, 1.875) {};
		\node [style=none] (33) at (-1.75, 1.875) {};
		\node [style=none] (34) at (-0.25, 1.875) {};
		\node [style=none] (35) at (0, 1.875) {};
		\node [style=none] (36) at (0.25, 1.875) {};
		\node [style=none] (37) at (1.75, 1.875) {};
		\node [style=none] (38) at (2, 1.875) {};
		\node [style=none] (39) at (-2.25, 1.5) {};
		\node [style=none] (40) at (-1.5, 1.5) {};
		\node [style=none] (41) at (-1.5, 0.75) {};
		\node [style=none] (42) at (-2.25, 0.75) {};
		\node [style=none] (43) at (-0.5, 1.5) {};
		\node [style=none] (44) at (0.5, 1.5) {};
		\node [style=none] (45) at (0.5, 0.75) {};
		\node [style=none] (46) at (-0.5, 0.75) {};
		\node [style=none] (47) at (1.5, 1.5) {};
		\node [style=none] (48) at (2.25, 1.5) {};
		\node [style=none] (49) at (2.25, 0.75) {};
		\node [style=none] (50) at (1.5, 0.75) {};
		\node [style=none] (51) at (-2, -0.875) {};
		\node [style=none] (52) at (-1.75, -0.875) {};
		\node [style=none] (53) at (-0.25, -0.875) {};
		\node [style=none] (54) at (0, -0.875) {};
		\node [style=none] (55) at (0.25, -0.875) {};
		\node [style=none] (56) at (1.75, -0.875) {};
		\node [style=none] (57) at (2, -0.875) {};
		\node [style=none] (58) at (-2, -1.875) {};
		\node [style=none] (59) at (-1.75, -1.875) {};
		\node [style=none] (60) at (-0.25, -1.875) {};
		\node [style=none] (61) at (0, -1.875) {};
		\node [style=none] (62) at (0.25, -1.875) {};
		\node [style=none] (63) at (1.75, -1.875) {};
		\node [style=none] (64) at (2, -1.875) {};
		\node [style=none] (65) at (-2.25, -1.5) {};
		\node [style=none] (66) at (-1.5, -1.5) {};
		\node [style=none] (67) at (-1.5, -0.75) {};
		\node [style=none] (68) at (-2.25, -0.75) {};
		\node [style=none] (69) at (-0.5, -1.5) {};
		\node [style=none] (70) at (0.5, -1.5) {};
		\node [style=none] (71) at (0.5, -0.75) {};
		\node [style=none] (72) at (-0.5, -0.75) {};
		\node [style=none] (73) at (1.5, -1.5) {};
		\node [style=none] (74) at (2.25, -1.5) {};
		\node [style=none] (75) at (2.25, -0.75) {};
		\node [style=none] (76) at (1.5, -0.75) {};
	\end{pgfonlayer}
	\begin{pgfonlayer}{edgelayer}
		\draw (13.center)
			 to (6.center)
			 to (32.center);
		\draw [in=-90, out=90] (14.center) to (8.center);
		\draw (33.center)
			 to (7.center)
			 to [in=90, out=-90] (15.center);
		\draw (16.center) to (9.center);
		\draw [in=90, out=-90] (10.center) to (18.center);
		\draw [in=-90, out=90] (17.center) to (11.center);
		\draw (12.center) to (19.center);
		\draw (34.center) to (8.center);
		\draw (35.center) to (9.center);
		\draw (36.center) to (10.center);
		\draw (37.center) to (11.center);
		\draw (38.center) to (12.center);
		\draw [style=graydash] (42.center)
			 to (39.center)
			 to (40.center)
			 to (41.center)
			 to cycle;
		\draw [style=graydash] (46.center)
			 to (43.center)
			 to (44.center)
			 to (45.center)
			 to cycle;
		\draw [style=graydash] (50.center)
			 to (47.center)
			 to (48.center)
			 to (49.center)
			 to cycle;
		\draw (51.center) to (58.center);
		\draw (59.center) to (52.center);
		\draw (60.center) to (53.center);
		\draw (61.center) to (54.center);
		\draw (62.center) to (55.center);
		\draw (63.center) to (56.center);
		\draw (64.center) to (57.center);
		\draw [style=graydash] (68.center)
			 to (65.center)
			 to (66.center)
			 to (67.center)
			 to cycle;
		\draw [style=graydash] (69.center)
			 to (70.center)
			 to (71.center)
			 to (72.center)
			 to cycle;
		\draw [style=graydash] (74.center)
			 to (75.center)
			 to (76.center)
			 to (73.center)
			 to cycle;
	\end{pgfonlayer}
\end{tikzpicture}

%% file: sections/canonical-circuit.tex
\section{The canonical circuit shape}\label{sec:canonical-circuit}
Given a relation $G \subseteq\labelsA\times\labelsB$, there generally exist many possible circuit shapes with connectivity $G$ (see e.g.\ \cref{ex:connectivity}).
As shown in Ref.~\cite{vdL25}, all of them can however be rewritten into a particular canonical shape that can be constructed directly from the relation $G$.
This is therefore the shape we should be looking for: if a {\unichan} $\U$ has a causal decomposition, then it necessarily has one of this canonical shape.

In this section we recall the construction of this circuit shape, which follows a construction in lattice theory first presented by \textcite{Birk67} and extensively studied in the field of formal concept analysis~\cite{GW24}.
We then rephrase the {\CEPtext} \PIC{} in terms of this shape.
Note that this section is entirely syntactic (except for the statement of \cref{thm:canonicity}, since stating also that result syntactically requires additional machinery from Ref.~\cite{vdL25} we will not discuss); we will consider the consequences for quantum semantics in \cref{sec:sufficiency,app:faithful}.

\subsection{Construction of the canonical circuit shape $\latt{G}$}\label{subsec:lattice-construction}
For a binary relation $G\subseteq \labelsA\times\labelsB$, we write
\begin{equation}
    \ChG(\lA) \coloneqq \{ \lB\in\labelsB \mid \lA \relG \lB \} \quadand \PaG(\lB) \coloneqq \{ \lA\in\labelsA \mid \lA\relG \lB \}.
\end{equation}
In the construction below, an important role is played by intersections of sets of the form $\ChG(\lA)$ and $\PaG(\lB)$.
Let us start by giving some intuition for why these intersections are relevant when constructing a circuit shape that is as general as possible.
Imagine that $\csP$ is a circuit shape with connectivity $G$ and consider a box $q\in\csP$.%
\footnote{Recall that we use `box' to refer to an element of a circuit shape, while reserving the term `gate' for its semantic counterpart (i.e.\ a quantum channel placed in the context of a quantum circuit).}
Let $\a\subseteq\labelsA$ be the set of overall input wires that lie to $q$'s past; that is, the input wires from which $q$ can be reached via upward-directed paths through the circuit shape.
Consider also the outputs $\b\subseteq\labelsB$ that lie in $q$'s future.
Because of the fact that $\csP$ has connectivity $G$, we have $\lA \relG \lB$ for each $\lA\in\a,\lB\in\b$: in other words, $\a\subseteq \bigcap_{\lB\in\b} \PaG(\lB)$.
Denote the latter set by $\gP(\b)$.
Suppose, now, that the inclusion $\a\subseteq \gP(\b)$ is unsaturated: that there is $\lA'\in \gP(\b)\setminus\a$.
Then $\lA' \relG \lB$ for all $\lB\in\b$.
This means that adding a path from $\lA'$ to the box $q$ yields a circuit shape whose connectivity is still $G$.
Doing this for all $\lA'\in \gP(\b)\setminus\a$, we obtain a new circuit shape $\csP'$ with respect to which $\a = \gP(\b)$ and which is at least as general as $\csP$ in terms of the overall channels it can realise.
By a dual construction, we can ensure that also the inclusion $\b \subseteq \bigcap_{\lA\in\a} \ChG(\lA) \eqqcolon \gC(\a)$ is saturated.
In agreement with this intuition, each box in the circuit shape constructed below corresponds precisely to a pair $\a\subseteq\labelsA,\b\subseteq\labelsB$ so that $\a = \gP(\b)$ and $\b = \gC(\a)$.

Formally, fix two finite sets $\labelsA$, $\labelsB$ and an arbitrary relation $G \subseteq \labelsA\times\labelsB$.
Denote by $\cP(\labelsA)$ and $\cP(\labelsB)$ the powersets of $\labelsA$ and $\labelsB$.
As anticipated, we consider the maps
\begin{align}
    \gC &: \cP(\labelsA) \to \cP(\labelsB) \dblcolon \a \mapsto \bigcap_{\lA\in\a} \ChG(\lA) = \{\lB \in \labelsB \mid \forall \lA \in\a: \lA \relG \lB \} \\
    \text{and\quad} \gP &: \cP(\labelsB) \to \cP(\labelsA) \dblcolon \b \mapsto \bigcap_{\lB\in\b} \PaG(\lB) = \{\lA \in \labelsA \mid \forall \lB \in\b: \lA \relG \lB \}; \label{eq:galois-connection-p}
\end{align}
in particular, $\gC(\emptyset) = \labelsB$ and $\gP(\emptyset) = \labelsA$.
When $\cP(\labelsA)$ and $\cP(\labelsB)$ are ordered under inclusion, these maps are order-reversing:
\begin{equation}
    \label{eq:p-c-order-reversing}
    \begin{split}
        \forall \a,\a'\subseteq\labelsA&: \a\subseteq\a'\implies \gC(\a)\supseteq \gC(\a') \\
        \text{ and \quad } \forall \b,\b'\subseteq\labelsB&: \b\subseteq\b'\implies \gP(\b)\supseteq \gP(\b')
    \end{split}
\end{equation}
They also satisfy the relations
\begin{equation}
    \label{eq:extensivity}
    \forall \a\subseteq\labelsA: \a \subseteq \gP\gC(\a) \qquad\text{and}\qquad \forall\b\subseteq\labelsB: \b\subseteq \gC\gP(\b).
\end{equation}
By definition, \cref{eq:p-c-order-reversing,eq:extensivity} make the pair $(\gC,\gP)$ into an (antitone) \defn{Galois connection} between the partially ordered sets $\cP(\labelsA)$ and $\cP(\labelsB)$.%
\footnote{In categorical terms, this is to say that they form a dual adjunction between the orders $\cP(\labelsA)$ and $\cP(\labelsB)$ seen as categories.
Moreover, the composites $\gC\gP$, $\gP\gC$ always form \emph{closure operators}: they are (i) monotone ($\a\subseteq\a'\Rightarrow\gP\gC(\a)\subseteq\gP\gC(\a')$, and similarly for $\gC\gP$); (ii) extensive (\cref{eq:extensivity}); and (iii) idempotent (as a consequence of \cref{eq:galois-connection}).}
A direct consequence of these equations is that
\begin{equation}
    \label{eq:galois-connection}
    \gP\gC\gP = \gP \qquad\text{and}\qquad \gC\gP\gC = \gC.
\end{equation}

For a set $\a\in\cP(\labelsA)$, we call $\gP\gC(\a)$ the \defn{closure} of $\a$ and we say that $\a$ is \defn{closed} if $\a = \gP\gC(\a)$.
One can see from \cref{eq:galois-connection} that $\a$ is closed if and only if it lies in the image of $\gP$.
We denote by $\closedA\subseteq\cP(\labelsA)$ the set of closed subsets of $\labelsA$.
This set is closed under intersections: this follows directly from the definition of $\gP$ in \cref{eq:galois-connection-p}.

Dually, we say that $\gC\gP(\b)$ is the \defn{closure} of $\b\subseteq\labelsB$, and that $\b$ is \defn{closed} if $\b = \gC\gP(\b)$, or, equivalently, if there is an $\a\subseteq\labelsA$ so that $\b = \gC(\a)$.
We denote by $\closedB \subseteq\cP(\labelsB)$ the set of closed subsets of $\labelsB$; also this set is closed under intersections.

By \cref{eq:galois-connection}, then, the maps $\gC$ and $\gP$ restrict to a pair of bijective, mutually inverse order-reversing maps $\gC : \closedA \to \closedB$ and $\gP : \closedB \to \closedA$.
The following combines $\closedA$ and $\closedB$ into one object.

\begin{definition}[\cite{GW24}]
    \label{def:concept-lattice-and-circuit}
    Let $G\subseteq \labelsA\times\labelsB$.
    The \defn{concept lattice} $\latt{G}$ is the set
    \begin{equation}
        \label{eq:concept-lattice}
        \latt{G} \coloneqq \{\<\a,\b\> \in \closedA\times\closedB \mid \a = \gP(\b) \text{ and } \b = \gC(\a)\}
    \end{equation}
    ordered by
    \begin{equation}
        \<\a,\b\> \leq \<\a',\b'\> \quad:\!\iff\quad \a \subseteq \a' \text{ or, equivalently, } \b \supseteq \b'.
    \end{equation}
    For an arbitrary element $v\in\latt{G}$, we will denote by $\a_v$ and $\b_v$ the closed subsets of $\labelsA$ and $\labelsB$ so that $v = \<\a_v,\b_v\>$.

    Moreover, defining the functions
    \label{eq:la-and-mu}
    \begin{align}
        \la_\latt{G}: \labelsA \to \latt{G}, \quad \lA &\mapsto \big\< \gP\gC(\{\lA\}), \gC(\{\lA\})\big\> \\
        \text{ and \quad }\mu_\latt{G}: \labelsB \to \latt{G}, \quad \lB &\mapsto \big\< \gP(\{\lB\}), \gC\gP(\{\lB\})\big\>,
    \end{align}
    the tuple $(\latt{G},\leq,\la_\latt{G}, \mu_\latt{G})$ forms a circuit shape in the sense of \cref{def:circuit-shape}, which we shall call the \defn{canonical circuit shape with connectivity} $G\subseteq\labelsA\times\labelsB$.
\end{definition}

The partial order $(\latt{G},\leq)$ is a complete lattice: that is, every subset has a greatest lower bound and least upper bound.
They are given, respectively, by
\begin{align}
    \bigmeet_{i\in I} \<\a_i,\b_i\> = \left\< \bigcap_{i\in I} \a_i, \gC\gP\left(\bigcup_{i\in I} \b_i\right) \right\> \qquad\text{and}\qquad
    \bigjoin_{i\in I} \<\a_i,\b_i\> = \left\< \gP\gC\left(\bigcup_{i\in I} \a_i\right), \bigcap_{i\in I} \b_i \right\>,
\end{align}
where $I$ is an arbitrary set and $\<\a_i,\b_i\>\in\latt{G}$ for all $i\in I$.

\begin{example}
    It may be verified that $\latt{\cccG}$ from \cref{ex:connectivity} is the canonical shape for $\cccG$ and $\latt{\picG}$ for $\picG$.
    Here we illustrate the construction of $\latt{G}$ in a slightly more involved case.
    Consider the relation between $\labelsA = \{1,2,3,4\}$ and $\labelsB = \{a,b,c,d,e\}$ given by
    \begin{equation}
        G\ \coloneqq \input{structures/lattice-example-relation.tikz}.
    \end{equation}
    We will abbreviate subsets like $\{2,3,4\}\subseteq\labelsA$ by `234'.
    Recall that the closed subsets of $\labelsA$ are those of the form $\gC(\b)$ for some $\b\subseteq\labelsB$; these are precisely the set $\labelsA$ itself (in this case $1234$), the sets of the form $\PaG(\lB)$ for $\lB\in\labelsB$ ($12$, $234$, $34$, and $4$), and all their intersections ($2$ and $\emptyset$).
    Taken together, we get $\closedA = \{\emptyset,2,12,4,34,234,1234\}$.
    We can now obtain $\closedB$ by applying the map $\gC$ to each of the closed sets $\a\in\closedA$; we get $\closedB = \{abcde,abc,ab,cde,cd,c,\emptyset\}$, respectively.

    Ordering $\closedA$ under $\subseteq$ and $\closedB$ under $\supseteq$, these sets are represented by the Hasse diagrams
    \begin{equation}
        \label{eq:hasse-diagrams-example}
        (\closedA, \subseteq) \quad=\input{structures/lattice-example-closedA.tikz} \quad\text{and}\qquad (\closedB, \supseteq) \quad=\input{structures/lattice-example-closedB.tikz}
    \end{equation}
    with $\gP,\gC$ acting as mutually inverse order-isomorphisms between the two posets.
    The corresponding concept lattice for $G$ is therefore
    \begin{equation}
        \label{eq:lattice-example-concept-lattice}
        \input{structures/lattice-example-concept-lattice.tikz}.
    \end{equation}

    In order to go from the concept lattice to the corresponding canonical circuit shape we additionally need the maps $\la_{\latt{G}}$ and $\mu_{\latt{G}}$ providing the location of the input and output wires.
    To calculate $\la_{\latt{G}}(3)$, for instance, one could explicitly compute the closure $\gP\gC(\{3\}) = \gP(\{c,d\}) = \{3,4\}$, or use the fact that $\la_{\latt{G}}(3)$ is the smallest element $v\in\latt{G}$ so that $3\in\a_v$ (cf.\ \cref{prop:useful-lattice-properties} below).
    From \cref{eq:lattice-example-concept-lattice} this is easily read off to be the pair $v = \<34,cd\>$.
    Similarly, for $\lB\in\labelsB$, $\mu_{\latt{G}}(\lB)$ is the largest element $w\in\latt{G}$ that satisfies $\lB\in\b_w$.
    The circuit diagram of the resulting canonical circuit shape is
    \begin{equation}
        \label{eq:canonical-circuit-example}
        \latt{G} \quad = \tikzfigpad{structures/lattice-example-canonical-circuit_v3}.
    \end{equation}
    (See \cref{rmk:latt-vs-hlatt} below for a comment about the bottom and top gates of this circuit shape.)
\end{example}

The following lists some properties useful to have at hand and which are all manifested as intuitive properties of the circuit diagram.
\begin{proposition}
    \label{prop:useful-lattice-properties}
    Let $G\subseteq\labelsA \times \labelsB$ be a binary relation and $\latt{G}$ the canonical circuit shape.
    \begin{enumerate}
        \item\label{itm:propo-one} For all $\lA\in\labelsA$, $\lB\in\labelsB$, and $v\in\lattK$, we have
        \begin{equation}
            \la_\lattK(\lA) \leq v \iff \lA\in\a_v \text{\quad and \quad} v \leq \mu_\lattK(\lB) \iff \lB\in\b_v.
        \end{equation}
        In other words, the inputs to the past of $v = \<\a_v,\b_v\>$ are precisely $\a_v$; the outputs to its future are $\b_v$.
        As a consequence, $\la_\lattK(\lA) = \min\{v \in \lattK \mid \lA\in \a_v\}$ and $\mu_\lattK(\lB) = \max\{v\in\lattK \mid \lB\in \b_v\}$.
        \item\label{itm:propo-faithful} $\lattK$ has connectivity $G$.
        \item\label{itm:propo-join-and-meet-dense} For all $v\in\lattK$, we have
        \begin{equation}
            \label{eq:join-and-meet-dense}
            v = \bigjoin_{\lA\in\a_v} \la_\latt{G}(\lA) = \bigmeet_{\lB\in\b_v} \mu_{\latt{G}}(\lB).
        \end{equation}
        \item\label{itm:propo-beta-union} For all $v\in\lattK$, we have
        \begin{equation}
            \label{eq:beta-union}
            \a_v = \la_{\lattK}^{-1}(v) \cup \bigcup_{u\lessdot v} \a_u \quadand
            \b_v = \mu_\lattK^{-1}(v) \cup \bigcup_{w \gtrdot v} \b_w.
        \end{equation}
        In other words, each $\lB\in\b_v$ is either an output wire of the box $v$ itself, or an output that can be reached from $v$ by a path passing through some $w \gtrdot v$ (and dually for $\lA\in\a_v$).
    \end{enumerate}
\end{proposition}
\begin{proof}
    See e.g.~\cite{vdL25}.
\end{proof}

\subsection{Canonicity of $\latt{G}$}\label{subsec:canonicity}
The reason for calling $\latt{G}$ `canonical' lies in a syntactical result from~\cite{vdL25}, which for us has the following relevant semantic consequence.
(This also holds for general quantum channels, but our focus lies on unitaries.)

\begin{theorem}
    \label{thm:canonicity}
    If a {\unichan} $\U$ has a (unitary) circuit decomposition $\circuit$ with connectivity $G_\circuit \subseteq G\subseteq\labelsA\times\labelsB$, then it has a (unitary) circuit decomposition with shape $\latt{G}$.
\end{theorem}
\begin{proof}
    This is a direct consequence of~\cite{vdL25}, which introduces a notion of morphisms between circuit shapes that formalise syntactical circuit rewrites.
    It shows that any circuit shape $\csP$ with connectivity $G_\csP\subseteq G$ admits a circuit morphism (`rewrite') $f: \csP \to \latt{G}$.
    Thus, if $\U$ has a circuit decomposition with shape $\csP$, then applying the rewrite provided by $f$ yields a circuit decomposition of $\U$ of shape $\latt{G}$.
\end{proof}

\begin{example}
    Any circuit with shape as on the left can be rewritten into one with the shape on the right by the syntactical operation of merging the highlighted pairs of boxes:
    \begin{equation}
        \tikzfigpad{circuits/ccc-classical-merge} \rightsquigarrow \tikzfigpad{circuits/ccc-diamond}.
    \end{equation}
    The shape on the right is the canonical circuit shape $\latt{\cccG}$ for the relation $\cccG$ from \cref{ex:connectivity,ex:ccc-has-no-unitary-deco}.
    What `rewriting' formally means in this context is explained in~\cite{vdL25}.
\end{example}

\begin{remark}
    \label{rmk:latt-vs-hlatt}
    The circuit shape $\latt{G}$ may in some cases be simplified without impacting the validity of \cref{thm:canonicity} by removing its top and/or bottom gate.
    \Cref{eq:canonical-circuit-example} is an example: its top gate $\<1234,\emptyset\>$ has no output wires and is, when the circuit shape is provided with a quantum semantics, therefore necessarily assigned a discard operation (the only trace-preserving quantum channel with trivial output).
    The discard however tensor-factorises, meaning it could just as well be absorbed into the two gates just below it.
    More generally, the top gate $\top \coloneqq \<\labelsA,\gC(\labelsA)\>$ may safely be removed from the circuit shape whenever it has no output wires and no overall input wires: i.e.\ when $\la^{-1}(\top) = \mu^{-1}(\top) = \emptyset$.

    Moreover, when specialising to \emph{unitary} circuit decompositions, the same may be said about the bottom gate $\bot \coloneqq \<\gP(\labelsB),\labelsB\>$: if $\la^{-1}(\bot) = \mu^{-1}(\bot) = \emptyset$, as for the example in \cref{eq:canonical-circuit-example}, then such a gate does not add to the expressivity of the circuit shape.
    After all, the only unitary channel with trivial input system is the identity channel between two trivial systems (i.e.\ a global phase).

    That said, although it will in special cases yield slightly unusual quantum circuit shapes, we will keep using the entire concept lattice $\latt{G}$ as it does not hurt and saves us from making case distinctions.
\end{remark}

\begin{remark}
    The fact that the maximally expressive circuit shape with a given connectivity, $\latt{G}$, turns out to be a lattice is natural from a physical point of view.
    To see this, consider the following two simple circuit shapes with identical connectivity relation $G_2$:
    \begin{equation}
        \latt{G_2} = \input{circuits/lg2.tikz};\quad \csP_2 \coloneqq \input{circuits/c2.tikz}; \quad G_2 \coloneqq \input{structures/g2.tikz}.
    \end{equation}
    $\latt{G_2}$ is the canonical circuit shape with connectivity $G_2$, and its underlying partial order is the trivial one-element lattice.
    Its single box, call it $p$, is, informally speaking, a place where all four systems can meet and interact: that is, it satisfies
    \begin{equation}
        \label{eq:interaction}
        \la(\lA_1) \leq p, \la(\lA_2)\leq p, p\leq \mu(\lB_1) \text{ and } p\leq\mu(\lB_2)
    \end{equation}
    (in this case all equalities).
    In fact, \emph{any} circuit shape with connectivity $G_2$ and whose underlying partial order is a lattice has a box $p$ satisfying \cref{eq:interaction}: indeed, having connectivity $G_2$ implies that $\mu(\lB_1)$ and $\mu(\lB_2)$ are both upper bounds to $\{\la(\lA_1),\la(\lA_2)\}$, so taking $p$ to be the least upper bound $\la(\lA_1)\join\la(\lA_2)$ does the trick.

    The circuit shape $\csP_2$, on the other hand, does not contain a box $p$ with this property, and is consequently not a lattice (indeed, the set $\{\la_{\csP_2}(\lA_1), \la_{\csP_2}(\lA_2) \}$ has no least upper bound).
    Classically, this does not impact its expressivity: any bipartite stochastic channel admits a decomposition into stochastic channels of shape $\csP_2$ due to the possibility of copying classical information.
    Quantumly, however, this is not the case: the two-qubit CNOT channel, for instance, cannot be expressed by a circuit of shape $\csP_2$---even when allowing all four gates to be general channels (for a proof, see~\cite{vanderlugt2026causalcompositional}).%
    \footnote{Notably, the CNOT \emph{can} be implemented by a circuit like $\csP_2$ but with an additional resource of pre-shared entanglement (i.e.\ by the circuit shape obtained by adjoining $\csP_2$ with a minimum element). In fact any bipartite quantum channel can be arbitrarily closely approximated (though not necessarily implemented exactly~\cite{MA24}) by circuits of this more general form using \emph{nonlocal quantum computation} protocols~\cite{Buhr+14}, which may thus be regarded as substituting interaction for shared entanglement.}

    The property of being a lattice thus guarantees the presence of certain \emph{interaction} gates which are relevant to quantum dynamics.
    This generalises: Ref.~\cite{vdL25} shows that in fact any circuit shape with connectivity $G$ whose underlying partial order is a complete lattice is, like $\latt{G}$, maximally expressive (with the concept lattice $\latt{G}$ being the smallest such example).
    Whether the full generality of the concept lattice, with its potentially many layers of interaction gates, is always strictly necessary to express certain quantum channels is however unknown.
    This question is tightly linked to that of the relevance of intermediate latents in quantum networks: see e.g.\ \cite{CW24}.
\end{remark}

\begin{remark}
    It is a central result of formal concept analysis~\cite{GW24} that any complete lattice $\latt{}$ may arise as the concept lattice of some relation $G\subseteq\labelsA\times\labelsB$.
    As a consequence, every finite lattice may arise as the underlying partial order of the canonical circuit shape for some relation $G\subseteq\labelsA\times\labelsB$.
\end{remark}

\subsection{The {\CEPtext} in terms of the canonical circuit shape}
\label{subsec:pic-in-terms-of-lattice}
The constructions in \cref{subsec:lattice-construction} and results in \cref{subsec:canonicity} go through whether or not $G$ satisfies the {\CEPtext} we introduced in \cref{sec:main-result}.
However, as we will see in the theorem below, the property manifests itself naturally in terms of properties of the canonical circuit shape $\latt{G}$.
It is useful to remind ourselves of $\cccG$ and its canonical circuit shape, which are
\begin{equation}
    \label{eq:ccc-lattice-reminder}
    \cccG\ = \input{structures/ccc-structure-numbers.tikz} \qquadand \latt\cccG = \input{circuits/ccc-diamond.tikz}.
\end{equation}
In the below, an \defn{order embedding} is a map $\phi: \csP\to\csQ$ between partial orders $\csP,\csQ$ satisfying $p\leq_\csP q \Leftrightarrow \phi(p)\leq_\csQ \phi(q)$.

\begin{theorem}
    \label{thm:pic-equivalent-conditions}
    Let $G\subseteq\labelsA\times\labelsB$.
    The following are equivalent.
    \begin{enumerate}
        \item\label{itm:pic-exclusion-of-c3} $G$ satisfies the {\CEPtext} \CEP.
        \item\label{itm:pic-exclusion-of-diamond}
        The circuit shape $\latt{\cccG}$ does not embed into $\latt{G}$, in the following sense: denoting by $\bot,\top\in\latt\cccG$ respectively the minimum and maximum of $\latt\cccG$, there is no order embedding $\phi: \latt\cccG \to \latt G$ such that both $\a_{\phi(\bot)}$ and $\b_{\phi(\top)}$ are nonempty.
        \item\label{itm:pic-unique-paths} The circuit shape $\latt{G}$ has, for each $\lA\in\labelsA$ and $\lB\in\labelsB$, no more than one path from $\lA$ to $\lB$. (Recall \cref{def:faithfulness} of a path.)
        \item\label{itm:pic-betas-disjoint} For all $v\in\latt{G}$ so that $\a_v\neq\emptyset$ and distinct $w,w'\in\Covp(v)$, $\b_w\cap\b_{w'} = \emptyset$: that is, the union on the right-hand side in \cref{eq:beta-union} is disjoint whenever $\a_v \neq \emptyset$. (Equivalently, the dual statement.)
    \end{enumerate}
\end{theorem}
Condition~\itmref{itm:pic-exclusion-of-diamond} is closely analogous to the \CEPtext{} itself in that it asserts the absence of a particular substructure, albeit now on the level of the canonical circuit shape. Condition~\itmref{itm:pic-unique-paths} is the condition that is graphically easiest to check.%
\footnote{
    Condition~\itmref{itm:pic-unique-paths} is also meaningful in the context of routed circuits.
    In a routed circuit, any nontrivial sectorial correlations (repeated indices) created at a bifurcation in the circuit must later be recombined (see for instance the circuits of \textcite{LB21}).
    If condition~\itmref{itm:pic-unique-paths} holds, no bifurcations are followed by recombinations in the circuit shape $\latt{G}$, except possibly at trivial top and bottom gates as in \cref{eq:canonical-circuit-example}---and therefore routed unitary circuits are in this case equally powerful to traditional unitary circuits.
    What this and \cref{thm:main-thm} thus tell us is that whenever the canonical circuit shape \emph{does} allow for genuinely routed circuits (i.e.\ in case non-unique paths exist), then traditional circuits do not suffice for representing the causal constraint.
    (In some such cases routed circuits may suffice, but this has not been proven in general; recall \cref{subsubsec:existing-results}).
}
Note that since $\latt{G}$ has connectivity $G$, another way of phrasing \itmref{itm:pic-unique-paths} is that $\latt{G}$ has exactly one path between each $\lA$ and $\lB$ such that $\lA\relG\lB$.
Finally, condition~\itmref{itm:pic-betas-disjoint} is most directly useful for our proof in \cref{sec:sufficiency}.

\begin{proof}
    Most implications are most straightforwardly proven in their contrapositive form.
    To avoid an overabundance of subscripts we shall write $\la$ and $\mu$ for the maps $\la_{\latt{G}}$ and $\mu_{\latt{G}}$ defined in \cref{eq:la-and-mu}, respectively.

    Our proof for the direction $\neg\itmref{itm:pic-exclusion-of-c3}\Rightarrow\neg\itmref{itm:pic-exclusion-of-diamond}$ builds on a general result in formal concept analysis~\cite[Proposition~38]{GW24} which states that if $G\subseteq\labelsA\times\labelsB$, $\labelsA'\subseteq\labelsA$, $\labelsB'\subseteq\labelsB$, and $G'\coloneqq G\cap(\labelsA'\times\labelsB')$, then the concept lattice $\latt{G'}$ embeds into $\latt{G}$.
    One such order embedding is given by $\phi:\latt{G'}\to\latt{G}, v \mapsto \<\gP\gC(\a_v),\gC(\a_v)\>$.
    Here $(\gC,\gP)$ is the Galois connection derived from $G$; the relation $G'$ comes with its own Galois connection $(\gC',\gP')$ between $\cP(\labelsA')$ and $\cP(\labelsB')$.
    For completeness we will show that $\phi$ is indeed an order embedding.
    This amounts to demonstrating that for all $\a_1,\a_2\subseteq\labelsA'$ that are closed with respect to $\gP'\gC'$, we have
    \begin{equation}
        \a_1\subseteq\a_2 \iff \gP\gC(\a_1) \subseteq \gP\gC(\a_2).
    \end{equation}
    The left-to-right direction is immediate from \cref{eq:p-c-order-reversing}.
    For the right-to-left direction, note that $\gP\gC(\a_1) \subseteq \gP\gC(\a_2)$ implies $\gC(\a_1) \supseteq \gC(\a_2)$ (by \cref{eq:p-c-order-reversing,eq:galois-connection}), which in particular means that $\gC(\a_1)\cap\labelsB' \supseteq \gC(\a_2) \cap \labelsB'$.
    Since $\gC'(-) = \gC(-)\cap\labelsB'$, which can be easily verified, this means that $\gC'(\a_1) \supseteq\gC'(\a_2)$, and therefore $\gP'\gC'(\a_1)\subseteq\gP'\gC'(\a_2)$.
    But since $\a_1,\a_2$ were closed with respect to $\gP'\gC'$, this just means that $\a_1\subseteq \a_2$.

    Now, assume $\neg\itmref{itm:pic-exclusion-of-c3}$, that is, that there exist $\labelsA'\coloneqq\{\lA_1,\lA_2,\lA_3\}\subseteq\labelsA$ and $\labelsB'\coloneqq\{\lB_1,\lB_2,\lB_3\}\subseteq\labelsB$ so that $G\cap(\labelsA'\times\labelsB') = \cccG$.
    To show $\neg\itmref{itm:pic-exclusion-of-diamond}$, it remains to show that $\a_{\phi(\bot)} \neq \emptyset$ and $\b_{\phi(\top)} \neq \emptyset$.
    For the former, note (e.g.\ from the diagram for $\latt\cccG$ above) that $\a_\bot \neq \emptyset$; therefore the closure $\gP\gC(\a_\bot)$ is nonempty too and thus $\a_{\phi(\bot)} = \gP\gC(\a_\bot) \neq \emptyset$.
    For the latter, note similarly that $\gC(\a_\top)\supseteq \gC'(\a_{\top}) = \b_\top \neq \emptyset$, so that $\b_{\phi(\top)} = \gC(\a_\top) \neq \emptyset$ either.

    Next, let us show that $\neg\itmref{itm:pic-exclusion-of-diamond}\Rightarrow\neg\itmref{itm:pic-exclusion-of-c3}$.
    $\neg\itmref{itm:pic-exclusion-of-diamond}$ tells us that there exist $v,w,w',x\in\latt{G}$ so that $\a_v\neq\emptyset\neq\b_x$, $v<w<x$, $v<w'<x$, and $w\incomp w'$ (the latter meaning $w$ and $w'$ are incomparable in the order $\leq$ on $\latt{G}$).
    Pick arbitrary $\lA_2\in\a_v$ and $\lB_2\in\b_x$.
    We claim that it is possible to pick $\lA_1\in\a_w$ and $\lB_3\in\b_{w'}$ so that $\la(\lA_1)\nleq \mu(\lB_3)$.
    Indeed, if all $\lA_1\in\a_w,\lB_3\in\b_{w'}$ satisfied $\la(\lA_1)\leq \mu(\lB_3)$, then we would have $\bigjoin_{\lA_1\in\a_w} \la(\lA_1) \leq \bigmeet_{\lB_3\in\b_{w'}} \mu(\lB_3)$, which by \cref{prop:useful-lattice-properties}\itmref{itm:propo-join-and-meet-dense} just means $w \leq w'$, in contradiction with the assumption that $w\incomp w'$.
    Similarly, it is possible to pick $\lA_3\in\a_{w'}$ and $\lB_1\in\b_w$ so that $\la(\lA_3)\nleq\mu(\lB_1)$.

    Proving $\neg\itmref{itm:pic-exclusion-of-c3}$ is now a matter of verifying that on the inputs $\lA_1,\lA_2,\lA_3$ and outputs $\lB_1,\lB_2,\lB_3$, $G$ restricts to the relation $\cccG$.
    First of all, $\lA_1\relG\lB_1$ follows from the facts that $\latt{G}$ has connectivity $G$ and that $\la(\lA_1) \leq w \leq \mu(\lB_1)$ (by \cref{prop:useful-lattice-properties}\itmref{itm:propo-faithful} and \itmref{itm:propo-one}).
    Likewise, $\lA_1\relG\lB_2$ since $\la(\lA_1)\leq w<x\leq \mu(\lB_2)$.
    Meanwhile, we have $\neg(\lA_1\relG\lB_3)$ by construction.
    Similarly it can be shown that $\lA_2\relG\lB_i$ for all of $i\in\{1,2,3\}$, and $\neg(\lA_3\relG\lB_1)$ while $\lA_3\relG\lB_2$ and $\lA_3\relG\lB_3$.
    This concludes the proof of $\neg\itmref{itm:pic-exclusion-of-diamond}\Rightarrow\neg\itmref{itm:pic-exclusion-of-c3}$.

    We omit the proof of equivalence of~\itmref{itm:pic-exclusion-of-diamond} and~\itmref{itm:pic-unique-paths}, which is straightforward.
    Let's prove equivalence of $\itmref{itm:pic-unique-paths}$ and $\itmref{itm:pic-betas-disjoint}$.
    To see that $\itmref{itm:pic-unique-paths}\Rightarrow\itmref{itm:pic-betas-disjoint}$, let $v\in\latt{G}$ be such that $\a_v \neq\emptyset$ and suppose $w,w'\in\Covp(v)$ are distinct. Pick $\lA\in\a_v$.
    If there were $\lB \in \b_w\cap\b_{w'}$, then there would be two distinct paths from $\lA$ to $\lB$: one through $v$ and $w$ and another through $v$ and $w'$.

    Conversely, to show that $\neg\itmref{itm:pic-unique-paths} \Rightarrow \neg\itmref{itm:pic-betas-disjoint}$, suppose that there are two distinct paths from $\lA\in\labelsA$ to $\lB\in\labelsB$ in $\latt{G}$; denote them by $\{p_i\}_{i=1}^s$ and $\{p'_i\}_{i=1}^{s'} \subseteq \latt{G}$.
    Let $k\in\{1,\dots,\min\{s,s'\}\}$ be the first index where they diverge, so that $p_k = p'_k$ but $p_{k+1} \neq p'_{k+1}$.
    Let $v \coloneqq p_k = p'_k$, $w \coloneqq p_{k+1}$, and $w'\coloneqq p'_{k+1}$.
    Note that $\a_v$ is non-empty, as it contains $\lA$. We also have $w = p_{k+1} \leq p_s = \mu(\lB)$, meaning $\lB\in\b_w$; and similarly, $\lB\in\b_{w'}$.
    This establishes $\neg\itmref{itm:pic-betas-disjoint}$.
\end{proof}

We conclude this section by stating a consequence of \itmref{itm:pic-betas-disjoint} above, which will be useful in \cref{sec:sufficiency}.
Here we write $\PaG(\b) \coloneqq \bigcup_{\lB\in\b} \PaG(\lB)$ for the relational preimage of $\b\subseteq\labelsB$.

\begin{lemma}
    \label{lma:overlap-only-on-alpha}
    Let $G$ satisfy \PIC{}, let $v\in\latt{G}$ be such that $\a_v\neq\emptyset$ and let $w,w'\in\Covp(v)$ be distinct.
    We have $\PaG(\b_w)\cap \PaG(\b_{w'}) = \a_v$.
\end{lemma}
\begin{proof}
    It suffices to show that for all $\lB\in\b_w$ and $\lB'\in\b_{w'}$, $\PaG(\lB)\cap\PaG(\lB') = \gP(\{\lB,\lB'\}) = \a_v$.
    To do so, note first of all that $\a_v\subseteq\a_w = \gP(\b_w) \subseteq \gP(\{\lB\})$; similarly, ${\a_v\subseteq \gP(\{\lB'\})}$.
    This implies that $\a_v\subseteq \gP(\{\lB,\lB'\})$, that is, $v\leq x$ where $x \coloneqq \< \gP(\{\lB,\lB'\}) , \gC\gP(\{\lB,\lB'\}) \> \in \latt{G}$.
    Suppose that $v<x$; then there is $w''\in\Covp(v)$ so that $v\cvr w''\leq x$.
    We have $w \neq w''$ or $w'\neq w''$; without loss of generality, say that the former holds.
    By \itmref{itm:pic-betas-disjoint} in \cref{thm:pic-equivalent-conditions}, we must have $\b_{w}\cap\b_{w''} = \emptyset$.
    However, both $\b_w$ and $\b_{w''}$ contain $\lB$ ($\lB\in\b_w$ by assumption and $\lB\in \gC\gP(\{\lB,\lB'\}) = \b_x \subseteq \b_{w''}$), yielding a contradiction.
    Therefore $v<x$ cannot hold, and we must have $v=x$, or in other words, $\a_v = \gP(\{\lB,\lB'\})$.
\end{proof}

%% file: figures/structures/lattice-example-relation.tikz
\begin{tikzpicture}
	\begin{pgfonlayer}{nodelayer}
		\node [style=label] (0) at (1, -1) {$1$};
		\node [style=label] (1) at (2.75, -1) {$2$};
		\node [style=label] (2) at (4.5, -1) {$3$};
		\node [style=label] (3) at (0, 1) {$a$};
		\node [style=label] (4) at (1.75, 1) {$b$};
		\node [style=label] (5) at (3.5, 1) {$c$};
		\node [style=label] (6) at (6.25, -1) {$4$};
		\node [style=label] (7) at (5.25, 1) {$d$};
		\node [style=label] (8) at (7, 1) {$e$};
	\end{pgfonlayer}
	\begin{pgfonlayer}{edgelayer}
		\draw [style=dir] (0) to (3);
		\draw [style=dir] (0) to (4);
		\draw [style=dir] (1) to (3);
		\draw [style=dir] (1) to (4);
		\draw [style=dir] (1) to (5);
		\draw [style=dir] (2) to (5);
		\draw [style=dir] (2) to (7);
		\draw [style=dir] (6) to (5);
		\draw [style=dir] (6) to (7);
		\draw [style=dir] (6) to (8);
	\end{pgfonlayer}
\end{tikzpicture}

%% file: figures/structures/lattice-example-closedA.tikz
\begin{tikzpicture}
	\begin{pgfonlayer}{nodelayer}
		\node [style=label] (0) at (-1.5, 1.5) {12};
		\node [style=label] (1) at (2.5, -2.5) {$\emptyset$};
		\node [style=label] (2) at (0, 0) {2};
		\node [style=label] (3) at (4, -1) {4};
		\node [style=label] (4) at (2.75, 0.25) {34};
		\node [style=label] (5) at (1.5, 1.5) {234};
		\node [style=label] (6) at (0, 3) {1234};
	\end{pgfonlayer}
	\begin{pgfonlayer}{edgelayer}
		\draw (1) to (3);
		\draw (3) to (4);
		\draw (4) to (5);
		\draw (5) to (6);
		\draw (6) to (0);
		\draw (0) to (2);
		\draw (1) to (2);
		\draw (5) to (2);
	\end{pgfonlayer}
\end{tikzpicture}

%% file: figures/structures/lattice-example-closedB.tikz
\begin{tikzpicture}
	\begin{pgfonlayer}{nodelayer}
		\node [style=label] (0) at (-1.5, 1.5) {$ab$};
		\node [style=label] (1) at (2.5, -2.5) {$abcde$};
		\node [style=label] (2) at (0, 0) {$abc$};
		\node [style=label] (3) at (4, -1) {$cde$};
		\node [style=label] (4) at (2.75, 0.25) {$cd$};
		\node [style=label] (5) at (1.5, 1.5) {$c$};
		\node [style=label] (6) at (0, 3) {$\emptyset$};
	\end{pgfonlayer}
	\begin{pgfonlayer}{edgelayer}
		\draw (1) to (3);
		\draw (3) to (4);
		\draw (4) to (5);
		\draw (5) to (6);
		\draw (6) to (0);
		\draw (0) to (2);
		\draw (1) to (2);
		\draw (5) to (2);
	\end{pgfonlayer}
\end{tikzpicture}

%% file: figures/structures/lattice-example-concept-lattice.tikz
\begin{tikzpicture}
	\begin{pgfonlayer}{nodelayer}
		\node [style=label] (0) at (-1.5, 1.5) {$\langle  12, ab \rangle$};
		\node [style=label] (1) at (3, -3) {$\langle \emptyset, abcde \rangle$};
		\node [style=label] (2) at (0, 0) {$\langle  2, abc \rangle$};
		\node [style=label] (3) at (4.5, -1.5) {$\langle  4, cde \rangle$};
		\node [style=label] (4) at (3, 0) {$\langle  34, cd \rangle$};
		\node [style=label] (5) at (1.5, 1.5) {$\langle 234, c \rangle$};
		\node [style=label] (6) at (0, 3) {$\langle1234, \emptyset \rangle$};
	\end{pgfonlayer}
	\begin{pgfonlayer}{edgelayer}
		\draw (1) to (3);
		\draw (3) to (4);
		\draw (4) to (5);
		\draw (5) to (6);
		\draw (6) to (0);
		\draw (0) to (2);
		\draw (1) to (2);
		\draw (5) to (2);
	\end{pgfonlayer}
\end{tikzpicture}

%% file: figures/circuits/lg2.tikz
\begin{tikzpicture}
	\begin{pgfonlayer}{nodelayer}
		\node [style=tiny map] (0) at (0, 0) {};
		\node [style=label] (1) at (-0.5, 1.25) {$\lB_1$};
		\node [style=label] (2) at (0.5, 1.25) {$\lB_2$};
		\node [style=label] (3) at (-0.5, -1.25) {$\lA_1$};
		\node [style=label] (4) at (0.5, -1.25) {$\lA_2$};
		\node [style=none] (5) at (-0.15, -0.25) {};
		\node [style=none] (6) at (0.15, -0.25) {};
		\node [style=none] (7) at (-0.15, 0.25) {};
		\node [style=none] (8) at (0.15, 0.25) {};
	\end{pgfonlayer}
	\begin{pgfonlayer}{edgelayer}
		\draw [in=90, out=-90, looseness=1.25] (2) to (8.center);
		\draw [in=90, out=-90] (1) to (7.center);
		\draw [in=90, out=-90, looseness=1.25] (5.center) to (3);
		\draw [in=90, out=-90, looseness=1.25] (6.center) to (4);
	\end{pgfonlayer}
\end{tikzpicture}

%% file: figures/circuits/c2.tikz
\begin{tikzpicture}
	\begin{pgfonlayer}{nodelayer}
		\node [style=tiny map] (0) at (-1.75, 1.125) {};
		\node [style=tiny map] (1) at (0, 1.125) {};
		\node [style=tiny map] (3) at (-1.75, -1.125) {};
		\node [style=tiny map] (4) at (0, -1.125) {};
		\node [style=none] (6) at (-1.9, 0.875) {};
		\node [style=none] (7) at (-1.6, 0.875) {};
		\node [style=none] (8) at (-0.15, 0.875) {};
		\node [style=none] (9) at (0.15, 0.875) {};
		\node [style=none] (13) at (-1.9, -0.875) {};
		\node [style=none] (14) at (-1.6, -0.875) {};
		\node [style=none] (15) at (-0.15, -0.875) {};
		\node [style=none] (16) at (0.15, -0.875) {};
		\node [style=none] (20) at (-1.75, 2.125) {};
		\node [style=none] (21) at (0, 2.125) {};
		\node [style=none] (24) at (0, -2.125) {};
		\node [style=none] (25) at (-1.75, -2.125) {};
		\node [style=label] (26) at (-1.75, -2.625) {$\lA_1$};
		\node [style=label] (27) at (0, -2.625) {$\lA_2$};
		\node [style=label] (29) at (-1.75, 2.625) {$\lB_1$};
		\node [style=label] (30) at (0, 2.625) {$\lB_2$};
	\end{pgfonlayer}
	\begin{pgfonlayer}{edgelayer}
		\draw (6.center) to (13.center);
		\draw [in=-90, out=90] (14.center) to (8.center);
		\draw [in=90, out=-90] (7.center) to (15.center);
		\draw (16.center) to (9.center);
		\draw (25.center) to (3);
		\draw (24.center) to (4);
		\draw (1) to (21.center);
		\draw (0) to (20.center);
	\end{pgfonlayer}
\end{tikzpicture}

%% file: figures/structures/g2.tikz
\begin{tikzpicture}
	\begin{pgfonlayer}{nodelayer}
		\node [style=label] (0) at (0, -1) {$\lA_1$};
		\node [style=label] (1) at (2, -1) {$\lA_2$};
		\node [style=label] (2) at (0, 1) {$\lB_1$};
		\node [style=label] (3) at (2, 1) {$\lB_2$};
	\end{pgfonlayer}
	\begin{pgfonlayer}{edgelayer}
		\draw [style=dir] (0) to (2);
		\draw [style=dir] (0) to (3);
		\draw [style=dir] (1) to (3);
		\draw [style=dir] (1) to (2);
	\end{pgfonlayer}
\end{tikzpicture}

%% file: figures/circuits/ccc-diamond.tikz
\begin{tikzpicture}
	\begin{pgfonlayer}{nodelayer}
		\node [style=tiny map] (0) at (1.75, 1.25) {};
		\node [style=none] (1) at (1.575, 1) {};
		\node [style=none] (2) at (1.925, 1) {};
		\node [style=none] (3) at (0.575, 0.25) {};
		\node [style=none] (4) at (0.925, 0.25) {};
		\node [style=none] (5) at (2.575, 0.25) {};
		\node [style=none] (6) at (2.925, 0.25) {};
		\node [style=label] (8) at (1.75, 2.5) {$\lB_2$};
		\node [style=label] (11) at (2.925, 2.5) {$\lB_3$};
		\node [style=label] (12) at (0.575, 2.5) {$\lB_1$};
		\node [style=tiny map] (13) at (1.75, -1.25) {};
		\node [style=tiny map] (14) at (0.75, 0) {};
		\node [style=tiny map] (15) at (2.75, 0) {};
		\node [style=none] (16) at (1.575, -1) {};
		\node [style=none] (17) at (1.925, -1) {};
		\node [style=none] (18) at (0.575, -0.25) {};
		\node [style=none] (19) at (0.925, -0.25) {};
		\node [style=none] (20) at (2.575, -0.25) {};
		\node [style=none] (21) at (2.925, -0.25) {};
		\node [style=label] (23) at (1.75, -2.5) {$\lA_2$};
		\node [style=label] (26) at (2.925, -2.5) {$\lA_3$};
		\node [style=label] (27) at (0.575, -2.5) {$\lA_1$};
	\end{pgfonlayer}
	\begin{pgfonlayer}{edgelayer}
		\draw [in=-90, out=90] (4.center) to (1.center);
		\draw [in=90, out=-90] (2.center) to (5.center);
		\draw [in=90, out=-90] (19.center) to (16.center);
		\draw [in=-90, out=90] (17.center) to (20.center);
		\draw (13) to (23);
		\draw (26) to (21.center);
		\draw (27) to (18.center);
		\draw (0) to (8);
		\draw (11) to (6.center);
		\draw (12) to (3.center);
	\end{pgfonlayer}
\end{tikzpicture}

%% file: sections/sufficiency.tex
\section{Sufficiency of the {\CEPtext}}\label{sec:sufficiency}
In this section we prove the direction $\itmref{itm:main-thm-pic}\Rightarrow\itmref{itm:main-thm-unitary-decos}$ in \cref{thm:main-thm}.
Having previously introduced the canonical circuit shape $\latt{G}$ with connectivity $G$, the task now is to provide $\latt{G}$ with suitable quantum semantics that implements a given unitary channel $\U : \aA_\labelsA \to \aB_\labelsB$ with causal structure $G_{\U}\subseteq G$.
For the theorem below, which achieves precisely that, recall the algebraic view on quantum systems and our notation from \cref{subsec:prelims-algebra}; in particular, we write $\aB_{\b}$, for $\b\subseteq\labelsB$, to denote the tensor product $\bigotimes_{\lB\in\b} \aB_\lB$ (or $\mathbb C$ if $\b=\emptyset$) but often also let it refer to the corresponding subalgebra $\aB_\b \tns \{\one_{\aB_{\labelsB\setminus\b}}\}$ of $\aB_\labelsB$ ($\mathbb C \one_{\aB_\labelsB}$ if $\beta=\emptyset$).
Furthermore, we write, for $v\in\latt{G}$ (recalling~\eqref{eq:covm-and-covp})
\begin{align}
    \Covm(v) &\coloneqq \{u \in \latt{G} \mid u \cvr v \}, \\
    \Covp(v) &\coloneqq \{w \in \latt{G} \mid v \cvr w \}, \\
    \down v &\coloneqq \{u \in \latt{G} \mid u \leq v\}, \\
    \strdown v &\coloneqq \{u\in \latt{G} \mid u < v\},
\end{align}
and we abbreviate $\la_{\latt{G}}$ by $\la$ and $\mu_{\latt{G}}$ by $\mu$, as before.
The theorem below is formulated in a somewhat roundabout way to anticipate its inductive proof.

\begin{theorem}[{entailing $\itmref{itm:main-thm-pic}\Rightarrow\itmref{itm:main-thm-unitary-decos}$ in \cref{thm:main-thm}}]
    \label{thm:inductive-semantics-thm}
    Let $G\subseteq\labelsA\times\labelsB$ be a binary relation satisfying \PIC{}, suppose that $\aA_\lA$ for $\lA\in\labelsA$ and $\aB_\lB$ for $\lB\in\labelsB$ are quantum systems, and let $\U : \aA_\labelsA \to \aB_\labelsB$ be a {\unichan} whose causal structure satisfies $G_\U \subseteq G$.
    Let $\latt{G}$ be the canonical circuit shape from \cref{def:concept-lattice-and-circuit}.
    There exist quantum systems
    \begin{equation}
        \aZ_v^w \text{ \quad for all } v,w\in \latt{G}\text{ such that } v \cvr w
    \end{equation}
    and {\unichan}s
    \begin{equation}
        \label{eq:Vv}
        \V_v : \aA_{\la^{-1}(v)} \aZ_{\Covm(v)}^v \xrightarrow{\sim} \aZ_v^{\Covp(v)} \aB_{\mu^{-1}(v)} \text{\quad for all } v\in \latt{G}
    \end{equation}
    such that, defining $\V_T$, for any $T\subseteq \latt{G}$, to be the composition of $\{\V_v\mid v\in T\}$ along the systems $\{\aZ_v^w \mid v,w\in T, v\cvr w\}$, we have that for all $v\in \latt{G}$,
    \begin{subequations}
        \label{eq:induction-hypothesis}
        \begin{alignat}{3}
            \label{eq:induction-hypothesis-b} \aB_\lB &= \V_{\down v}\U^\dagger (\aB_\lB) &&\text{\quad for } \lB\in\mu^{-1}(v); \\
            \label{eq:induction-hypothesis-z} \aZ_v^w &\subseteq \V_{\down v} \U^\dagger(\aB_{\b_w}) &&\text{\quad for } w\in\Covp v.
        \end{alignat}
    \end{subequations}

    As a result of \cref{eq:induction-hypothesis-b}, the complete {\unichan} $\V_{\latt{G}}$ is identical to $\U$ up to local {\unichan}s on the output systems $\{\aB_\lB\}_{\lB\in\labelsB}$.
    Therefore, $\U$ has a unitary circuit decomposition of shape $\latt{G}$.
\end{theorem}

\begin{remark}
    Given a relation $G\subseteq\labelsA\times\labelsB$, recall from \cref{sec:canonical-circuit} that the canonical circuit shape $\latt{G}$ may contain either or both of $\< \emptyset, \labelsB \>$ (with no input wires) as minimum and $\< \labelsA, \emptyset \>$ (with no output wires) as maximum---see e.g.\ the minimum and maximum in \cref{eq:canonical-circuit-example}.
    As noted in \cref{rmk:latt-vs-hlatt}, bookkeeping is simpler if we leave those gates in the circuit shape $\latt{G}$ even though in a unitary circuit they necessarily correspond to identity transformations between trivial (one-dimensional) quantum systems and could therefore in principle be removed.
    Indeed, the proof of \cref{thm:inductive-semantics-thm} will automatically ensure that in the circuit decomposition of $\U$ with shape $\latt{G}$ asserted to exist by the theorem, the unitary channels $\V_{\<\emptyset,\labelsB\>}$ and $\V_{\<\labelsA,\emptyset\>}$ (if they exist) are identity channels on trivial systems.
    If desired, these trivial gates can at the end be dropped from the obtained unitary causal decomposition.
\end{remark}

We will prove \cref{thm:inductive-semantics-thm} by induction on $\latt{G}$, providing it with a quantum semantics from the bottom upwards.
At each box $v\in\latt G$, we are given incoming systems $\aA_{\la^{-1}(v)}\aZ_{\Covm(v)}^v$ which need to be refactored into appropriate outgoing systems $\aZ_v^{\Covp(v)} \aB_{\mu^{-1}(v)}$ by making use of the fact that $G_\U\subseteq G$.
(All of these algebras may be trivial if $v$ is $\latt{G}$'s minimal or maximal element.)
Since the latter can be expressed in terms of commutation relations between the sets of algebras $\{\aA_\lA\}_\lA$ and $\{\U^\dagger(\aB_\lB)\}_\lB$, the proof of each induction step reduces to a problem of representing subalgebras subject to commutation relations on appropriate tensor product factors.
Specifically, we will rely on \cref{lma:the-algebraic-lemma} below.
As this result exposes the operator-algebraic core of our methods, we temporarily adopt the terminology of `finite-dimensional factor {\cstaralg}' and `*-isomorphism' where we normally say `quantum system' and `{\unichan}', respectively.
This terminology is also used in \cref{app:algebra}, where \cref{lma:the-algebraic-lemma} is proven.

\begin{restatable}{lemma}{algebraiclemma}
    \label{lma:the-algebraic-lemma}
    Let $\aA,\aX_1,\dots,\aX_n$ be finite-dimensional factor {\cstaralg}s and let $\aB_k \subseteq \aA\aX_1\aX_2\cdots\aX_n$ for $k\in\{1,\dots,n\}\eqqcolon[n]$ be unital factor *-subalgebras of their tensor product so that
    \begin{enumerate}
        \item\label{itm:the-algebraic-lemma-commuting-assumption} $\aB_k\subseteq \aB_l'$ for all distinct $k,l\in[n]$;
        \item\label{itm:the-algebraic-lemma-overlap-assumption} $\aB_k \subseteq \aA\aX_k = (\aX_{[n]\setminus\{k\}})'$ for all $k\in[n]$; and
        \item\label{itm:the-algebraic-lemma-crucial-assumption} $\aA \subseteq \bigjoin_{k\in[n]} \aB_k$, the {\cstaralg} generated by the $\aB_k$s.
    \end{enumerate}
    Then there are finite-dimensional factor {\cstaralg}s $\aZ_k$ for each $k\in[n]$ and a *-isomorphism
    \begin{equation}
        \label{eq:the-algebraic-lemma-isomorphism}
        \V : \aA \xrightarrow{\sim} \aZ_1\aZ_2\cdots\aZ_n
    \end{equation}
    so that, writing $\V$ for $\V\tns\id_{\aX_1\cdots\aX_n}$, for each $k$ we have
    \begin{subequations}
        \begin{align}
            \label{eq:the-algebraic-lemma-conclusion-1} \V(\aB_k) &\subseteq \aZ_k \aX_k \text{ and } \\
            \label{eq:the-algebraic-lemma-conclusion-2} \aZ_k &= \V(\aA\cap\aB_k).
        \end{align}
    \end{subequations}
\end{restatable}
\begin{proof}
    See \cref{app:algebra}.
    Each algebra $\aZ_k$ ends up being given by the \emph{relative bicommutant} of $\aB_k \subseteq \aA\aX_1\aX_2\cdots\aX_n$ within $\aA$, which can be understood as the algebra generated by the support of $\aB_k$ on the factor $\aA$.
\end{proof}

Assumption~\itmref{itm:the-algebraic-lemma-crucial-assumption} is crucial here.
Without it, a similar representation can be given, with the exception that the right-hand side of \cref{eq:the-algebraic-lemma-isomorphism} will generally be an algebra of operators on a \emph{direct sum over} tensor products of Hilbert spaces.
Such a representation can be used, in some cases, to provide proofs of the \emph{routed} unitary circuit decompositions shown in \textcite{LB21}.
The assumption of \PIC, however, allows us to focus entirely on cases where assumption~\itmref{itm:the-algebraic-lemma-crucial-assumption} is satisfied, therefore yielding the tensor product factorisations in \cref{eq:the-algebraic-lemma-isomorphism} required to construct our (non-routed) unitary circuit.
The resulting proof of \cref{thm:inductive-semantics-thm} below forms, in a sense, the core of this work, combining the combinatorial characterisation of \PIC{} in \cref{sec:canonical-circuit} with the operator-algebraic result of \cref{lma:the-algebraic-lemma} to show the existence of unitary causal decompositions.

\begin{proof}[Proof of \cref{thm:inductive-semantics-thm}]
    \input{sections/sufficiency-proof}
\end{proof}

%% file: sections/sufficiency-proof.tex
We proceed by induction on $\latt{G}$.
The base case---i.e.\ when $v$ is the minimal element of $\latt{G}$---separates into two cases, where $\a_v = \emptyset$ or $\a_v \neq\emptyset$.

Suppose first that $\a_v = \emptyset$.
In this case $\aA_{\la^{-1}(v)} = \aA_\emptyset = \C$ is trivial, as is each $\aB_\lB$ for $\lB \in \mu^{-1}(v)$, since $G^{-1}(\lB)=\a_v$ so that $\U^\dagger(\aB_\lB) \subseteq \aA_{\a_v} = \aA_\emptyset = \C$.
Therefore it suffices to set $\aZ_v^w \coloneqq \C$ for each $w\in\Ch(v)$ and to let $\V_v$ be the trivial transformation between trivial systems.
This choice satisfies \cref{eq:induction-hypothesis}.

The base case where $\a_v \neq\emptyset$ is covered by the induction case (apart from some of the algebras below being trivial), which we turn to now.

So let $v\in\latt{G}$ be arbitrary such that $\a_v \neq \emptyset$; and assume that there exist systems $\aZ_u^{u'}$ for all $u,u' \in \latt{G}$ such that $u\cvr u' \leq v$ and, for each $u<v$, a {\unichan} $\V_u : \aA_{\la^{-1}(u)} \aZ_{\Covm(u)}^u \xrightarrow{\sim} \aZ_u^{\Covp(u)} \aB_{\mu^{-1}(u)}$ such that $\V_{\down u}$ satisfies~\eqref{eq:induction-hypothesis}.
Consider the composition of these {\unichan}s $\V_u$ for $u<v$,
\begin{equation}
    \label{eq:induction-step-V}
    \V_{\strdown v} : \aA_{\a_v\setminus\la^{-1}(v)} \to \aZ_{\Covm(v)}^v \aW,
\end{equation}
where
\begin{equation}
    \aW \coloneqq \bigotimes_{u<v} \aZ_{u}^{\Covp(u)\setminus\down v} \aB_{\mu^{-1}(u)}.
\end{equation}
(In the base case, let $\V_{\strdown v}$ denote the identity channel on the trivial system $\aA_{\a_v\setminus\la^{-1}(v)} = \aA_{\emptyset} =\mathbb C$.)
From now on we will, abusing notation as always, let $\V_{\strdown v}$ denote instead the extended unitary
\begin{equation}
    \label{eq:vanx-typing}
    \V_{\strdown v} \equiv \V_{\strdown v} \tns \id_{\labelsA\setminus(\a_v\setminus\la^{-1}(v))} : \aA_\labelsA
    \xrightarrow{\sim} \aA_{\la^{-1}(v)} \aZ_{\Covm(v)}^{v} \aW \aA_{\labelsA\setminus \a_v}.
\end{equation}
We wish to construct systems $\aZ_v^w$ for $w\in\Covp(v)$ and a {\unichan} $\V_v : \aA_{\la^{-1}(v)} \aZ_{\Covm(v)}^v \xrightarrow{\sim} \aZ_v^{\Covp(v)} \aB_{\mu^{-1}(v)}$ so that \cref{eq:induction-hypothesis} is satisfied for $\V_{\down v}$.
We will do this by an application of \cref{lma:the-algebraic-lemma}.
To this end, consider the following subalgebras of $\aA_{\la^{-1}(v)} \aZ_{\Covm(v)}^{v} \aW \aA_{\labelsA\setminus \a_v}$:
\begin{subequations}
    \label{eq:tilde-b-defs}
    \begin{alignat}{3}
        \tilde \aB_{\beta_w} &\coloneqq \V_{\strdown v}\U^{\dagger}(\aB_{\beta_w}) &&\text{\quad for } w\in\Covp(v) \\
        \text{and \quad}  \tilde \aB_\lB &\coloneqq \V_{\strdown v}\U^{\dagger}(\aB_\lB) &&\text{\quad for } \lB\in\mu^{-1}(v).
    \end{alignat}
\end{subequations}
We prove the following claims necessary for the application of \cref{lma:the-algebraic-lemma}.
Keep in mind that $\aA_{\la^{-1}(v)}\aZ_{\Covm(v)}^v$ is to be the input to the unitary $\V_v$ that we are looking to construct, and thus plays the role of the factor $\aA$ in \cref{lma:the-algebraic-lemma}.
\begin{enumerate}
    \item \label{itm:condition-i} All subalgebras defined in \cref{eq:tilde-b-defs} commute pairwise:
    \begin{subequations}
        \label{eq:condition-i}
        \begin{align}
            \label{eq:condition-i-1} \tilde \aB_\lB &\subseteq (\tilde \aB_{\lB'})' \text{\quad for all distinct } \lB,\lB'\in\mu^{-1}(v); \\
            \label{eq:condition-i-2} \tilde \aB_\lB &\subseteq (\tilde \aB_{\beta_w})' \text{\quad for all } \lB\in\mu^{-1}(v) \text{ and } w\in\Covp(v); \\
            \label{eq:condition-i-3} \text{and\quad} \tilde \aB_{\beta_w} &\subseteq (\tilde \aB_{\beta_{w'}})' \text{\quad for all distinct } w,w'\in\Covp(v).
        \end{align}
    \end{subequations}

    \item \label{itm:condition-ii} We have
    \begin{subequations}
        \begin{alignat}{3}
            \label{eq:condition-ii-bj} \tilde \aB_\lB &\subseteq \aA_{\la^{-1}(v)} \aZ_{\Covm(v)}^{v} = \left( \aW \aA_{\labelsA\setminus\a_v} \right)' &&\text{\quad for } \lB\in\mu^{-1}(v); \\
            \label{eq:condition-ii-bbw} \tilde \aB_{\beta_w} &\subseteq \aA_{\la^{-1}(v)} \aZ_{\Covm(v)}^{v} \aA_{\PaG(\b_w)\setminus\a_v} = \left( \aW  \aA_{\labelsA\setminus\PaG(\b_w)} \right)' &&\text{\quad for } w\in\Covp(v),
        \end{alignat}
    \end{subequations}
    where we again write $\PaG(\b_w) \coloneqq \bigcup_{\lB\in\b_w} \PaG(\lB)$.
    Moreover, for any two distinct $w,w'\in\Covp(v)$, the sets $\PaG(\b_w)\setminus\a_v$ and $\PaG(\b_{w'})\setminus\a_v$ are disjoint.
    Therefore, the only tensor factors of $\aA_{\la^{-1}(v)} \aZ_{\Covm(v)}^{v} \aW \aA_{\labelsA\setminus \a_v}$ on which more than one of the subalgebras from \cref{eq:tilde-b-defs} act non-trivially are $\aA_{\la^{-1}(v)} \aZ_{\Covm(v)}^{v}$.

    \item \label{itm:condition-iii} Our algebra of interest $\aA_{\la^{-1}(v)} \aZ_{\Covm(v)}^{v}$ is generated by the subalgebras under consideration:
    \begin{equation}
        \label{eq:algebra-is-spanned}
        \aA_{\la^{-1}(v)} \aZ_{\Covm(v)}^{v} \subseteq \left(\bigjoin_{\lB\in\mu^{-1}(v)} \tilde \aB_\lB\right) \join \left(\bigjoin_{w\in\Covp(v)} \tilde \aB_{\beta_w}\right).
    \end{equation}
\end{enumerate}

Once \itmref{itm:condition-i}--\itmref{itm:condition-iii} are established, \cref{lma:the-algebraic-lemma} will give quantum systems $\aZ_v^w$ for $w\in\Covp(v)$ and $\hat \aB_\lB$ for $\lB\in\mu^{-1}(v)$, as well as a {\unichan} $\V_v: \aA_{\la^{-1}(v)} \aZ_{\Covm(v)}^v \to \aZ_v^{\Covp(v)} \hat \aB_{\mu^{-1}(v)}$.
\cref{eq:the-algebraic-lemma-conclusion-2} entails that for $\lB\in\mu^{-1}(v)$,
\begin{equation}
    \hat\aB_\lB = \V_v(\aA_{\la^{-1}(v)} \aZ_{\Covm(v)}^v \cap \tilde\aB_\lB) = \V_v(\tilde\aB_\lB) = \V_{\down v}\U^{\dagger}(\aB_\lB),
\end{equation}
where in the second equality we have applied \cref{eq:condition-ii-bj}.
This means in particular that $\hat\aB_\lB \cong \aB_\lB$, so we can choose $\hat\aB_\lB$ and $\V_v$ so that in fact $\hat\aB_\lB = \aB_\lB$.
This establishes the induction hypothesis \cref{eq:induction-hypothesis-b}.

To establish the other part~\eqref{eq:induction-hypothesis-z}, note that according to \cref{eq:the-algebraic-lemma-conclusion-2},
\begin{equation}
    \label{eq:almost-induction-hypothesis}
    \aZ_v^w = \V_v(\aA_{\la^{-1}(v)} \aZ_{\Covm(v)}^v \cap \tilde \aB_{\b_w}) \subseteq \V_v(\tilde \aB_{\b_w}) = \V_{\down v}\U^{\dagger} (\aB_{\b_w}).
\end{equation}

It remains to prove \itmref{itm:condition-i}--\itmref{itm:condition-iii} and the final statement of \cref{thm:inductive-semantics-thm}.

\begin{innerproof}
    [Proof of $\itmref{itm:condition-i}$]
    Note that since $\V_{\strdown v}\U^\dagger$ is a *-isomorphism, the algebras in \cref{eq:condition-i} commute iff the same holds for the respective algebras without tildes (see \cref{eq:tilde-b-defs}), considered as subalgebras of $\aB_\labelsB$.
    It is clear that $\aB_\lB \subseteq (\aB_{\lB'})' = \aB_{\labelsB\setminus\{\lB'\}}$ for distinct $\lB,\lB'\in\mu^{-1}(v)$, establishing~\eqref{eq:condition-i-1}.
    For~\eqref{eq:condition-i-2}, it suffices to note that if $\lB\in\mu^{-1}(v)$ and $w\in\Covp(v)$, then $\lB\notin \b_w$.
    This follows from \cref{prop:useful-lattice-properties}\itmref{itm:propo-one}, which entails that $\lB\in\mu^{-1}(v)$ iff $v$ is the largest element of $\latt{G}$ so that $\lB\in\b_v$; for any $w\in\Covp(v)$, we have $w>v$, which implies that $\lB$ cannot be in $\b_w$.
    Finally, for \cref{eq:condition-i-3} it suffices to show that $\b_w\cap\b_{w'} = \emptyset$ for distinct $w,w'\in\Covp(v)$.
    This follows by assumption of \PIC{} and \cref{thm:pic-equivalent-conditions}\itmref{itm:pic-betas-disjoint}.
\end{innerproof}

\begin{innerproof}
    [Proof of $\itmref{itm:condition-ii}$]
	If $v$ is the minimum of $\latt{G}$, it is straightforward to verify $\itmref{itm:condition-ii}$ by using that $\la^{-1}(v) = \a_v$ and $G_\U \subseteq G$.
    Assume therefore that $v$ is not the minimum.
    Recall that
    \begin{equation}
        \aW = \bigotimes_{u<v} \aZ_{u}^{\Covp(u)\setminus\down v} \aB_{\mu^{-1}(u)}.
    \end{equation}
    First we show that
    \begin{equation}
        \label{eq:W-claim}
        \aW \subseteq \V_{\strdown v}\U^\dagger(\aB_{\labelsB\setminus\b_v}).
    \end{equation}
    Fix $u\in\latt{G}$ so that $u<v$ and $\a_u \neq \emptyset$. (If $\a_u = \emptyset$ then $\aZ_{u}^{\Covp(u)\setminus\down v} \aB_{\mu^{-1}(u)}$ is trivial so need not be considered.)
    It follows from the induction hypotheses~\eqref{eq:induction-hypothesis} that
    \begin{equation}
        \aZ_u^{\Covp(u)\setminus\down v} \aB_{\mu^{-1}(u)} \subseteq \V_{\down u} \U^{\dagger}\left(\aB_{\b_{\Covp(u)\setminus\down v} \cup \mu^{-1}(u)}\right)
    \end{equation}
    where $\b_{\Covp(u)\setminus\down v} \coloneqq \bigcup\{\b_w \mid w\in\Covp(u)\setminus\down v \}$.
    Now, $\V_{\strdown v}\U^{\dagger} = \V_{\strdown v\setminus\down u}\circ\V_{\down u}\circ\U^\dagger$, but $\V_{\strdown v\setminus\down u}$ acts trivially on $\aZ_u^{\Covp(u)\setminus\down v} \aB_{\mu^{-1}(u)}$.
    Therefore we also have
    \begin{equation}
        \aZ_u^{\Covp(u)\setminus\down v} \aB_{\mu^{-1}(u)} \subseteq \V_{\strdown v} \U^{\dagger}\left(\aB_{\b_{\Covp(u)\setminus\down v} \cup \mu^{-1}(u)}\right).
    \end{equation}
    To establish \cref{eq:W-claim} it remains to show that
    \begin{equation}
        \label{eq:asldfj}
        \b_{\Covp(u)\setminus\down v} \cap \b_v = \emptyset \text{\quad and\quad} \mu^{-1}(u) \cap \b_v = \emptyset.
    \end{equation}
    The first statement relies on \PIC{}.
    Let $w \in \Covp(u)\setminus\down v$.
    Let $w'$ be such that $u\lessdot w' \leq v$.
    By \PIC{}, in particular the formulation of \cref{thm:pic-equivalent-conditions}\itmref{itm:pic-betas-disjoint}, it follows that $\b_w\cap\b_{w'} = \emptyset$, and since $w'\leq v$, we have $\b_v\subseteq \b_{w'}$ so $\b_w \cap \b_v = \emptyset$ too.
    This holds for all $w\in\Covp(u)\setminus\down v$, which implies that $\b_{\Covp(u)\setminus\down v} \cap \b_v = \emptyset$.

    To show the second equation in~\eqref{eq:asldfj}, suppose that $\lB\in\mu^{-1}(u)$.
    In this case, $u$ is the greatest element of $\latt{G}$ satisfying $\lB\in\b_u$ (see \cref{prop:useful-lattice-properties}\itmref{itm:propo-one}).
    Since $v > u$, we can't have $\lB\in\b_v$.
    This completes the proof of~\eqref{eq:asldfj} and hence of~\eqref{eq:W-claim}.

    To show \cref{eq:condition-ii-bj}, it suffices to show that $\tilde \aB_\lB$, for $\lB\in\mu^{-1}(v)$, commutes with $\aW$ and with $\aA_{\labelsA\setminus\a_v}$.
    Commutation with $\aW$ follows from \cref{eq:W-claim} and the fact that $\lB\in\mu^{-1}(v)\subseteq\b_v$.
    To show commutation with $\aA_{\labelsA\setminus\a_v}$, note that since $\U$ has causal structure $G_\U\subseteq G$, we have $\U^\dagger(\aB_\lB) \subseteq \aA_{\PaGarg{G_\U}(\lB)} \subseteq \aA_{\PaG(\lB)} = \aA_{\gP(\{\lB\})} = \aA_{\a_{\mu(\lB)}} = \aA_{\a_v}$.
    Applying $\V_{\strdown v}$, which acts trivially on $\aA_{\labelsA\setminus\a_v}$, to both sides yields $\tilde \aB_\lB = \V_{\strdown v}\U^\dagger (\aB_\lB) \subseteq \aA_{\a_v}$, so $\tilde \aB_\lB$ must commute with $\aA_{\labelsA\setminus\a_v}$.

    Similarly, for \cref{eq:condition-ii-bbw} we need to show that $\tilde \aB_{\b_w}$ commutes with $\aW$ and with $\aA_{\labelsA\setminus\PaG(\b_w)}$.
    The first fact follows from \cref{eq:W-claim} and the fact that $\b_w\subseteq\b_v$ for $w\in\Covp(v)$.
    For the second fact, note that since $G_\U\subseteq G$, $\U^\dagger(\aB_{\b_w}) \subseteq \aA_{\PaGarg{G_\U}(\b_w)} \subseteq \aA_{\PaG(\b_w)}$.
    This implies that $\tilde \aB_{\b_w}$ commutes with $\aA_{\labelsA\setminus\PaG(\b_w)}$ by an argument similar to the above.

    The final claim that $\PaG(\b_w)\setminus\a_v$ and $\PaG(\b_{w'})\setminus\a_v$ are disjoint for distinct $w,w'\in\Covp(v)$ follows directly from \cref{lma:overlap-only-on-alpha}, which established that $\PaG(\b_w)\cap\PaG(\b_{w'}) = \a_v$. 
\end{innerproof}

\begin{innerproof}
    [Proof of $\itmref{itm:condition-iii}$]
    Observe that by~\cref{eq:beta-union} in \cref{prop:useful-lattice-properties}\itmref{itm:propo-beta-union}, the algebra on the right-hand side of \cref{eq:algebra-is-spanned} is just $\tilde \aB_{\b_v} \coloneqq \V_{\strdown v}\U^\dagger (\aB_{\b_v})$.
    For each $\lA\in\la^{-1}(v)$, $\ChG(\lA) = \b_v$ and therefore $\aA_{\la^{-1}(v)} \subseteq \U^\dagger(\aB_{\b_v})$.
    $\V_{\strdown v}$ acts trivially on $\aA_{\la^{-1}(v)}$, so we get $\aA_{\la^{-1}(v)} \subseteq \V_{\strdown v} \U^\dagger(\aB_{\b_v}) = \tilde \aB_{\b_v}$.
    It remains to show that $\aZ_u^v\subseteq \tilde \aB_{\b_v}$ for $u\in\Covm(v)$.
    This follows from the induction hypothesis~\eqref{eq:induction-hypothesis-z} applied to $\aZ_u^v$ and the fact that $\V_{\strdown v\setminus\down u}$ acts trivially on $\aZ_u^v$.
\end{innerproof}

Let us now prove the final statement of the theorem.
Consider the unitary $\V_{\latt{G}} : \aA_\labelsA \to \aB_\labelsB$ and let $\lB\in\labelsB$.
Note that $\V_{\latt{G}} = \V_{\latt{G}\setminus\down\mu(\lB)} \circ \V_{\down\mu(\lB)}$ and that $\V_{\latt{G}\setminus\down\mu(\lB)}$ acts trivially on $\aB_\lB$.
Therefore, by \cref{eq:induction-hypothesis-b},
\begin{equation}
    \V_{\latt{G}}\U^\dagger(\aB_\lB) = \V_{\latt{G}\setminus\down\mu(\lB)}\left(\V_{\down\mu(\lB)}\U^\dagger(\aB_\lB)\right) = \V_{\latt{G}\setminus\down\mu(\lB)}(\aB_\lB) = \aB_\lB.
\end{equation}
Let $\W_\lB : \aB_\lB \to \aB_\lB$ be the *-automorphism defined by $\W_\lB \coloneqq \left(\U\V_{\latt{G}}^\dagger\right) \big|_{\aB_\lB}$;
then $\W_\lB\V_{\latt{G}}\U^\dagger\big|_{\aB_\lB} = \id_{\aB_\lB}$.
Since this holds for each $\lB\in\labelsB$, we have $\left(\bigotimes_{\lB\in\labelsB} \W_\lB\right)\V_{\latt{G}} \U^\dagger = \id_{\aB_\labelsB}$, so that
\begin{equation}
    \left(\bigotimes_{\lB\in\labelsB} \W_\lB\right)\V_{\latt{G}} = \U.
\end{equation}
By construction $\V_{\latt{G}}$ has a unitary circuit decomposition of shape $\latt{G}$; by incorporating the local unitaries $\W_\lB$ into the existing gates, we see that $\U$ itself also admits a unitary circuit decomposition of the same shape, completing the proof.

%% file: sections/discussion.tex
\section{Discussion}\label{sec:discussion}
This work has studied causal decompositions of unitary transformations and, like Ref.~\cite{LB21}, asked: when does a causal constraint $G\subseteq\labelsA\times\labelsB$---as a purely combinatorial object---imply the existence of causal decompositions for \emph{all} unitaries satisfying the constraint $G_\U \subseteq G$?
Answering an open question from~\cite{LB21} we have fully characterised those structures for which \emph{traditional} (non-routed) \emph{unitary} circuit decompositions suffice, that is, ones consisting only of tensor products and sequential compositions of unitary gates.
Put into more algebraic terms, as is indeed natural when studying causal decompositions, we have characterised the class of causal constraints that can be explained by viewing the unitary as a sequence of re-factorisations of \emph{factor} algebras into other \emph{factor} algebras.
\cref{app:generic-channels} considers decompositions of generic (non-unitary) quantum channels and discusses our motivation for focussing on unitary channels in the main text.

The main results of this paper (\cref{thm:main-thm} and \cref{thm:pic-equivalent-conditions}) can be summarised by the equivalence of the following four conditions for any given binary relation $G \subseteq \labelsA\times\labelsB$.
\begin{enumerate}[label=(\arabic*)]
    \item $G$ satisfies \PIC{}, i.e.\ it restricts nowhere to the `forbidden' relation
    \begin{equation*}
        \cccG\ = \input{structures/ccc-structure-numbers.tikz}.
    \end{equation*}
    \item \label{itm:discussion-unique-paths} The canonical circuit shape $\latt{G}$ contains no copy of $\latt{\cccG}$, in the sense formally stated in \cref{thm:pic-equivalent-conditions}\itmref{itm:pic-exclusion-of-diamond}. Equivalently, $\latt{G}$ has no more than one path between each input and output.
    \item\label{itm:discussion-causaldecos} $G$ is compositionally representable in unitary circuits. That is, every {\unichan} $\U$ satisfying $G_\U \subseteq G$ admits a decomposition into a unitary circuit $\circuit$ with connectivity $G_\circuit \subseteq G$.
    \item \label{itm:discussion-faithful} Every {\unichan} $\U$ satisfying $G_\U = G$ admits a \emph{causally faithful} unitary circuit decomposition, i.e.\ a decomposition into a unitary circuit $\circuit$ with connectivity $G_\circuit = G_\U$ (proven in \cref{app:faithful}).
\end{enumerate}

Behind condition \itmref{itm:discussion-unique-paths} is a central aspect of this work, which leverages a purely syntactic result from Ref.~\cite{vdL25} (cf.\ \cref{thm:canonicity}): a relation $G$ induces a lattice $\latt{G}$, which in turn induces a canonical, most expressive circuit shape with connectivity $G$; that is, one such that any circuit with connectivity (at most) $G$ can be rewritten into it.
Being able to refer to the concept lattice $\latt{G}$ was instrumental in two ways.
First, it allowed us to rephrase \PIC{} in terms that are useful for proving its sufficiency for \itmref{itm:discussion-causaldecos} and \itmref{itm:discussion-faithful}.
Second, together with the operator-algebraic \cref{lma:the-algebraic-lemma}, it provided the very blueprint for the construction of a decomposition for any given unitary satisfying the appropriate causal constraint.

Both of these two aspects are promising looking into the future, as they offer a view to a systematic lattice-based treatment of causal decompositions more generally.
We now have the right kind of syntactic object, namely $\latt{G}$, in terms of which one can hope to identify a combinatorial condition---a weakening of \PIC{}---that is necessary and sufficient for $G$ to be representable in terms of \emph{routed} unitary causal decompositions.
This would provide a complete answer to the main open question in Ref.~\cite{LB21}.
Moreover, one can expect that also here the concept lattice does not only provide the terms for such a condition, but also the syntax for the \emph{routed} circuit and the skeleton for an inductive proof of the existence of appropriate semantics, generalising the one for sufficiency of \PIC{} here.
Indeed, the same result of Ref.~\cite{vdL25} used here implies that all routed unitary circuits can be rewritten into the concept lattice form; and one may verify that all routed unitary causal decompositions thus far found in the literature~\cite{LB21,VMA25b} have been of that form.\footnote{The only exception is Theorem~8 in~\cite{LB21}, which can however be simplified into the canonical form.}

Finally, we note that there are other lines of attack to make progress towards a complete understanding of causal decompositions.
Rather than studying classes of {\unichan}s with a given causal structure or that satisfy a given causal constraint, as here and in Ref.~\cite{LB21}, one could investigate classes of {\unichan}s characterised by other properties that may imply the existence of causal decompositions.
An example of this kind is a result in Refs.~\cite{vanderlugt2026causalcompositional,losec-pirsa}, showing that any Clifford unitary---no matter what its causal structure happens to be---has a causally faithful circuit decomposition, which however is in general not unitary as it relies on ancillary systems that start out in an entangled state.

%% file: sections/generic-channels.tex
\section{Decompositions of generic, non-unitary channels}\label{app:generic-channels}
The main text of this work focusses on circuit decompositions of unitary channels.
One might wonder whether an appropriate generalisation of causal decompositions exists for the case of generic channels (CPTP maps) $\E:\aA_\labelsA\to\aB_\labelsB$.
Indeed, many of the definitions given for unitary channels in \cref{sec:prelims} have natural analogues for generic channels.
For instance, \cref{eq:no-influence} defines a notion $\aA_\a\ninfl_\E \aB_\b$ usually referred to as \defn{no-signalling} through $\E$, with the terminology of \emph{no-influence} being reserved for the unitary case (following e.g.~\cite{BLO19,LB21,ABH17}).
The signalling relations $\aA_\lA\infl_\E\aB_\lB$ for individual elements $\lA\in\labelsA,\lB\in\labelsB$ define a relation $G_\E\subseteq\labelsA\times\labelsB$ that one might call the channel's \emph{single-system signalling structure}.
Additionally, one may consider circuit decompositions $\circuit$ of generic channels (\cref{def:circuit-and-circuit-decomposition}) and their connectivity $G_\circuit$.
By analogy to \cref{prop:soundness}, absence-of-paths is then sound for no-signalling; that is, if $\E$ has circuit decomposition $\circuit$ then $G_\E \subseteq G_\circuit$.
One might thus wonder whether, or in what special cases, there exist circuit decompositions $\circuit$ of $\E$ that explain a given collection of no-signalling relations $G_\E \subseteq G$ by factoring the inclusion through as $G_\E \subseteq G_\circuit \subseteq G$.

One can generally however not expect such decompositions to exist.
One reason is that signalling through generic channels is not \emph{atomic} in the sense of \cref{prop:causal-atomicity}: if $\E$, for instance, satisfies the no-signalling relations $\aA_{\lA} \ninfl_\E \aB_{\lB_1}$ and $\aA_{\lA} \ninfl_\E \aB_{\lB_2}$ for some $\lA\in\labelsA$ and $\lB_1,\lB_2\in\labelsB$, then it may still exhibit signalling to the composite system $\aA_{\lA} \infl_\E \aB_{\{\lB_1,\lB_2\}}$.
An example is the qubit channel
\begin{equation}
    \label{eq:failure-of-atomicity-example}
    \input{circuits/failure-of-atomicity-example-pic.tikz},
\end{equation}
where $\ket{\Phi^+} \coloneqq \left(\ket{00}+\ket{11}\right)/\sqrt2$ is the maximally entangled state.
The connectivity of a quantum circuit, on the other hand, \emph{is} atomic, in the sense that the absence of paths from $\lA$ to $\lB_1$ and from $\lA$ to $\lB_2$ through a circuit decomposition of $\E$ implies the absence of a path from $\lA$ to the joint system $\{\lB_1,\lB_2\}$---which in turn necessarily implies the no-signalling relation $\aA_{\lA} \ninfl_\E \aB_{\{\lB_1,\lB_2\}}$.
Equality $G_\E=G_\circuit$ for the example above can thus not be reached: any circuit decomposition $\circuit$ satisfies the strict inclusion $G_\E\subsetneq G_\circuit$.

A more fundamental understanding of this failure can be reached by considering unitary dilations of $\E$---unitary channels $\U:\aA_\labelsA\aE\to\aB_\labelsB\aF$ satisfying $\E(-) = \Tr_F \U(-\tns\rho)$ for some state $\rho\in\aE$---and their causal structure $G_\U \subseteq (\labelsA\cup\{\lE\})\times(\labelsB\cup\{\lF\})$.
Note first of all the following, which can be straightforwardly verified.

\begin{proposition}
    \label{prop:signalling-v-influence}
    If $\E$ has a unitary dilation $\U$, then for any $\a\subseteq\labelsA$ and $\b\subseteq\labelsB$, $\aA_\a \ninfl_\U \aB_\b$ implies $\aA_\a \ninfl_\E \aB_\b$.
    When specialising to singletons $\a,\b$, this entails that $G_\E \subseteq G_\U \cap (\labelsA\times\labelsB)$.
\end{proposition}

\noindent In other words, the absence of \emph{influence} on the fundamental, unitary level implies the absence of \emph{signalling} on the operational level.
The converse however generally fails: influences on the unitary level may become unobservable on the operational level due to a fine-tuned choice of dilation state $\rho$.
In fact, \emph{every} unitary dilation $\U$ of the example channel in \cref{eq:failure-of-atomicity-example} satisfies the strict inclusion $G_\E \subsetneq G_\U\cap(\labelsA\times\labelsB)$.
Indeed, suppose that $G_\E = G_\U\cap(\labelsA\times\labelsB)$, so that in particular $\aA_{\lA} \ninfl_\U \aB_{\lB_1}$ and $\aA_{\lA} \ninfl_\U \aB_{\lB_2}$; causal atomicity, which does hold for $\U$ (\cref{prop:causal-atomicity}), would then imply that also $\aA_{\lA} \ninfl_\U \aB_{\{\lB_1,\lB_2\}}$.
This would in turn mean that $\aA_{\lA} \ninfl_\E \aB_{\{\lB_1,\lB_2\}}$ (\cref{prop:signalling-v-influence}), contradicting the definition given in \cref{eq:failure-of-atomicity-example}.

Now, the connectivity of a circuit decomposition does in fact not only impose no-signalling relations through $\E$, but also tells us about no-influence relations through its unitary dilations:
\begin{proposition}
    \label{prop:circuit-deco-implies-dilation}
    If a channel $\E: \aA_\labelsA\to\aB_\labelsB$ has a circuit decomposition $\circuit$ with connectivity $G_{\circuit}$, then it admits of a unitary dilation $\U: \aA_\labelsA\aE\to\aB_\labelsB\aF$ that itself has a unitary circuit decomposition $\circuit_\U$ whose connectivity relation $G_{\circuit_\U} \subseteq (\labelsA\cup\{\lE\})\times(\labelsB\cup\{\lF\})$ satisfies $G_{\circuit_\U} \cap (\labelsA\times\labelsB) = G_{\circuit}$.
    Overall, then, we have
    \begin{equation}
        \label{eq:signalling-influence-connectivity-relations}
        G_\E \subseteq G_{\U}\cap(\labelsA\times\labelsB) \subseteq G_{\circuit_\U} \cap (\labelsA\times\labelsB) = G_{\circuit}.
    \end{equation}
\end{proposition}
\begin{proof}
    For each $p\in\csP$, the circuit shape underlying $\circuit$, dilate the gate $\E_p: \aA_{\la^{-1}(p)} \aZ_{\Covm(p)}^p \to \aZ_p^{\Covp(p)} \aB_{\mu^{-1}(p)}$ to a unitary $\U_p : \aA_{\la^{-1}(p)} \aZ_{\Covm(p)}^p \aE_p \to \aZ_p^{\Covp(p)} \aB_{\mu^{-1}(p)} \aF_p$ with dilating systems $\aE_p$ and $\aF_p$ and dilation state $\rho_{\aE_p} \in \aE_p$.
    Collect all dilating systems together into $\aE \coloneqq \bigotimes_{p\in\csP} \aE_p$ and $\aF \coloneqq \bigotimes_{p\in\csP} \aF_p$ and define the state $\rho_\aE \coloneqq \bigotimes_{p\in\csP} \rho_{\aE_p}$.
    Let $\circuit_\U$ be the circuit consisting of the unitary gates $\U_p$ and $\U: \aA_\labelsA\aE \to\aB_\labelsB\aF$ the unitary it implements; then $\U$ is a unitary dilation of $\E$ and $\circuit_\U$ satisfies the required connectivity constraint.
\end{proof}

The observation about the example $\E$ in \cref{eq:failure-of-atomicity-example} we made above---that every circuit decomposition $\circuit$ of $\E$ necessarily satisfies $G_\E \subsetneq G_\circuit$---can thus be understood as a consequence of the combination of two facts: first, that all of $\E$'s unitary dilations satisfy $G_\E \subsetneq G_\U \cap (\labelsA\times\labelsB)$, and second, that the circuit decomposition $\circuit$ implies the existence of a unitary dilation with $G_\U \cap (\labelsA\times\labelsB) \subseteq G_\circuit$.
More generally speaking, \emph{signalling} and \emph{influence} are notions that should not be conflated, and out of the two, it is influence that is more closely related to compositional structure (see \cref{eq:signalling-influence-connectivity-relations}).

Finally, even in cases where equality $G_\E = G_\circuit$ cannot be reached, one may nevertheless be interested in determining whether $\E$ admits a circuit decomposition with some other given connectivity structure $G_\circuit$.
However, as \cref{prop:circuit-deco-implies-dilation} shows, the existence of such a decomposition is equivalent to the existence of an appropriate circuit decomposition of one of $\E$'s unitary dilations.
It is for these reasons that this work focusses on the purely unitary case.

%% file: figures/circuits/failure-of-atomicity-example-pic.tikz
\begin{tikzpicture}
	\begin{pgfonlayer}{nodelayer}
		\node [style=medium map] (0) at (0, 0) {$\E$};
		\node [style=none] (2) at (-0.625, 0.25) {};
		\node [style=none] (3) at (0.625, 0.25) {};
		\node [style=none] (4) at (-0.625, 1.25) {};
		\node [style=none] (5) at (0.625, 1.25) {};
		\node [style=none] (6) at (0, -1.25) {};
		\node [style=label] (7) at (0.5, -1.125) {$\aA_\lA$};
		\node [style=label] (8) at (0, 1) {$\aB_{\lB_1}$};
		\node [style=label] (9) at (1.25, 1) {$\aB_{\lB_2}$};
		\node [style=none] (10) at (2.5, 0) {$\coloneqq$};
		\node [style=wide point] (11) at (5, -0.75) {\small$\Phi^+$};
		\node [style=none] (12) at (4.25, -0.5) {};
		\node [style=none] (13) at (5.75, -0.5) {};
		\node [style=none] (14) at (4.25, 1.5) {};
		\node [style=none] (15) at (5.75, 1.5) {};
		\node [style=upground] (16) at (7.25, 1.5) {};
		\node [style=none] (17) at (7.25, -1.5) {};
		\node [style=cnot1] (18) at (7.25, 0.5) {};
		\node [style=none] (19) at (5.5, 0.5) {};
		\node [style=label] (20) at (7.75, -1.375) {$\aA_\lA$};
		\node [style=label] (21) at (4.875, 1.25) {$\aB_{\lB_1}$};
		\node [style=label] (22) at (6.375, 1.25) {$\aB_{\lB_2}$};
		\node [style=cnot2] (23) at (5.75, 0.5) {};
	\end{pgfonlayer}
	\begin{pgfonlayer}{edgelayer}
		\draw (5.center) to (3.center);
		\draw (4.center) to (2.center);
		\draw (0) to (6.center);
		\draw (14.center) to (12.center);
		\draw (15.center) to (13.center);
		\draw (17.center) to (16);
		\draw (18) to (19.center);
	\end{pgfonlayer}
\end{tikzpicture}

%% file: sections/app-algebra.tex
\section{Relative bicommutants and proof of \texorpdfstring{\cref{lma:the-algebraic-lemma}}{Lemma~\ref{lma:the-algebraic-lemma}}}\label{app:algebra}

\subsection{{\cstaralg}s and basic results}

In \cref{subsec:prelims-algebra} we defined a quantum system as the algebra of operators $\L(\H)$ on a given finite-dimensional Hilbert space.
In what follows, an algebra-first approach is more useful, even though we will stick to the finite-dimensional case throughout.
Most {\cstaralg}ic texts focus on the intricacies arising in infinite dimensions; see however~\textcite{Fare01} for a text on the finite-dimensional case.
Here we just recall the bare necessities.

In the finite-dimensional case a {\cstaralg} is nothing more than a complex *-algebra $\aA$ whose involution $(\cdot)^*:\aA\to\aA$ is \defn{positive}, meaning that $\forall a\in\aA: a^* a = 0 \Leftrightarrow a = 0$.
(More precisely, there is a unique norm making such a *-algebra into a {\cstaralg}~\cite{Fare01}).
A \defn{*-subalgebra} $\aX$ of $\aA$ is a subalgebra closed under the involution, and a \defn{unital} *-subalgebra is one that contains $\one_\aA$.
The \defn{commutant} of $\aX$, understood as a *-subalgebra of $\aA$, is $\aX' \coloneqq \{a\in\aA \mid \forall x\in\aX: ax = xa\}$.
The \defn{centre} of $\aX$ is $\centre(\aX) \coloneqq \aX \cap \aX'$, and $\aX$ is a \defn{factor} if it has trivial centre: $\centre(\aX) = \mathbb C \one_{\aX}$.
If $\aA$ itself is a factor and $\aX$ is an arbitrary *-subalgebra of it, we have $\aX'' = \aX$. (For this latter fact finite-dimensionality is essential.)
A \defn{*-isomorphism} is an algebra isomorphism that preserves the involution.
Finally, for subsets $S_1,\dots,S_n\subseteq \aA$, we write $S_1 \lor S_2 \lor\dots\lor S_n$ or $\bigvee_{k=1}^n S_k$ for the {\cstaralg} generated by $\bigcup_{k=1}^n S_k$---i.e.\ the smallest *-subalgebra containing all $S_k$s.

Every algebra $\L(\H)$ of linear operators on a finite-dimensional Hilbert space $\H$ forms a {\cstaralg} under the Hermitian adjoint, and conversely, every finite-dimensional {\cstaralg} $\aA$ is *-isomorphic to a *-subalgebra of $\L(\H)$ for some $\H$ (the finite-dimensional Gelfand-Naimark theorem).
For the purposes of this paper we only need concern ourselves with factor algebras.
$\aA$ is a factor algebra precisely if it is *-isomorphic to a \emph{full} operator algebra $\L(\K)$.
Moreover, if $\aA \subseteq \L(\H)$ is a unital factor *-subalgebra then there are $\K,\J$ and a unitary $U: \H \to \K \otimes \J$ such that
\begin{equation}
    \label{eq:rep-of-factor}
    U\aA U^* = \L(\K)\otimes\one_\J \qquadand U\aA' U^* = \one_\K\otimes\L(\J).
\end{equation}

The following tells us that wherever we said `unitary channel' in the main text, we could just as well have said `*-isomorphism'.
\begin{proposition}
    \label{prop:isomorphisms-are-unitaries}
    A map between factors $\phi:\L(\H)\to\L(\K)$ is a *-isomorphism iff is it of the form $\phi(a) = Ua U^*$ for some unitary map $U:\H\to\K$.
\end{proposition}
\begin{proof}
    For $U:\H\to\K$, write $\operatorname{Ad}_U:\L(\H)\to\L(\K)$ for the mapping $a\mapsto UaU^*$.
    It is clear that this is always a *-isomorphism.
    For the converse direction, suppose $\phi: \L(\H) \to \L(\K)$ is a *-isomorphism, pick an arbitrary unitary map $V: \K\to\H$ and let $\psi \coloneqq \operatorname{Ad}_V \circ \phi$, so that $\psi$ is a *-automorphism of $\L(\H)$.
    Every *-automorphism of a factor is \emph{inner}, i.e.\ is of the form $\psi = \operatorname{Ad}_W$ for some unitary $W\in\L(\H)$.
    A proof of this fact for the finite-dimensional case follows, for instance, from~\cite[Proposition~5.5]{Fare01}.
    We now have $\phi = \operatorname{Ad}_{V^*} \circ \operatorname{Ad}_W = \operatorname{Ad}_{V^* W}$, where $V^* W: \H\to\K$.
\end{proof}

Finally, the following correspondence between commuting factors and tensor products will be useful.
\begin{proposition}
    \label{prop:commuting-factors-vs-tns}
    Let $\H$ be finite-dimensional and $\aM_1,\aM_2,\dots,\aM_n \subseteq \L(\H)$ be unital *-subalgebras that are factors and that commute pairwise.
    There is a *-isomorphism
    \begin{equation}
        \V : \bigjoin_{k=1}^n \aM_k \ \xrightarrow{\sim}\ \bigotimes_{k=1}^n \aM_k
    \end{equation}
    such that for each $k=1,\dots,n$,
    \begin{equation}
        \label{eq:required-property}
        \V(\aM_k) = \one_{\aM_1}\tns\cdots\tns\one_{\aM_{k-1}}\tns\aM_k\tns\one_{\aM_{k+1}}\tns\cdots\tns\one_{\aM_n}.
    \end{equation}
\end{proposition}
\begin{proof}
    We give the proof for $n=2$; the general case follows by induction.
    By \cref{eq:rep-of-factor} and since $\aM_1\subseteq\L(\H)$ is a unital factor, there is an isomorphism $\H \cong \K_1 \tns \J_1$ under which
    \begin{equation}
        \aM_1 = \L(\K_1)\tns\one_{\J_1} \qquadand \aM_2 \subseteq \aM_1' = \one_{\K_1}\tns\L(\J_1).
    \end{equation}
    The fact that $\aM_2$ is a unital factor in turn implies the existence of an isomorphism $\J_1 \cong \K_2 \tns \J_2$ under which
    \begin{equation}
        \aM_1 = \L(\K_1)\tns\one_{\K_2}\tns\one_{\J_2} \qquadand \aM_2 = \one_{\K_1}\tns\L(\K_2)\tns\one_{\J_2}.
    \end{equation}
    Now it is clear that there exists an isomorphism
    \begin{equation}
        \V : \aM_1 \lor \aM_2 = \L(\K_1)\tns\L(\K_2)\tns\one_{\J_2} \xrightarrow{\sim} \L(\K_1)\tns\one_{\K_2}\tns\one_{\J_2}\tns\one_{\K_1}\tns\L(\K_2)\tns\one_{\J_2} = \aM_1\tns\aM_2
    \end{equation}
    satisfying the required property~\eqref{eq:required-property}.
\end{proof}

\subsection{Relative bicommutants}\label{subsec:rel-bics}
We will now build up to a proof of \cref{lma:the-algebraic-lemma}.
Recall that the aim of the lemma is to decompose the algebra $\aA$ into subsystems in a way dictated by the pairwise commuting subalgebras $\aB_k \subseteq \aA\aX_1\cdots\aX_n$.
The proof can be simplified by finding a way to restrict our attention to subalgebras of $\aA$ instead, allowing us to ignore the ambient algebras $\aX_1,\dots,\aX_n$.
The relevant subalgebras of $\aA$ to consider are the \emph{relative bicommutants} of the $\aB_k$s within $\aA$, which we now discuss.
Relative bicommutants have also been called \emph{support algebras} in~\cite{SW04}, where they were used for a similar purpose.
They also have a physical interpretation, in that they are the algebras of \emph{accessible} observables from~\cite{Orm25,ormrod2026decoherencestate}.

If $\aC$ is some global {\cstaralg}, $\aA\subseteq\aC$ is a *-subalgebra and $S\subseteq\aC$ is just any subset, the \defn{relative commutant}, \defn{relative bicommutant}, and \defn{relative tricommutant} of $S$ within $\aA$ are defined, respectively, by
\begin{align}
    S_{\aA}' &\coloneqq S' \cap \aA = \{ a\in \aA : as = sa\ \forall s\in S \}; \\
    S_{\aA}'' &\coloneqq (S' \cap \aA)' \cap \aA = \{ a\in \aA : at = ta\ \forall t\in S_\aA' \} \\
    \text{ and }
    S_{\aA}''' &\coloneqq ((S' \cap \aA)' \cap \aA)' \cap \aA = \{ a\in \aA : at = ta\ \forall t\in S_\aA'' \}.
\end{align}
It is well-known that if $\aC$ is a finite-dimensional factor and $S$ is *-closed (i.e.\ closed under $\aC$'s involution) then the \emph{absolute} bicommutant $S''$ coincides with the unital *-subalgebra generated by $S$.
This suggests that if $\aA$ is a factor too, we can think of the relative bicommutant $S_\aA''$ as the algebra `generated by $S$ within $\aA$'.
The following theorem makes that intuition more precise.
(Here we write $\aX \coloneqq \aA'$, so that $\aC \cong \aA \tns \aX$.)

\begin{theorem}
    \label{thm:rel-bics}
    Let $\aA$ and $\aX$ be finite-dimensional factor {\cstaralg}s and $S\subseteq \aA\tns\aX$ a *-closed subset.
    The following are identical as *-subalgebras of $\aA$:
    \begin{enumerate}
        \item\label{itm:rel-bics-original} the relative bicommutant $S_\aA'' \coloneqq (S'\cap \aA)'\cap \aA$;
        \item\label{itm:rel-bics-smallest} the smallest unital *-subalgebra $\aM\subseteq\aA$ such that $S\subseteq \aM\otimes\aX$;
        \item\label{itm:rel-bics-support} the \emph{support algebra} of~\cite{SW04}: i.e.\ the unital *-subalgebra of $\aA$ generated by the support of $S$ on $\aA$, seen as a vector space; that is, the *-algebra generated by $\one_\aA$ and $a_j^{(s)}$ for all $j\in \{1,\dots,\dim(\aX)\}$ and $s\in S$, where $a_j^{(s)}$ are the unique $\aA$-elements occurring in the expansion $s = \sum_{j=1}^{\dim(\aX)} a_j^{(s)} \otimes x_j$ for some fixed choice of basis $(x_j)_j$ of $\aX$.
        The resulting algebra is independent of this choice of basis;
        \item\label{itm:rel-bics-slices} the unital *-subalgebra of $\aA$ generated by the `slices' $\{(\id_\aA\otimes \phi)(b) \mid b\in S, \phi\in \aX^*\}$.
    \end{enumerate}
\end{theorem}
\begin{proof}
    $\mathrm{\ref{itm:rel-bics-original}} \subseteq \mathrm{\ref{itm:rel-bics-smallest}}$:
    It suffices to show that $S_\aA'' \subseteq \aM$ for any unital *-subalgebra $\aM\subseteq\aA$ satisfying $S \subseteq \aM\tns\aX$.
    Since $\aA$ is a finite-dimensional factor, von Neumann's bicommutant theorem implies that $\aM = \aM''_\aA$.
    Thus
    \begin{equation}
        S_\aA'' \subseteq \aM \iff S_\aA'' \subseteq \aM''_\aA \iff S_\aA' \supseteq \aM'_\aA \iff S'\cap \aA \supseteq \aM'\cap \aA,
    \end{equation}
    so it remains to show that $\aM'\cap\aA \subseteq S'$.
    This follows from the assumption that $S\subseteq \aM\tns\aX$, since that implies $S' \supseteq (\aM\tns\aX)' = \aM' \cap \aX' = \aM' \cap \aA$.

    $\mathrm{\ref{itm:rel-bics-smallest}} \subseteq \mathrm{\ref{itm:rel-bics-support}}$:
    Denoting by $\aN$ the support algebra defined in~\ref{itm:rel-bics-support}, we have $S\subseteq \aN\otimes\aX$; this is immediate from the construction of $\aN$.
    $\aM$ from~\ref{itm:rel-bics-smallest} being smallest, it must be contained in $\aN$.

    $\mathrm{\ref{itm:rel-bics-support}} \subseteq \mathrm{\ref{itm:rel-bics-original}}$:
    Denote again by $\aN$ the support algebra from~\ref{itm:rel-bics-support}.
    We first show that
    \begin{equation}
        \label{eq:sdlfsddesdkjf}
        S'\cap\aA \subseteq \aN' \cap \aA.
    \end{equation}
    Let $a\in S'\cap\aA$ and pick a basis $(x_j)_j$ of $\aX$.
    An arbitrary $s\in S$ can be expanded as $s = \sum_j a_j^{(s)} \tns x_j$.
    We have
    \begin{equation}
        a \in S'\cap\aA
        \implies \forall s\in S: [a,s] = 0
        \implies \forall s\in S: \sum_j [a,a_j^{(s)}] \tns x_j = 0
        \implies \forall s\in S \forall j: [a,a_j^{(s)}] = 0.
    \end{equation}
    Thus $a$ commutes with all of $\aN$'s generators, and therefore with all of $\aN$, too, establishing \cref{eq:sdlfsddesdkjf}.
    Taking commutants and intersecting with $\aA$ gives
    \begin{equation}
    (S'\cap\aA)
        '\cap\aA \supseteq (\aN' \cap \aA)'\cap\aA = \aN,
    \end{equation}
    where for the last equality we have used that $\aN$ is a unital *-subalgebra of $\aA$.

    The equivalence between~\ref{itm:rel-bics-slices} and the rest is left to the reader; we won't make use of it.
\end{proof}

Here are some fairly immediate but useful properties.
\begin{proposition}
    \label{prop:useful-rel-bic-properties}
    Let $\aA,\aX$ be factors and $S,T\subseteq\aA\tns\aX$ be *-closed subsets.
    \begin{enumerate}
        \item\label{itm:monotonicity} $S\subseteq T \implies S_{\aA}'' \subseteq T_{\aA}''$.
        \item\label{itm:trinity} $S_\aA''' = S_\aA'$.
        \item\label{itm:span-commutes-with-rel-bic} $(S \lor T)_{\aA}'' = S_{\aA}'' \lor T_{\aA}''$.
        \item\label{itm:rel-bic-vs-intersection} $S\cap\aA \subseteq S''_\aA$, and we have equality if and only if $S''_\aA \subseteq S$.
    \end{enumerate}
\end{proposition}
\begin{proof}
    \cref{itm:monotonicity} follows from definitions; \cref{itm:trinity} from von Neumann's bicommutant theorem applied within the factor $\aA$.
    Because it's fun, let's prove \cref{itm:span-commutes-with-rel-bic} by using the characterisation of relative bicommutants as in \cref{thm:rel-bics}\ref{itm:rel-bics-smallest} (alternatively one may use their original definition).
    For the inclusion $(S \lor T)_{\aA}'' \subseteq S_{\aA}'' \lor T_{\aA}''$, note that by \cref{thm:rel-bics}\ref{itm:rel-bics-smallest}, $S \subseteq S_\aA'' \tns \aX$ and $T \subseteq T_\aA'' \tns \aX$.
    This implies that $S\lor T \subseteq (S_\aA'' \lor T_\aA'')\tns\aX$.
    But \cref{thm:rel-bics}\ref{itm:rel-bics-smallest} also tells us that $(S\lor T)_\aA''$ is the \emph{smallest} $\aM$ such that $S\lor T \subseteq \aM \tns \aX$.
    Thus, $(S\lor T)_\aA'' \subseteq (S_\aA'' \lor T_\aA'')$.
    The opposite inclusion follows from monotonicity~\ref{itm:monotonicity} and the fact that $(S\lor T)_\aA''$ is an algebra.

    For \cref{itm:rel-bic-vs-intersection}, note first that $S'\cap \aA \subseteq S'$ and that this implies, by anti-monotonicity of the commutant, that $S \subseteq (S')' \subseteq (S'\cap \aA)'$.
    Taking intersections with $\aA$ on both sides yields the inclusion $S\cap \aA\subseteq S''_\aA$.
    The equality condition follows immediately.
\end{proof}

The following shows that relevant properties of the $\aB_k$s in \cref{lma:the-algebraic-lemma} are preserved once we move to considering their relative bicommutants $(\aB_k)''_\aA$.
For this we need the additional assumption that they `only overlap on $\aA$'.

\begin{proposition}
    \label{prop:more-useful-properties}
    Let $\aA,\aX_1,\aX_2$ be factors and $S_1\subseteq \aA\aX_1, S_2\subseteq\aA\aX_2$ be *-closed subsets.
    \begin{enumerate}
        \item If $S_1$ and $S_2$ mutually commute (as subsets of $\aA\aX_1\aX_2$) then so do their relative bicommutants~\cite[Lemma~8]{SW04}:
        \begin{equation}
            S_1 \subseteq S_2' \quad \implies \quad (S_1)''_\aA \subseteq (S_2)'''_\aA.
        \end{equation}
        \item If $\aA\subseteq S_1\lor S_2$ then $\aA = (S_1)''_\aA \lor (S_2)''_\aA$.
    \end{enumerate}
\end{proposition}
The obvious generalisations to more than two $S_k$s and $\aX_k$s as required for \cref{lma:the-algebraic-lemma} follow immediately.
\begin{proof}
    The first statement is perhaps easiest to prove when treating the relative bicommutants in the guise of support algebras (\cref{thm:rel-bics}\ref{itm:rel-bics-support}); see~\cite[Lemma~8]{SW04}.
    The second statement follows from \cref{prop:useful-rel-bic-properties}\ref{itm:span-commutes-with-rel-bic}.
\end{proof}

\begin{remark}
    A lot changes if we let go of the assumption that $\aA$ and $\aC$ are factors.
    Even when it comes to absolute bicommutants, it is then no longer true that $S''$ is the unital {\cstaralg} generated by $S$.
    (Instead, it is generated by $S$ and the centre of the ambient algebra; see e.g.~\cite[Prop.~B.3]{VMA25b}.)
\end{remark}

\subsection{Proof of \texorpdfstring{\cref{lma:the-algebraic-lemma}}{Lemma~\ref{lma:the-algebraic-lemma}}}
Recall from \cref{subsec:prelims-algebra} that $[n] \coloneqq \{1,2,\dots,n\}$, that the notation $\aZ_1\aZ_2\cdots\aZ_n$ or $\aZ_{[n]}$ refers to the tensor product $\aZ_1\tns\aZ_2\tns\cdots\tns\aZ_n$, and that depending on the context, $\aZ_k$ may also refer to the unital *-subalgebra $\aZ_k \tns \{\one_{\aZ_{[n]\setminus\{k\}}}\} \subseteq \aZ_{[n]}$.

\algebraiclemma*  
\begin{proof}
    For each $k\in [n]$, let $\aZ_k \coloneqq (\aB_k)''_{\aA}$ be the relative bicommutant.
    By \cref{prop:more-useful-properties}, assumptions~\ref{itm:the-algebraic-lemma-commuting-assumption} and~\ref{itm:the-algebraic-lemma-crucial-assumption} imply that as subalgebras of $\aA$, all $\aZ_k$s mutually commute and generate all of $\aA$:
    \begin{equation}
        \aZ_k \subseteq \aZ_l' \text{ for distinct } k,l\in [n] \qquadand \bigjoin_{k\in[n]} \aZ_k = \aA.
    \end{equation}
    These two conditions imply that each $\aZ_k$ is a factor algebra.
    Indeed, let $k\in [n]$ and suppose that $m\in\centre(\aZ_k)$.
    Then $m\in\aZ_k \subseteq \aA$; but also $m\in\aZ_l'$ for all $l\in [n]$, and therefore $m\in\aA'$, so that $m\in\centre(\aA)$.
    Concluding, $\centre(\aZ_k) \subseteq \centre(\aA) = \mathbb C \one_\aA$, since $\aA$ is a factor; and hence $\aZ_k$ is a factor, too.

    We can therefore apply \cref{prop:commuting-factors-vs-tns} and obtain a *-isomorphism
    \begin{equation}
        \V : \aA \to \aZ_1\tns\aZ_2\tns\cdots\tns\aZ_n
    \end{equation}
    so that for all $k$,
    \begin{equation}
        \label{eq:tautology}
        \V(\aZ_k) = \one_{\aZ_1}\tns\cdots\tns\one_{\aZ_{k-1}}\tns\aZ_k\tns\one_{\aZ_{k+1}}\tns\cdots\tns\one_{\aZ_{n}}.
    \end{equation}
    Employing our usual abuse of notation, we write the right-hand side as $\aZ_k$ and write tensor products as juxtaposition.
    Extending $\V$ to
    \begin{equation}
        \V \equiv \V\tns\id_{\aX_1\cdots\aX_n} : \aA\aX_1\cdots\aX_n \xrightarrow{\sim} \aZ_1\cdots\aZ_n\aX_1\cdots\aX_n,
    \end{equation}
    we see that
    \begin{equation}
        \V(\aB_k) \subseteq \V(\aZ_k\aX_k) = \aZ_k \aX_k,
    \end{equation}
    appealing for the first inclusion to \cref{thm:rel-bics}\ref{itm:rel-bics-smallest} about relative bicommutants.
    This establishes the first conclusion of the lemma, \cref{eq:the-algebraic-lemma-conclusion-1}.

    It remains to prove \cref{eq:the-algebraic-lemma-conclusion-2}, which due to \cref{eq:tautology} is equivalent to $\aZ_k = \aA \cap \aB_k$.
    By \cref{prop:useful-rel-bic-properties}\ref{itm:rel-bic-vs-intersection} this is in turn equivalent to
    \begin{equation}
        \label{eq:final-thing-to-prove}
        \aZ_k \subseteq \aB_k.
    \end{equation}
    First, from assumption~\ref{itm:the-algebraic-lemma-crucial-assumption} it follows that
    \begin{equation}
        \aZ_k \subseteq \aA \subseteq \bigjoin_{l\in[n]} \aB_l.
    \end{equation}
    For $l$ distinct from $k$, we also have
    \begin{equation}
        \aB_l \subseteq \aZ_l \aX_l \subseteq \aZ_k',
    \end{equation}
    or in other words, $\aZ_k \subseteq \aB_l'$.
    Combining this we get
    \begin{equation}
        \aZ_k \subseteq \left( \bigjoin_{l\in[n]} \aB_l \right) \cap \bigcap_{l\neq k} \aB_l'.
    \end{equation}
    The right-hand side is just $\aB_k$; this is clear when regarding it under the isomorphism from \cref{prop:commuting-factors-vs-tns}.
    This establishes \cref{eq:final-thing-to-prove} and concludes the proof of \cref{lma:the-algebraic-lemma}.
\end{proof}

%% file: sections/necessity.tex
\section{Causally faithful unitary circuit decompositions}\label{app:faithful}
This appendix concerns causally faithful circuit decompositions, which were introduced in \cref{subsec:faithful-decos} and feature in \cref{thm:main-thm}\itmref{itm:main-thm-causally-faithful}.
In the main text we have shown that \cref{itm:main-thm-pic,itm:main-thm-unitary-decos} of \cref{thm:main-thm} are equivalent, and it is immediate that $\itmref{itm:main-thm-unitary-decos} \Rightarrow \itmref{itm:main-thm-causally-faithful}$.
Here, we finish by showing that $\neg\itmref{itm:main-thm-pic} \Rightarrow \neg\itmref{itm:main-thm-causally-faithful}$.
That is, we show that whenever $G \subseteq \labelsA\times\labelsB$ fails the condition \PIC{}, then there exists a choice of quantum systems $\aA_\lA$ for $\lA\in\labelsA$ and $\aB_\lB$ for $\lB\in\labelsB$ and a {\unichan} $\U : \aA_\labelsA\to\aB_\labelsB$ which has causal structure exactly $G_\U = G$, yet no unitary circuit decomposition with connectivity $G$---that is, $\U$ has no causally faithful unitary circuit decomposition.

To see the challenge in proving this, recall for a moment our proof of $\neg\itmref{itm:main-thm-pic} \Rightarrow \neg\itmref{itm:main-thm-unitary-decos}$ on page \pageref{page:proof-of-ii-implies-i}.
There, we simply considered the unitary $\cccU$ from \cref{ex:ccc-has-no-unitary-deco}, padded appropriately by trivial systems.
This padded unitary $\U$ satisfied the condition $G_\U \subseteq G$ but had no unitary circuit decomposition with $G_\circuit \subseteq G$, thus providing the required counterexample.

But to prove $\neg\itmref{itm:main-thm-pic} \Rightarrow \neg\itmref{itm:main-thm-causally-faithful}$, we need our counterexample unitary to satisfy the saturated condition $G_\U = G$.
Padding by trivial systems generally won't do, since there is no influence between trivial systems.
In the proof below, we therefore pad $\cccU$ instead with some other, nontrivial unitary $\V$ that already has the desired causal structure $G$.
(Such a choice of $\V$ always exists for appropriate choices of quantum systems; however, $G$ may constrain the dimensions of these systems. See e.g.\ \cref{eq:some-loose-wires} and Ref.~\cite[§7.2]{LB21}.)




\begin{proposition}[{proving $\neg\itmref{itm:main-thm-pic} \Rightarrow \neg\itmref{itm:main-thm-causally-faithful}$ in \cref{thm:main-thm}}]
    \label{thm:necessity}
    Suppose $G\subseteq\labelsA\times\labelsB$ fails \PIC{}; without loss of generality, write $\labelsA = \{1,\dots,\nA\}$ and $\labelsB = \{1,\dots,\nB\}$ so that $G \cap \left(\{1,2,3\}\times\{1,2,3\}\right) = \cccG$.
    Let $\V: \aA_1\cdots\aA_{\nA} \to \aB_1\cdots\aB_{\nB}$ be a {\unichan} between quantum systems $\aA_1,\dots,\aA_{\nA}$ and $\aB_1,\dots,\aB_{\nB}$ so that $G_\V = G$.
    In addition, let $\bar\aA_i = \bar\aB_i = \L(\mathbb{C}^2)$ for $i\in\{1,2,3\}$ be qubits and $\cccU: \bar\aA_1\bar\aA_2\bar\aA_3 \to \bar\aB_1\bar\aB_2\bar\aB_3$ be the {\unichan}
    \begin{equation}
        \label{eq:ccc-unitary-again}
        \input{circuits/u-bar-bar.tikz} \coloneqq \quad \input{circuits/u-bar-definition.tikz},
    \end{equation}
    Finally, define
    \begin{equation}
        \U \coloneqq \cccU\tns\V : (\bar\aA_1\aA_1)(\bar\aA_2\aA_2)(\bar\aA_3\aA_3)\aA_4\cdots\aA_{\nA} \to (\bar\aB_1\aB_1)(\bar\aB_2\aB_2)(\bar\aB_3\aB_3)\aB_4\cdots\aB_{\nB},
    \end{equation}
    regarded as a {\unichan} with $\nA$ inputs and $\nB$ outputs:
    \begin{equation}
        \label{eq:construction-of-u}
		\input{necessity/construction-of-u_v2.tikz}.
    \end{equation}
    Then $\U$ has causal structure $G$, yet no unitary circuit decomposition with connectivity $G$.
\end{proposition}

\begin{lemma}
    \label{lma:supports}
    Let $\ket\psi \in \H_X\tns\H_Y$ be a bipartite state with marginals $\rho_X \coloneqq \Tr_Y \ket{\psi}\bra{\psi}$ and $\rho_Y = \Tr_X \ket{\psi}\bra{\psi}$, supported on $\supp\rho_X\subseteq\H_X$ and $\supp\rho_Y\subseteq\H_Y$, respectively.
    Then $\ket\psi \in \supp\rho_X \otimes \supp\rho_Y$.
\end{lemma}
\begin{proof}
    Let $\ket\psi = \sum_{i=1}^r c_i \ket{\alpha_i}\tns\ket{\beta_i}$ be the Schmidt decomposition, where all coefficients $c_i$ are nonzero.
    Then $\supp\rho_X$ is the span of the vectors $\{\ket{\a_i}\}_{i=1}^r$ and $\supp\rho_Y$ the span of $\{\ket{\b_i}\}_{i=1}^r$.
    Clearly, then, $\ket\psi\in\supp\rho_X\tns\supp\rho_Y$.
\end{proof}

\begin{proof}[Proof of \cref{thm:necessity}]
    It is immediate from the construction that $\U$ (under the appropriate grouping of systems) has causal structure $G$.
    Suppose, for contradiction, that $\U$ has a unitary causally faithful circuit decomposition: i.e.\ a circuit decomposition $\circuit$ with connectivity $G$.
    Then $\circuit$ has, in particular, no path from $\bar\aA_1\aA_1$ to $\bar\aB_3\aB_3$, and no path from $\bar\aA_3\aA_3$ to $\bar\aB_1\aB_1$.

    Now abbreviate $\hat\aA_4 \coloneqq \aA_2\aA_4\cdots\aA_{\nA}$ and $\hat\aB_4 \coloneqq \aB_2\aB_4\cdots\aB_{\nB}$ and regard $\U$ as a four-input, four-output {\unichan}
    \begin{equation}
        \U : (\bar\aA_1\aA_1)(\bar\aA_2)(\hat\aA_4)(\bar\aA_3\aA_3) \to (\bar\aB_1\aB_1)(\bar\aB_2)(\hat\aB_4)(\bar\aB_3\aB_3).
    \end{equation}
    The circuit $\circuit$, now interpreted as a four-input, four-output circuit, then satisfies
    \begin{equation}
        G_\circuit \subseteq \necessityG\ \coloneqq \input{necessity/four-partite-causal-structure.tikz}.
    \end{equation}
    By \cref{thm:canonicity}, then, $\U$ also has a unitary circuit decomposition of the canonical shape $\latt{\necessityG}$.
    One may compute this shape, much like $\latt\cccG$ in \cref{eq:ccc-lattice-reminder}, to be
    \begin{equation}
        \latt{\necessityG} = \input{necessity/four-partite-shape.tikz}.
    \end{equation}
    In other words, there are quantum systems $\aC,\aD,\aE,\aF$ and {\unichan}s $\uW,\uX,\uY,\uZ$ so that
    \begin{equation}
        \label{eq:four-partite-causal-deco}
        \input{necessity/four-partite-causal-deco_b.tikz}.
    \end{equation}

    Recall that $\bar\aA_2 = \L(\mathbb C^2)$; moreover, let $\H_{\hat\aA_4}$, $\H_{\aC}, \H_{\aD}$ be the finite-dimensional Hilbert spaces so that $\hat\aA_4 = \L(\H_{\hat\aA_4})$, $\aC = \L(\H_{\aC})$, $\aD = \L(\H_{\aD})$.
    Let $d \coloneqq \dim\H_{\hat\aA_4}$.
    Our aim is to derive a contradiction on the dimensions of the Hilbert spaces $\H_\aC$ and $\H_\aD$.

    Denote by $\{\ket0,\ket1\}$ the basis of $\mathbb C^2$ with respect to which the CNOTs in the construction of $\cccU$ are defined.
    Also choose a basis $\{\ket j\}_{j=0}^{d-1}$ for the Hilbert space $\H_{\hat\aA_4}$.
    For $i\in\{0,1\}$ and $j\in\{0,\dots,d-1\}$, define the states
    \begin{equation}
        \label{eq:a-state}
        \rho_{\aC\aD}^{ij} \coloneqq \input{necessity/a-state.tikz}; \quad \rho_{\aC}^{ij} \coloneqq \Tr_{\aD}\rho_{\aC\aD}^{ij}; \quadand \rho_{\aD}^{ij} \coloneqq \Tr_{\aC}\rho_{\aC\aD}^{ij}.
    \end{equation}
    Let us first consider the states $\rho_{\aC}^{ij}$ on system $\aC$.
    We claim that for any $j,j'\in\{0,\dots,d-1\}$, the states $\rho_\aC^{0j}$ and $\rho_\aC^{1j'}$ have orthogonal supports, i.e.\ are perfectly distinguishable.
    To see this intuitively, consider the qubits $\bar\aA_1,\bar\aA_2$, and $\bar\aB_1$.
    From the CNOTs in the definition of $\U$ in \cref{eq:construction-of-u}, it is clear that if $\bar\aA_1$ is prepared in state $\ket0$, the information about the computational degree of freedom $i$ of $\bar\aA_2$ ends up on $\bar\aB_1$.
    Since every path from $\bar\aA_2$ to $\bar\aB_1$ in \cref{eq:four-partite-causal-deco} passes through $\aC$, that information must also be perfectly recoverable from $\aC$.

    To see it formally, notice that, for any (irrelevant) choice for the unlabelled states,
    \begin{equation}
        \label{eq:proof-of-orthogonality}
		\input{necessity/proof-of-orthogonality.tikz}.
    \end{equation}
    The circuit fragment enclosed in a dashed box on the left-hand side forms a CPTP map $\E: \aC\to\bar\aB_1$.
    Being CPTP, $\E$ cannot increase distinguishability; in particular, if $\rho_\aC^{0j}$ and $\rho_\aC^{1j'}$ had non-orthogonal supports, then $\E(\rho_\aC^{0j})$ and $\E(\rho_\aC^{1j'})$ would have non-orthogonal supports.
    In light of the right-hand side of \cref{eq:proof-of-orthogonality}, the latter is obviously untrue; therefore, $\rho_\aC^{0j}$ and $\rho_\aC^{1j'}$ must have orthogonal supports.

    Let $\H_\aC^0 \coloneqq \left(\bigcup_{j=0}^{d-1} \supp\rho_\aC^{0j}\right)^{\perp\perp} \subseteq \H_\aC$ be the subspace of $\H_\aC$ spanned by the supports of $\rho_\aC^{0j}$ for $j\in\{0,\dots,d-1\}$.
    Moreover, let $\H_\aC^1 \coloneqq (\H_\aC^0)^\perp$.
    It follows from the claim we just proved that then
    \begin{equation}
        \label{eq:support-C}
        \supp\rho_{\aC}^{ij} \subseteq \H_\aC^i
    \end{equation}
    for all $i\in\{0,1\}$ and $j\in\{0,\dots,d-1\}$. 
    By a symmetric argument concerning instead the states $\rho_\aD^{ij}$ on $\aD$, we can construct orthogonal subspaces $\H_\aD^0,\H_\aD^1\subseteq \H_\aD$ so that
    \begin{equation}
        \label{eq:support-D}
        \supp\rho_{\aD}^{ij} \subseteq \H_\aD^i
    \end{equation}
    for $i\in\{0,1\}$ and $j\in\{0,\dots,d-1\}$.

    By \cref{lma:supports,eq:support-C,eq:support-D} and the fact that $\rho_{\aC\aD}^{ij}$ is pure,
    \begin{equation}
        \supp\rho_{\aC\aD}^{ij} \subseteq \H_{\aC}^i\tns\H_{\aD}^i.
    \end{equation}
    Moreover, for fixed $i$, the $d$ states $\rho_{\aC\aD}^{i0},\rho_{\aC\aD}^{i1},\dots,\rho_{\aC\aD}^{i(d-1)}$ are all pairwise orthogonal by their construction in \cref{eq:a-state} (and the fact that $\uW$ is a {\unichan}).
    These two facts imply that
    \begin{equation}
        \dim(\H_\aC^i\tns\H_\aD^i) \geq d
    \end{equation}
    for each $i\in\{0,1\}$.
    We conclude that
    \begin{align}
        \dim (\H_\aC\tns\H_\aD) &= \dim\H_{\aC}^0 \dim\H_{\aD}^0 + \dim\H_{\aC}^0 \dim\H_{\aD}^1 + \dim\H_{\aC}^1 \dim\H_{\aD}^0 + \dim\H_{\aC}^1 \dim\H_{\aD}^1 \nonumber \\
        &\geq 2d + \dim\H_{\aC}^0 \dim\H_{\aD}^1 + \dim\H_{\aC}^1 \dim\H_{\aD}^0 > 2d,
    \end{align}
    which is in contradiction with the fact that $\uW$ is a unitary channel and $\dim(\mathbb C^2\tns\H_{\hat\aA_4}) = 2d$.
    $\U$ can therefore have no causally faithful unitary decomposition.
\end{proof}

%% file: figures/circuits/u-bar-bar.tikz
\begin{tikzpicture}
	\begin{pgfonlayer}{nodelayer}
		\node [style=large map] (0) at (1.5, 0) {$\cccU$};
		\node [style=none] (6) at (3, 1.25) {};
		\node [style=label] (18) at (0.5, 1) {$\bar\aB_1$};
		\node [style=label] (19) at (2, 1) {$\bar\aB_2$};
		\node [style=label] (20) at (3.5, 1) {$\bar\aB_3$};
		\node [style=label] (21) at (0.5, -1) {$\bar\aA_1$};
		\node [style=label] (22) at (2, -1) {$\bar\aA_2$};
		\node [style=label] (23) at (3.5, -1) {$\bar\aA_3$};
		\node [style=none] (34) at (1.5, 1.25) {};
		\node [style=none] (36) at (0, 1.25) {};
		\node [style=none] (40) at (3, -1.25) {};
		\node [style=none] (48) at (1.5, -1.25) {};
		\node [style=none] (50) at (0, -1.25) {};
	\end{pgfonlayer}
	\begin{pgfonlayer}{edgelayer}
		\draw (40.center) to (6.center);
		\draw (36.center) to (50.center);
		\draw (34.center) to (48.center);
	\end{pgfonlayer}
\end{tikzpicture}

%% file: figures/circuits/u-bar-definition.tikz
\begin{tikzpicture}
	\begin{pgfonlayer}{nodelayer}
		\node [style=none] (6) at (3, 1.5) {};
		\node [style=label] (18) at (0.5, 1.25) {$\bar\aB_1$};
		\node [style=label] (19) at (2, 1.25) {$\bar\aB_2$};
		\node [style=label] (20) at (3.5, 1.25) {$\bar\aB_3$};
		\node [style=label] (21) at (0.5, -1) {$\bar\aA_1$};
		\node [style=label] (22) at (2, -1) {$\bar\aA_2$};
		\node [style=label] (23) at (3.5, -1) {$\bar\aA_3$};
		\node [style=none] (34) at (1.5, 1.5) {};
		\node [style=none] (36) at (0, 1.5) {};
		\node [style=none] (40) at (3, -1.25) {};
		\node [style=none] (48) at (1.5, -1.25) {};
		\node [style=none] (50) at (0, -1.25) {};
		\node [style=none] (51) at (-0.2, -0.25) {};
		\node [style=none] (52) at (1.5, -0.25) {};
		\node [style=none] (53) at (1.5, 0.5) {};
		\node [style=none] (54) at (3.2, 0.5) {};
		\node [style=small black dot] (56) at (1.5, -0.25) {};
		\node [style=small black dot] (57) at (1.5, 0.5) {};
		\filldraw[fill=none](0, -0.25) circle (0.2);
		\filldraw[fill=none](3, 0.5) circle (0.2);
	\end{pgfonlayer}
	\begin{pgfonlayer}{edgelayer}
		\draw (40.center) to (6.center);
		\draw (36.center) to (50.center);
		\draw (34.center) to (48.center);
		\draw (51.center) to (52.center);
		\draw (53.center) to (54.center);
	\end{pgfonlayer}
\end{tikzpicture}

%% file: figures/necessity/construction-of-u_v2.tikz
\begin{tikzpicture}
	\begin{pgfonlayer}{nodelayer}
		\node [style=none] (6) at (4.5, 2) {};
		\node [style=large map] (7) at (8.75, 0) {\hspace*{2.5cm}};
		\node [style=none] (8) at (6.4, 4) {};
		\node [style=none] (9) at (6.4, -4) {};
		\node [style=none] (10) at (7.4, 4) {};
		\node [style=none] (11) at (7.4, -4) {};
		\node [style=none] (12) at (8.4, -4) {};
		\node [style=none] (13) at (8.4, 4) {};
		\node [style=none] (14) at (9.25, 4) {};
		\node [style=none] (16) at (11, 4) {};
		\node [style=label] (18) at (2, 1.75) {$\bar\aB_1$};
		\node [style=label] (19) at (3.5, 1.75) {$\bar\aB_2$};
		\node [style=label] (20) at (5, 1.75) {$\bar\aB_3$};
		\node [style=label] (21) at (2, -1.75) {$\bar\aA_1$};
		\node [style=label] (22) at (3.5, -1.75) {$\bar\aA_2$};
		\node [style=label] (23) at (5, -1.75) {$\bar\aA_3$};
		\node [style=label] (24) at (6.775, 1.75) {$\aB_1$};
		\node [style=label] (25) at (7.775, 1.75) {$\aB_2$};
		\node [style=label] (26) at (8.775, 1.75) {$\aB_3$};
		\node [style=none] (29) at (10.25, 2.8) {$\cdots$};
		\node [style=label] (31) at (9.75, 1.75) {$\aB_4$};
		\node [style=label] (32) at (11.75, 1.75) {$\aB_\nB$};
		\node [style=none] (33) at (8.25, 4) {};
		\node [style=none] (34) at (3, 2) {};
		\node [style=none] (35) at (7.25, 4) {};
		\node [style=none] (36) at (1.5, 2) {};
		\node [style=none] (37) at (6.25, 4) {};
		\node [style=none] (38) at (8.75, 0) {};
		\node [style=none] (39) at (8.75, 0) {$\V$};
		\node [style=none] (40) at (4.5, -2) {};
		\node [style=none] (41) at (6.4, -4) {};
		\node [style=none] (42) at (7.4, -4) {};
		\node [style=none] (43) at (8.4, -4) {};
		\node [style=none] (44) at (9.25, -4) {};
		\node [style=none] (45) at (11, -4) {};
		\node [style=none] (46) at (10.25, -2.8) {$\cdots$};
		\node [style=none] (47) at (8.25, -4) {};
		\node [style=none] (48) at (3, -2) {};
		\node [style=none] (49) at (7.25, -4) {};
		\node [style=none] (50) at (1.5, -2) {};
		\node [style=none] (51) at (6.25, -4) {};
		\node [style=label] (52) at (6.75, -1.75) {$\aA_1$};
		\node [style=label] (53) at (7.75, -1.75) {$\aA_2$};
		\node [style=label] (54) at (8.75, -1.75) {$\aA_3$};
		\node [style=label] (55) at (9.725, -1.75) {$\aA_4$};
		\node [style=label] (56) at (11.725, -1.75) {$\aA_\nA$};
		\node [style=large map] (57) at (-6.85, 0) {\hspace*{3cm}};
		\node [style=none] (58) at (-9.725, 1.5) {};
		\node [style=none] (60) at (-8.4, 1.5) {};
		\node [style=none] (62) at (-7.075, -1.5) {};
		\node [style=none] (63) at (-7.075, 1.5) {};
		\node [style=none] (64) at (-6, 1.5) {};
		\node [style=none] (65) at (-4.025, 1.5) {};
		\node [style=none] (69) at (-4.925, 1.25) {$\cdots$};
		\node [style=none] (76) at (-6.85, 0) {$\U$};
		\node [style=none] (77) at (-9.725, -1.5) {};
		\node [style=none] (78) at (-8.4, -1.5) {};
		\node [style=none] (80) at (-6, -1.5) {};
		\node [style=none] (81) at (-4.025, -1.5) {};
		\node [style=none] (82) at (-4.925, -1.25) {$\cdots$};
		\node [style=small label] (86) at (-9.725, -2) {$\bar\aA_1\aA_1$};
		\node [style=small label] (89) at (-6, -2) {$\aA_4$};
		\node [style=small label] (90) at (-4.025, -2) {$\aA_\nA$};
		\node [style=small label] (91) at (-8.4, -2) {$\bar\aA_2\aA_2$};
		\node [style=small label] (92) at (-7.075, -2) {$\bar\aA_3\aA_3$};
		\node [style=small label] (98) at (-9.725, 1.75) {$\bar\aB_1\aB_1$};
		\node [style=small label] (99) at (-6, 1.75) {$\aB_4$};
		\node [style=small label] (100) at (-4.025, 1.75) {$\aB_\nB$};
		\node [style=small label] (101) at (-8.4, 1.75) {$\bar\aB_2\aB_2$};
		\node [style=small label] (102) at (-7.075, 1.75) {$\bar\aB_3\aB_3$};
		\node [style=none] (103) at (-2, 0) {$\coloneqq$};
		\node [style=none] (104) at (3, -0.5) {};
		\node [style=none] (105) at (1.3, -0.5) {};
		\node [style=none] (106) at (4.7, 0.5) {};
		\node [style=none] (107) at (3, 0.5) {};
		\node [style=small black dot] (108) at (3, 0.5) {};
		\node [style=small black dot] (110) at (3, -0.5) {};
		\node [style=none] (112) at (0.95, 0.5) {{\small \color{gray} $\cccU$}};
		\node [style=none] (113) at (1.5, -0.5) {};
		\node [style=none] (114) at (4.5, 0.5) {};
	\end{pgfonlayer}
	\begin{pgfonlayer}{edgelayer}
		\draw[dashed, color=gray] (0.3,1.15) rectangle (5.2,-1.15);
		\filldraw[fill=none](113) circle (0.2);
		\filldraw[fill=none](114) circle (0.2);
		\draw (8.center) to (9.center);
		\draw (10.center) to (11.center);
		\draw (12.center) to (13.center);
		\draw [in=-90, out=90, looseness=0.75] (6.center) to (33.center);
		\draw [in=-90, out=90, looseness=0.50] (34.center) to (35.center);
		\draw [in=-90, out=90, looseness=0.50] (36.center) to (37.center);
		\draw [in=90, out=-90, looseness=0.75] (40.center) to (47.center);
		\draw [in=90, out=-90, looseness=0.50] (48.center) to (49.center);
		\draw [in=90, out=-90, looseness=0.50] (50.center) to (51.center);
		\draw (40.center) to (6.center);
		\draw (16.center) to (45.center);
		\draw (44.center) to (14.center);
		\draw (36.center) to (50.center);
		\draw (34.center) to (48.center);
		\draw (62.center) to (63.center);
		\draw (65.center) to (81.center);
		\draw (80.center) to (64.center);
		\draw (58.center) to (77.center);
		\draw (60.center) to (78.center);
		\draw (104.center) to (105.center);
		\draw (106.center) to (107.center);
	\end{pgfonlayer}
\end{tikzpicture}

%% file: figures/necessity/four-partite-causal-structure.tikz
\begin{tikzpicture}
	\begin{pgfonlayer}{nodelayer}
		\node [style=label] (0) at (0, -1) {$1$};
		\node [style=label] (1) at (2, -1) {$2$};
		\node [style=label] (2) at (6, -1) {$3$};
		\node [style=label] (3) at (0, 1) {$1$};
		\node [style=label] (4) at (2, 1) {$2$};
		\node [style=label] (5) at (6, 1) {$3$};
		\node [style=label] (6) at (4, -1) {4};
		\node [style=label] (7) at (4, 1) {4};
	\end{pgfonlayer}
	\begin{pgfonlayer}{edgelayer}
		\draw [style=dir] (0) to (3);
		\draw [style=dir] (0) to (4);
		\draw [style=dir] (1) to (3);
		\draw [style=dir] (1) to (4);
		\draw [style=dir] (1) to (5);
		\draw [style=dir] (2) to (5);
		\draw [style=dir] (2) to (4);
		\draw [style=dir] (6) to (3);
		\draw [style=dir] (6) to (7);
		\draw [style=dir] (6) to (5);
		\draw [style=dir] (6) to (4);
		\draw [style=dir] (1) to (7);
		\draw [style=dir] (2) to (7);
		\draw [style=dir] (0) to (7);
	\end{pgfonlayer}
\end{tikzpicture}

%% file: figures/necessity/four-partite-shape.tikz
\begin{tikzpicture}
	\begin{pgfonlayer}{nodelayer}
		\node [style=tiny map] (0) at (1.75, 1.25) {};
		\node [style=none] (1) at (1.575, 1) {};
		\node [style=none] (2) at (1.925, 1) {};
		\node [style=none] (4) at (0.925, 0.25) {};
		\node [style=none] (5) at (2.575, 0.25) {};
		\node [style=label] (8) at (1.45, 2.25) {2};
		\node [style=label] (11) at (2.925, 2.25) {3};
		\node [style=label] (12) at (0.575, 2.25) {1};
		\node [style=tiny map] (13) at (1.75, -1.25) {};
		\node [style=tiny map] (14) at (0.75, 0) {};
		\node [style=tiny map] (15) at (2.75, 0) {};
		\node [style=none] (16) at (1.575, -1) {};
		\node [style=none] (17) at (1.925, -1) {};
		\node [style=none] (19) at (0.925, -0.25) {};
		\node [style=none] (20) at (2.575, -0.25) {};
		\node [style=label] (26) at (2.925, -2.25) {3};
		\node [style=label] (27) at (0.575, -2.25) {1};
		\node [style=none] (28) at (1.575, 1.5) {};
		\node [style=none] (29) at (1.925, 1.5) {};
		\node [style=label] (30) at (2.05, 2.25) {4};
		\node [style=label] (31) at (1.45, -2.25) {2};
		\node [style=none] (32) at (1.575, -1.5) {};
		\node [style=none] (33) at (1.925, -1.5) {};
		\node [style=label] (34) at (2.05, -2.25) {4};
	\end{pgfonlayer}
	\begin{pgfonlayer}{edgelayer}
		\draw [in=-90, out=90] (4.center) to (1.center);
		\draw [in=90, out=-90] (2.center) to (5.center);
		\draw [in=90, out=-90] (19.center) to (16.center);
		\draw [in=-90, out=90] (17.center) to (20.center);
		\draw (12) to (27);
		\draw (11) to (26);
		\draw [in=105, out=-90] (8) to (28.center);
		\draw [in=75, out=-90] (30) to (29.center);
		\draw [in=255, out=90] (31) to (32.center);
		\draw [in=285, out=90] (34) to (33.center);
	\end{pgfonlayer}
\end{tikzpicture}

%% file: figures/necessity/four-partite-causal-deco_b.tikz
\begin{tikzpicture}
	\begin{pgfonlayer}{nodelayer}
		\node [style=none] (103) at (-1, 0) {$=$};
		\node [style=medium map] (104) at (1.2, 0) {$\uX$};
		\node [style=medium map] (105) at (6.2, 0) {$\uY$};
		\node [style=medium map] (106) at (3.7, 3) {$\uZ$};
		\node [style=medium map] (107) at (3.7, -3) {$\uW$};
		\node [style=none] (108) at (3.075, 2.75) {};
		\node [style=none] (109) at (4.325, 2.75) {};
		\node [style=none] (110) at (1.7, 0.25) {};
		\node [style=none] (111) at (5.7, 0.25) {};
		\node [style=none] (112) at (3.075, -2.75) {};
		\node [style=none] (113) at (4.325, -2.75) {};
		\node [style=none] (114) at (5.7, -0.25) {};
		\node [style=none] (115) at (1.7, -0.25) {};
		\node [style=none] (116) at (3.075, -4.25) {};
		\node [style=none] (117) at (3.075, -3.25) {};
		\node [style=none] (118) at (4.325, -3.25) {};
		\node [style=none] (119) at (4.325, -4.25) {};
		\node [style=none] (120) at (0.575, -4.25) {};
		\node [style=none] (121) at (0.575, 4.25) {};
		\node [style=none] (122) at (0.825, 4.25) {};
		\node [style=none] (123) at (0.825, -4.25) {};
		\node [style=none] (124) at (6.575, -4.25) {};
		\node [style=none] (125) at (6.575, 4.25) {};
		\node [style=none] (126) at (6.825, 4.25) {};
		\node [style=none] (127) at (6.825, -4.25) {};
		\node [style=none] (128) at (3.075, 3.25) {};
		\node [style=none] (129) at (3.075, 4.25) {};
		\node [style=none] (130) at (4.325, 4.25) {};
		\node [style=none] (131) at (4.325, 3.25) {};
		\node [style=small label] (132) at (0.25, -4) {$\bar\aA_1$};
		\node [style=small label] (133) at (1.175, -4) {$\aA_1$};
		\node [style=small label] (134) at (6.225, -4) {$\bar\aA_3$};
		\node [style=small label] (135) at (7.15, -4) {$\aA_3$};
		\node [style=small label] (136) at (6.225, 4) {$\bar\aB_3$};
		\node [style=small label] (137) at (7.15, 4) {$\aB_3$};
		\node [style=small label] (138) at (0.25, 4) {$\bar\aB_1$};
		\node [style=small label] (139) at (1.175, 4) {$\aB_1$};
		\node [style=small label] (140) at (2.575, -1.25) {$\aC$};
		\node [style=small label] (141) at (4.825, -1.25) {$\aD$};
		\node [style=small label] (142) at (3.4, -4) {$\bar\aA_2$};
		\node [style=small label] (143) at (4.65, -4) {$\hat\aA_4$};
		\node [style=small label] (144) at (3.4, 4) {$\bar\aB_2$};
		\node [style=small label] (145) at (4.65, 4) {$\hat\aB_4$};
		\node [style=small label] (146) at (2.2, 1.75) {$\aE$};
		\node [style=small label] (147) at (5.2, 1.75) {$\aF$};
		\node [style=large map] (148) at (-4.9, 0) {\hspace*{2cm}};
		\node [style=none] (149) at (-4.9, 0) {$\U$};
		\node [style=small label] (150) at (-6.75, -2.25) {$\bar\aA_1\aA_1$};
		\node [style=small label] (151) at (-3, -2.25) {$\bar\aA_3\aA_3$};
		\node [style=small label] (154) at (-6.75, 1.75) {$\bar\aB_1\aB_1$};
		\node [style=small label] (155) at (-3, 1.75) {$\bar\aB_3\aB_3$};
		\node [style=none] (158) at (-6.875, 1.5) {};
		\node [style=none] (159) at (-6.625, 1.5) {};
		\node [style=none] (160) at (-6.875, -1.75) {};
		\node [style=none] (161) at (-6.625, -1.75) {};
		\node [style=none] (162) at (-3.125, 1.5) {};
		\node [style=none] (163) at (-2.875, 1.5) {};
		\node [style=none] (164) at (-3.125, -1.75) {};
		\node [style=none] (165) at (-2.875, -1.75) {};
		\node [style=small label] (167) at (-4.25, -2.25) {$\hat\aA_4$};
		\node [style=small label] (168) at (-5.5, -2.25) {$\bar\aA_2$};
		\node [style=small label] (169) at (-4.25, 1.75) {$\hat\aB_4$};
		\node [style=small label] (170) at (-5.5, 1.75) {$\bar\aB_2$};
		\node [style=none] (171) at (-5.6, 1.5) {};
		\node [style=none] (172) at (-4.1, 1.5) {};
		\node [style=none] (173) at (-5.6, -1.75) {};
		\node [style=none] (174) at (-4.1, -1.75) {};
	\end{pgfonlayer}
	\begin{pgfonlayer}{edgelayer}
		\draw [in=-90, out=90] (110.center) to (108.center);
		\draw [in=90, out=-90] (109.center) to (111.center);
		\draw [in=-90, out=90] (112.center) to (115.center);
		\draw (117.center) to (116.center);
		\draw (118.center) to (119.center);
		\draw [in=90, out=-90] (114.center) to (113.center);
		\draw (120.center) to (121.center);
		\draw (122.center) to (123.center);
		\draw (124.center) to (125.center);
		\draw (126.center) to (127.center);
		\draw (129.center) to (128.center);
		\draw (130.center) to (131.center);
		\draw (160.center) to (158.center);
		\draw (159.center) to (161.center);
		\draw (164.center) to (162.center);
		\draw (163.center) to (165.center);
		\draw (173.center) to (171.center);
		\draw (172.center) to (174.center);
	\end{pgfonlayer}
\end{tikzpicture}

%% file: figures/necessity/a-state.tikz
\begin{tikzpicture}
	\begin{pgfonlayer}{nodelayer}
		\node [style=medium map] (0) at (0.15, 0) {$\uW$};
		\node [style=none] (1) at (-0.6, 1.5) {};
		\node [style=state] (2) at (-0.6, -1.5) {$i$};
		\node [style=none] (3) at (0.9, 1.5) {};
		\node [style=state] (4) at (0.9, -1.5) {$j$};
		\node [style=label] (5) at (-1, 1.25) {$\aC$};
		\node [style=label] (6) at (0.5, 1.25) {$\aD$};
	\end{pgfonlayer}
	\begin{pgfonlayer}{edgelayer}
		\draw (4) to (3.center);
		\draw (1.center) to (2);
	\end{pgfonlayer}
\end{tikzpicture}

%% file: figures/necessity/proof-of-orthogonality.tikz
\begin{tikzpicture}
	\begin{pgfonlayer}{nodelayer}
		\node [style=medium map] (50) at (5.9, 0) {$\uX$};
		\node [style=medium map] (51) at (10.9, 0) {$\uY$};
		\node [style=medium map] (52) at (8.4, 3) {$\uZ$};
		\node [style=medium map] (53) at (8.4, -3) {$\uW$};
		\node [style=none] (54) at (7.775, 2.75) {};
		\node [style=none] (55) at (9.025, 2.75) {};
		\node [style=none] (56) at (6.4, 0.25) {};
		\node [style=none] (57) at (10.4, 0.25) {};
		\node [style=none] (58) at (7.775, -2.75) {};
		\node [style=none] (59) at (9.025, -2.75) {};
		\node [style=none] (60) at (10.4, -0.25) {};
		\node [style=none] (61) at (6.4, -0.25) {};
		\node [style=none] (66) at (5.275, -3.75) {};
		\node [style=none] (67) at (5.275, 5.25) {};
		\node [style=none] (68) at (5.525, 3.375) {};
		\node [style=none] (69) at (5.525, -3.75) {};
		\node [style=none] (70) at (11.275, -3.75) {};
		\node [style=upground] (71) at (11.275, 4.25) {};
		\node [style=upground] (72) at (11.525, 4.25) {};
		\node [style=none] (73) at (11.525, -3.75) {};
		\node [style=none] (74) at (7.775, 3.25) {};
		\node [style=upground] (75) at (7.775, 4.25) {};
		\node [style=upground] (76) at (9.025, 4.25) {};
		\node [style=none] (77) at (9.025, 3.25) {};
		\node [style=small label] (78) at (4.95, -3.5) {$\bar\aA_1$};
		\node [style=small label] (79) at (5.875, -3.5) {$\aA_1$};
		\node [style=small label] (80) at (10.925, -3.5) {$\bar\aA_3$};
		\node [style=small label] (81) at (11.85, -3.5) {$\aA_3$};
		\node [style=small label] (82) at (10.925, 3.75) {$\bar\aB_3$};
		\node [style=small label] (83) at (11.85, 3.75) {$\aB_3$};
		\node [style=small label] (84) at (4.95, 5) {$\bar\aB_1$};
		\node [style=small label] (85) at (5.875, 3.125) {$\aB_1$};
		\node [style=small label] (86) at (7.3, -1.25) {$\aC$};
		\node [style=small label] (87) at (9.525, -1.25) {$\aD$};
		\node [style=small label] (90) at (8.1, 3.75) {$\bar\aB_2$};
		\node [style=small label] (91) at (9.35, 3.75) {$\hat\aB_4$};
		\node [style=state] (92) at (4.75, -4.5) {$0$};
		\node [style=state] (93) at (6, -4.5) {};
		\node [style=none] (94) at (11.275, -3.75) {};
		\node [style=none] (95) at (11.525, -3.75) {};
		\node [style=state] (96) at (10.75, -4.5) {};
		\node [style=state] (97) at (12, -4.5) {};
		\node [style=none] (98) at (6, 4.125) {};
		\node [style=upground] (99) at (6, 4.25) {};
		\node [style=medium map] (100) at (-0.075, 0.875) {$\uX$};
		\node [style=medium map] (103) at (1.9, -3) {$\uW$};
		\node [style=none] (106) at (0.525, 1.125) {};
		\node [style=none] (108) at (1.275, -2.75) {};
		\node [style=none] (109) at (2.525, -2.75) {};
		\node [style=upground] (110) at (2.525, -1.5) {};
		\node [style=none] (111) at (0.525, 0.375) {};
		\node [style=state] (112) at (1.275, -4.75) {$i$};
		\node [style=none] (113) at (1.275, -3.25) {};
		\node [style=none] (114) at (2.525, -3.25) {};
		\node [style=state] (115) at (2.525, -4.75) {$j$};
		\node [style=none] (116) at (-0.7, -0.625) {};
		\node [style=none] (117) at (-0.7, 4.25) {};
		\node [style=none] (118) at (-0.45, 2.25) {};
		\node [style=none] (119) at (-0.45, -0.625) {};
		\node [style=small label] (128) at (-1.025, -0.375) {$\bar\aA_1$};
		\node [style=small label] (129) at (-0.1, -0.375) {$\aA_1$};
		\node [style=small label] (134) at (-1.025, 4) {$\bar\aB_1$};
		\node [style=small label] (135) at (0.25, 2.575) {$\aB_1$};
		\node [style=small label] (136) at (1.05, -0.125) {$\aC$};
		\node [style=small label] (137) at (2.9, -2) {$\aD$};
		\node [style=small label] (138) at (1.6, -4.15) {$\bar\aA_2$};
		\node [style=small label] (139) at (2.85, -4.15) {$\hat\aA_4$};
		\node [style=state] (142) at (-1.225, -1.375) {$0$};
		\node [style=state] (143) at (0.025, -1.375) {};
		\node [style=none] (148) at (0.025, 3) {};
		\node [style=upground] (149) at (0.025, 3.125) {};
		\node [style=none] (150) at (0.525, 1.875) {};
		\node [style=upground] (151) at (0.525, 2) {};
		\node [style=none] (152) at (-1.85, -2.25) {};
		\node [style=none] (153) at (1.775, -2.25) {};
		\node [style=none] (154) at (1.775, 3.5) {};
		\node [style=none] (155) at (-1.85, 3.5) {};
		\node [style=small label] (156) at (6.9, 1.75) {$\aE$};
		\node [style=small label] (157) at (9.9, 1.75) {$\aF$};
		\node [style=none] (158) at (3.75, 0) {=};
		\node [style=none] (159) at (13, 0) {=};
		\node [style=large map] (160) at (16.85, 0) {\hspace*{2cm}};
		\node [style=none] (166) at (16.85, 0) {$\U$};
		\node [style=upground] (175) at (18.75, 1.75) {};
		\node [style=upground] (176) at (16.325, 1.75) {};
		\node [style=upground] (177) at (17.425, 1.75) {};
		\node [style=none] (178) at (14.875, 3.25) {};
		\node [style=none] (179) at (15.125, 1.5) {};
		\node [style=none] (182) at (18.625, 1.625) {};
		\node [style=none] (183) at (18.875, 1.625) {};
		\node [style=none] (188) at (18.65, -1.5) {};
		\node [style=none] (189) at (18.9, -1.5) {};
		\node [style=small label] (190) at (18.3, -1.25) {$\bar\aA_3$};
		\node [style=small label] (191) at (19.225, -1.25) {$\aA_3$};
		\node [style=none] (192) at (18.65, -1.5) {};
		\node [style=none] (193) at (18.9, -1.5) {};
		\node [style=state] (194) at (18.375, -2.25) {};
		\node [style=state] (195) at (19.625, -2.25) {};
		\node [style=none] (196) at (14.9, -1.5) {};
		\node [style=none] (197) at (15.15, -1.5) {};
		\node [style=small label] (198) at (14.575, -1.25) {$\bar\aA_1$};
		\node [style=small label] (199) at (15.5, -1.25) {$\aA_1$};
		\node [style=state] (200) at (14.125, -2.25) {$0$};
		\node [style=state] (201) at (15.375, -2.25) {};
		\node [style=none] (203) at (15.125, 1.5) {};
		\node [style=small label] (204) at (14.55, 2.875) {$\bar\aB_1$};
		\node [style=small label] (205) at (15.475, 1.25) {$\aB_1$};
		\node [style=none] (206) at (15.6, 2.25) {};
		\node [style=upground] (207) at (15.6, 2.375) {};
		\node [style=none] (208) at (20.5, 0) {=};
		\node [style=none] (210) at (1.275, -1.125) {};
		\node [style=state] (211) at (7.75, -5.25) {$i$};
		\node [style=none] (212) at (7.75, -3.25) {};
		\node [style=none] (213) at (9, -3.25) {};
		\node [style=state] (214) at (9, -5.25) {$j$};
		\node [style=small label] (215) at (8.075, -4.65) {$\bar\aA_2$};
		\node [style=small label] (216) at (9.325, -4.65) {$\hat\aA_4$};
		\node [style=state] (217) at (16.325, -3.3) {$i$};
		\node [style=state] (220) at (17.425, -3.3) {$j$};
		\node [style=small label] (221) at (16.65, -2.7) {$\bar\aA_2$};
		\node [style=small label] (222) at (17.75, -2.7) {$\hat\aA_4$};
		\node [style=state] (223) at (26.5, -0.75) {$i$};
		\node [style=none] (224) at (26.5, 0.5) {};
		\node [style=none] (225) at (25, 0) {=};
		\node [style=small label] (226) at (26.175, 0.125) {$\bar\aB_1$};
		\node [style=state] (251) at (23.5, -1.25) {$i$};
		\node [style=state] (252) at (22, -1.25) {$0$};
		\node [style=cnot1] (253) at (23.5, 0) {};
		\node [style=cnot2] (254) at (22, 0) {};
		\node [style=upground] (255) at (23.5, 1.25) {};
		\node [style=none] (256) at (22, 1.25) {};
		\node [style=none] (257) at (21.75, 0) {};
		\node [style=small label] (258) at (21.675, -0.625) {$\bar\aA_1$};
		\node [style=small label] (259) at (23.85, -0.625) {$\bar\aA_2$};
		\node [style=small label] (260) at (21.625, 1.025) {$\bar\aB_1$};
		\node [style=small label] (261) at (23.85, 0.625) {$\bar\aB_2$};
	\end{pgfonlayer}
	\begin{pgfonlayer}{edgelayer}
		\draw [in=-90, out=90] (56.center) to (54.center);
		\draw [in=90, out=-90] (55.center) to (57.center);
		\draw [in=-90, out=90] (58.center) to (61.center);
		\draw [in=90, out=-90] (60.center) to (59.center);
		\draw (66.center) to (67.center);
		\draw (68.center) to (69.center);
		\draw (70.center) to (71);
		\draw (72) to (73.center);
		\draw (75) to (74.center);
		\draw (76) to (77.center);
		\draw [in=90, out=-90] (66.center) to (92);
		\draw [in=90, out=-90] (69.center) to (93);
		\draw [in=90, out=-90] (94.center) to (96);
		\draw [in=90, out=-90] (95.center) to (97);
		\draw [in=-90, out=90] (68.center) to (98.center);
		\draw (113.center) to (112);
		\draw (114.center) to (115);
		\draw (110) to (109.center);
		\draw (116.center) to (117.center);
		\draw (118.center) to (119.center);
		\draw [in=90, out=-90] (116.center) to (142);
		\draw [in=90, out=-90] (119.center) to (143);
		\draw [in=-90, out=90] (118.center) to (148.center);
		\draw [in=-90, out=90] (106.center) to (150.center);
		\draw [style=graydash] (153.center)
			 to (154.center)
			 to (155.center)
			 to (152.center)
			 to cycle;
		\draw [in=90, out=-90] (192.center) to (194);
		\draw [in=90, out=-90] (193.center) to (195);
		\draw (192.center) to (182.center);
		\draw (183.center) to (193.center);
		\draw [in=90, out=-90] (196.center) to (200);
		\draw [in=90, out=-90] (197.center) to (201);
		\draw (196.center) to (178.center);
		\draw (179.center) to (197.center);
		\draw [in=-90, out=90] (203.center) to (206.center);
		\draw (108.center) to (210.center);
		\draw [in=90, out=-90] (111.center) to (210.center);
		\draw (212.center) to (211);
		\draw (213.center) to (214);
		\draw (176) to (217);
		\draw (220) to (177);
		\draw (223) to (224.center);
		\draw (255) to (251);
		\draw (256.center) to (252);
		\draw (253) to (257.center);
	\end{pgfonlayer}
\end{tikzpicture}

%% file: sections/app-proof-of-soundness.tex
\section{Proof of \texorpdfstring{\cref{prop:soundness}}{Proposition~\ref{prop:soundness}}}\label{app:proof-of-soundness}
\soundness* 

This is a standard fact that has been noted many times; see e.g.\ \cite{KHC17,LB21}.
Our order-theoretic approach allows for a compact (if perhaps tedious) way to state its proof formally.

\begin{proof}
    Let $\csP \equiv (\csP, \leq, \la,\mu)$ be the circuit shape underlying the quantum circuit $\circuit$ and denote by $\E_T$, for $T\subseteq\csP$, the channel obtained by composing $\circuit$'s gates $\E_p$ for $p\in T$ along the $\aZ_q^r$ systems they share.
    Suppose that $\lA\in\labelsA$ and $\lB\in\labelsB$ are such that $(\lA,\lB)\notin G_\circuit$; we aim to show that $(\lA,\lB)\notin G_\U$.
    Consider the set
    \begin{equation}
        \down \mu(\lB) \coloneqq \{ p \in \csP \mid p \leq \mu(\lB)\}
    \end{equation}
    of gates below output $\lB$ and the two channels $\E_{\down \mu(\lB)}$ and $\E_{\csP\setminus \down\mu(\lB)}$.
    Since $\down\mu(\lB)$ is downward-closed and $\csP\setminus\down\mu(\lB)$ is upward-closed, the domains of these channels are
    \begin{align}
        \E_{\down\mu(\lB)} &: \aA_{\la^{-1}(\down\mu(\lB))} \to \aZ \aB_{\mu^{-1}(\down\mu(\lB))} \\
        \quadand \E_{\csP\setminus\down\mu(\lB)} &: \aA_{\labelsA\setminus \la^{-1}(\down\mu(\lB))} \aZ \to \aB_{\labelsB\setminus\mu^{-1}(\down\mu(\lB))},
    \end{align}
    where $\aZ \coloneqq \bigotimes_{p\in\down\mu(\lB); q\notin\down\mu(\lB);p\cvr q} \aZ_p^q$.
    Since $\circuit$ is a circuit decomposition of $\U$, we have $\U = \E_{\csP\setminus\down\mu(\lB)} \circ \E_{\down\mu(\lB)}$:
    \begin{equation}
        \label{eq:soundness-deco}
        \input{circuits/soundness-proof.tikz}.
    \end{equation}
    Note that since $(\lA,\lB)\notin G_\circuit$,
    \begin{equation}
        \label{eq:sfdd}
        \lA \in \labelsA\setminus\la^{-1}(\down\mu(\lB)) \quadand \lB\in\mu^{-1}(\down\mu(\lB)).
    \end{equation}

    From the fact that $\E_{\csP\setminus\down\mu(\lB)}$ is trace-preserving, we get
    \begin{equation}
        \label{eq:trace-drop}
        \input{circuits/soundness-general.tikz}.
    \end{equation}
    Taking the Hilbert-Schmidt adjoint of this equation and using \cref{eq:sfdd} yields
    \begin{equation}
        \U^\dagger(\aB_\lB) \subseteq \U^{\dagger} (\aB_{\mu^{-1}(\down\mu(\lB))}) \subseteq \aA_{\la^{-1}(\down\mu(\lB))} \subseteq \aA_{\labelsA\setminus\lA}.
    \end{equation}
    By definition this means $(\lA,\lB)\notin G_\U$, completing the proof.
\end{proof}

%% file: figures/circuits/soundness-proof.tikz
\begin{tikzpicture}
	\begin{pgfonlayer}{nodelayer}
		\node [style=medium map] (0) at (-3, 0) {$\U$};
		\node [style=none] (3) at (-3.7, 1) {};
		\node [style=none] (6) at (-3.7, -1) {};
		\node [style=none] (9) at (-0.5, 0) {$=$};
		\node [style=none] (31) at (-2.3, -1) {};
		\node [style=none] (32) at (-2.3, 1) {};
		\node [style=none] (33) at (-3, -0.85) {\smallcdots};
		\node [style=none] (34) at (-3, 0.8) {\smallcdots};
		\node [style=medium map] (35) at (5.375, -1.1) {$\E_{\down\mu(\lB)}$};
		\node [style=none] (37) at (1.175, -2.1) {};
		\node [style=medium label] (39) at (1.875, -2.6) {$\aA_{\labelsA\setminus \la^{-1}(\down\mu(\lB))}$};
		\node [style=none] (40) at (2.575, -2.1) {};
		\node [style=none] (42) at (1.925, -1.95) {\smallcdots};
		\node [style=medium map] (52) at (2.15, 1.15) {$\E_{\csP\setminus\down\mu(\lB)}$};
		\node [style=none] (53) at (1.175, 0.65) {};
		\node [style=none] (54) at (2.575, 0.65) {};
		\node [style=none] (55) at (3.125, 0.65) {};
		\node [style=none] (56) at (4.475, -0.6) {};
		\node [style=none] (57) at (1.3, 2.15) {};
		\node [style=medium label] (58) at (2, 2.775) {$\aB_{\labelsB\setminus\mu^{-1}(\down\mu(\lB))}$};
		\node [style=none] (59) at (2.7, 2.15) {};
		\node [style=none] (60) at (2.05, 1.95) {\smallcdots};
		\node [style=none] (61) at (1.3, 1.65) {};
		\node [style=none] (62) at (2.7, 1.65) {};
		\node [style=none] (63) at (4.675, -2.1) {};
		\node [style=medium label] (64) at (5.375, -2.6) {$\aA_{\la^{-1}(\down\mu(\lB))}$};
		\node [style=none] (65) at (6.075, -2.1) {};
		\node [style=none] (66) at (5.425, -1.95) {\smallcdots};
		\node [style=none] (67) at (4.675, -1.6) {};
		\node [style=none] (68) at (6.075, -1.6) {};
		\node [style=none] (69) at (4.8, -0.6) {};
		\node [style=medium label] (70) at (5.5, 2.775) {$\aB_{\mu^{-1}(\down\mu(\lB))}$};
		\node [style=none] (71) at (6.2, -0.6) {};
		\node [style=none] (72) at (5.55, 1.95) {\smallcdots};
		\node [style=none] (73) at (4.8, 2.15) {};
		\node [style=none] (74) at (6.2, 2.15) {};
		\node [style=label] (75) at (3.525, -0.35) {$\aZ$};
	\end{pgfonlayer}
	\begin{pgfonlayer}{edgelayer}
		\draw (6.center) to (3.center);
		\draw (32.center) to (31.center);
		\draw (53.center) to (37.center);
		\draw (54.center) to (40.center);
		\draw [in=90, out=-90, looseness=0.75] (55.center) to (56.center);
		\draw (57.center) to (61.center);
		\draw (59.center) to (62.center);
		\draw (67.center) to (63.center);
		\draw (68.center) to (65.center);
		\draw (73.center) to (69.center);
		\draw (74.center) to (71.center);
	\end{pgfonlayer}
\end{tikzpicture}

%% file: figures/circuits/soundness-general.tikz
\begin{tikzpicture}
	\begin{pgfonlayer}{nodelayer}
		\node [style=large map] (0) at (-4.25, 0) {\hspace*{1cm}$\U$\hspace*{1cm}};
		\node [style=upground] (3) at (-6.45, 1.25) {};
		\node [style=none] (6) at (-6.45, -1.25) {};
		\node [style=none] (9) at (-0.5, 0) {$=$};
		\node [style=medium label] (15) at (-6, 1.875) {$\aB_{\labelsB\setminus\mu^{-1}(\down\mu(\lB))}$};
		\node [style=medium label] (17) at (-6, -1.75) {$\aA_{\labelsA\setminus \la^{-1}(\down\mu(\lB))}$};
		\node [style=none] (23) at (-5.05, -1.25) {};
		\node [style=upground] (24) at (-5.05, 1.25) {};
		\node [style=none] (25) at (-5.7, -1.1) {\smallcdots};
		\node [style=none] (26) at (-5.7, 0.9) {\smallcdots};
		\node [style=none] (27) at (-3.45, 1.25) {};
		\node [style=none] (28) at (-3.45, -1.25) {};
		\node [style=medium label] (29) at (-2.5, 1.875) {$\aB_{\mu^{-1}(\down\mu(\lB))}$};
		\node [style=medium label] (30) at (-2.5, -1.75) {$\aA_{\la^{-1}(\down\mu(\lB))}$};
		\node [style=none] (31) at (-2.05, -1.25) {};
		\node [style=none] (32) at (-2.05, 1.25) {};
		\node [style=none] (33) at (-2.7, -1.1) {\smallcdots};
		\node [style=none] (34) at (-2.7, 1.05) {\smallcdots};
		\node [style=medium map] (35) at (5.375, -1.1) {$\E_{\down\mu(\lB)}$};
		\node [style=none] (37) at (0.55, -2.1) {};
		\node [style=label] (39) at (1.5, -2.6) {$\aA_{\labelsA\setminus \la^{-1}(\down\mu(\lB))}$};
		\node [style=none] (40) at (1.95, -2.1) {};
		\node [style=none] (42) at (1.3, -1.95) {\smallcdots};
		\node [style=medium map] (52) at (1.65, 1.15) {$\E_{\csP\setminus\down\mu(\lB)}$};
		\node [style=none] (53) at (0.55, 0.65) {};
		\node [style=none] (54) at (1.95, 0.65) {};
		\node [style=none] (55) at (2.625, 0.65) {};
		\node [style=none] (56) at (4.575, -0.6) {};
		\node [style=upground] (57) at (0.675, 2.15) {};
		\node [style=label] (58) at (1.625, 2.775) {$\aB_{\labelsB\setminus\mu^{-1}(\down\mu(\lB))}$};
		\node [style=upground] (59) at (2.575, 2.15) {};
		\node [style=none] (60) at (1.675, 1.8) {\smallcdots};
		\node [style=none] (61) at (0.675, 1.65) {};
		\node [style=none] (62) at (2.575, 1.65) {};
		\node [style=none] (63) at (4.925, -2.1) {};
		\node [style=label] (64) at (5.375, -2.6) {$\aA_{\la^{-1}(\down\mu(\lB))}$};
		\node [style=none] (65) at (5.825, -2.1) {};
		\node [style=none] (66) at (5.425, -1.95) {\smallcdots};
		\node [style=none] (67) at (4.925, -1.6) {};
		\node [style=none] (68) at (5.825, -1.6) {};
		\node [style=none] (69) at (5.05, -0.6) {};
		\node [style=label] (70) at (5.5, 2.775) {$\aB_{\mu^{-1}(\down\mu(\lB))}$};
		\node [style=none] (71) at (5.95, -0.6) {};
		\node [style=none] (72) at (5.55, 1.95) {\smallcdots};
		\node [style=none] (73) at (5.05, 2.15) {};
		\node [style=none] (74) at (5.95, 2.15) {};
		\node [style=label] (75) at (3.45, -0.4) {$\aZ$};
		\node [style=none] (105) at (7.6, 0) {$=$};
		\node [style=medium map] (129) at (13.575, -1.1) {$\E_{\down\mu(\lB)}$};
		\node [style=none] (130) at (9.25, -2.1) {};
		\node [style=label] (131) at (9.95, -2.6) {$\aA_{\labelsA\setminus \la^{-1}(\down\mu(\lB))}$};
		\node [style=none] (132) at (10.65, -2.1) {};
		\node [style=none] (133) at (10, -1.95) {\smallcdots};
		\node [style=upground] (135) at (9.25, 0.65) {};
		\node [style=upground] (136) at (10.65, 0.65) {};
		\node [style=none] (137) at (11.95, 0.525) {};
		\node [style=none] (138) at (12.75, -0.6) {};
		\node [style=none] (145) at (13.125, -2.1) {};
		\node [style=label] (146) at (13.575, -2.6) {$\aA_{\la^{-1}(\down\mu(\lB))}$};
		\node [style=none] (147) at (14.025, -2.1) {};
		\node [style=none] (148) at (13.625, -1.95) {\smallcdots};
		\node [style=none] (149) at (13.125, -1.6) {};
		\node [style=none] (150) at (14.025, -1.6) {};
		\node [style=none] (151) at (13.25, -0.6) {};
		\node [style=label] (152) at (13.7, 2.775) {$\aB_{\mu^{-1}(\down\mu(\lB))}$};
		\node [style=none] (153) at (14.15, -0.6) {};
		\node [style=none] (154) at (13.75, 1.95) {\smallcdots};
		\node [style=none] (155) at (13.25, 2.15) {};
		\node [style=none] (156) at (14.15, 2.15) {};
		\node [style=label] (157) at (12, -0.275) {$\aZ$};
		\node [style=upground] (158) at (11.95, 0.65) {};
	\end{pgfonlayer}
	\begin{pgfonlayer}{edgelayer}
		\draw (6.center) to (3);
		\draw (24) to (23.center);
		\draw (28.center) to (27.center);
		\draw (32.center) to (31.center);
		\draw (53.center) to (37.center);
		\draw (54.center) to (40.center);
		\draw [in=90, out=-90, looseness=0.75] (55.center) to (56.center);
		\draw (57) to (61.center);
		\draw (59) to (62.center);
		\draw (67.center) to (63.center);
		\draw (68.center) to (65.center);
		\draw (73.center) to (69.center);
		\draw (74.center) to (71.center);
		\draw (135) to (130.center);
		\draw (136) to (132.center);
		\draw [in=90, out=-90] (137.center) to (138.center);
		\draw (149.center) to (145.center);
		\draw (150.center) to (147.center);
		\draw (155.center) to (151.center);
		\draw (156.center) to (153.center);
	\end{pgfonlayer}
\end{tikzpicture}